%\input psfig
%%%%%%%%%%%%%%%%%%%%%%%%%%%%%%%%%%%%%%%
%%%%%%%%%%%%%%%%%%  tex macros for preprints, cm version %%%%%%%%%%%%%%
%         (P. Ginsparg <ginsparg@lanl.gov>, last updated 7/94)
%                if confused, type `b' in response to query 
%           hypertex extensions (still provisional), 7/26/94
%
%---------------------------------------------------------------------%
%\input hyperbasics %comment out this line to restore non-hyper functionality
%
%% site dependent options:
%% \unredoffs and \redoffs define horizontal and vertical offsets
%% respectively for unreduced and reduced modes. \speclscape defines
%% the \special{} call that sets printer to landscape (sideways) mode.
%% from standard set below, leave uncommented as appropriate or redefine
%
%%% next 400dpi
\def\unredoffs{} \def\redoffs{\voffset=-.31truein\hoffset=-.48truein}
\def\speclscape{}
%\def\speclscape{\special{papersize=11in,8.5in}}
%
%%% apple lw
%\def\unredoffs{} \def\redoffs{\voffset=-.31truein\hoffset=-.59truein}
%\def\speclscape{\special{ps: landscape}}
%
%%% qms lasergrafix:
%\def\unredoffs{} \def\redoffs{\voffset=-.4truein\hoffset=.125truein}
%\def\speclscape{\special{qms: landscape}}
%
%%% saclay A4 paper:
%\def\unredoffs{\hoffset-.14truein\voffset-.2truein}
%\def\redoffs{\voffset=-.45truein\hoffset=-.21truein}
%\def\speclscape{\special{landscape}}
%
%---------------------------------------------------------------------%
%
\newbox\leftpage \newdimen\fullhsize \newdimen\hstitle \newdimen\hsbody
\tolerance=1000\hfuzz=2pt
\catcode`\@=11 % This allows us to modify PLAIN macros.
\ifx\hyperdef\UNd@FiNeD\def\hyperdef#1#2#3#4{#4}\def\hyperref#1#2#3#4{#4}\fi
\def\bigans{b }
\def\answ{b }
%\message{ big or little (b/l)? }\read-1 to\answ
%
\ifx\answ\bigans\message{(This will come out unreduced.}
\magnification=1200\unredoffs\baselineskip=16pt plus 2pt minus 1pt
\hsbody=\hsize \hstitle=\hsize %take default values for unreduced format
\else\message{(This will be reduced.} \let\l@r=L
\magnification=1000\baselineskip=16pt plus 2pt minus 1pt \vsize=7truein
\redoffs \hstitle=8truein\hsbody=4.75truein\fullhsize=10truein\hsize=\hsbody
\output={\ifnum\pageno=0 %%% This is the HUTP version
  \shipout\vbox{\speclscape{\hsize\fullhsize\makeheadline}
    \hbox to \fullhsize{\hfill\pagebody\hfill}}\advancepageno
  \else
  \almostshipout{\leftline{\vbox{\pagebody\makefootline}}}\advancepageno
  \fi}
\def\almostshipout#1{\if L\l@r \count1=1 \message{[\the\count0.\the\count1]}
      \global\setbox\leftpage=#1 \global\let\l@r=R
 \else \count1=2
  \shipout\vbox{\speclscape{\hsize\fullhsize\makeheadline}
      \hbox to\fullhsize{\box\leftpage\hfil#1}}  \global\let\l@r=L\fi}
\fi
%---------------------------------------------------------------------
%
\newcount\yearltd\yearltd=\year\advance\yearltd by -2000

\def\Title#1#2{\nopagenumbers\abstractfont\hsize=\hstitle\rightline{#1}%
\vskip 1in\centerline{\titlefont #2}\abstractfont\vskip .5in\pageno=0}
\def\Date#1{\vfill\leftline{#1}\tenpoint\supereject\global\hsize=\hsbody%
\footline={\hss\tenrm\hyperdef\hypernoname{page}\folio\folio\hss}}%
% (restores pagenumbers)
%
%       use following instead of \Date on the preliminary draft,
%       puts date/time on each page in big mode, writes labels in margins

\def\draftmode{\message{ DRAFTMODE }\def\draftdate{{\rm preliminary draft:
\number\month/\number\day/\number\yearltd\ \ \hourmin}}%
\headline={\hfil\draftdate}\writelabels\baselineskip=20pt plus 2pt minus 2pt
 {\count255=\time\divide\count255 by 60 \xdef\hourmin{\number\count255}
  \multiply\count255 by-60\advance\count255 by\time
  \xdef\hourmin{\hourmin:\ifnum\count255<10 0\fi\the\count255}}}
%       use \nolabels to get rid of eqn, ref, and fig labels in draft mode
\def\nolabels{\def\wrlabeL##1{}\def\eqlabeL##1{}\def\reflabeL##1{}}
\def\writelabels{\def\wrlabeL##1{\leavevmode\vadjust{\rlap{\smash%
{\line{{\escapechar=` \hfill\rlap{\sevenrm\hskip.03in\string##1}}}}}}}%
\def\eqlabeL##1{{\escapechar-1\rlap{\sevenrm\hskip.05in\string##1}}}%
\def\reflabeL##1{\noexpand\llap{\noexpand\sevenrm\string\string\string##1}}}
\nolabels
%
% tagged sec numbers
\global\newcount\secno \global\secno=0
\global\newcount\meqno \global\meqno=1
\def\s@csym{}
\def\newsec#1{\global\advance\secno by1%
{\toks0{#1}\message{(\the\secno. \the\toks0)}}%
%\ifx\answ\bigans \vfill\eject \else \bigbreak\bigskip \fi  %if desired
\global\subsecno=0\eqnres@t\let\s@csym\secsym\xdef\secn@m{\the\secno}\noindent
{\bf\hyperdef\hypernoname{section}{\the\secno}{\the\secno.} #1}%
\writetoca{{\string\hyperref{}{section}{\the\secno}{\it\the\secno.}} {{\it #1} }}%
\par\nobreak\medskip\nobreak}
\def\eqnres@t{\xdef\secsym{\the\secno.}\global\meqno=1\bigbreak\bigskip}
\def\sequentialequations{\def\eqnres@t{\bigbreak}}\xdef\secsym{}
\global\newcount\subsecno \global\subsecno=0
\def\subsec#1{\global\advance\subsecno by1%
{\toks0{#1}\message{(\s@csym\the\subsecno. \the\toks0)}}%
\ifnum\lastpenalty>9000\else\bigbreak\fi       \global\subsubsecno=0
\noindent{\it\hyperdef\hypernoname{subsection}{\secn@m.\the\subsecno}%
{\secn@m.\the\subsecno.} #1}\writetoca{\string\quad
{\string\hyperref{}{subsection}{\secn@m.\the\subsecno}{\secn@m.\the\subsecno.}}
{#1}}\par\nobreak\medskip\nobreak}
\def\appendix#1#2{\global\meqno=1\global\subsecno=0\xdef\secsym{\hbox{#1.}}%
\bigbreak\bigskip\noindent{\bf Appendix \hyperdef\hypernoname{appendix}{#1}%
{#1.} #2}{\toks0{(#1. #2)}\message{\the\toks0}}%
\xdef\s@csym{#1.}\xdef\secn@m{#1}%
\writetoca{\string\hyperref{}{appendix}{#1}{{\it Appendix} {\it #1.}} {\it #2}}%
\par\nobreak\medskip\nobreak}
%
%       \eqn\label{a+b=c}	gives displayed equation, numbered
%				consecutively within sections.
%     \eqnn and \eqna define labels in advance (of eqalign?)
%
\def\checkm@de#1#2{\ifmmode{\def\f@rst##1{##1}\hyperdef\hypernoname{equation}%
{#1}{#2}}\else\hyperref{}{equation}{#1}{#2}\fi}
\def\eqnn#1{\DefWarn#1\xdef #1{(\noexpand\relax\noexpand\checkm@de%
{\s@csym\the\meqno}{\secsym\the\meqno})}%
\wrlabeL#1\writedef{#1\leftbracket#1}\global\advance\meqno by1}
\def\f@rst#1{\c@t#1a\em@ark}\def\c@t#1#2\em@ark{#1}
\def\eqna#1{\DefWarn#1\wrlabeL{#1$\{\}$}%
\xdef #1##1{(\noexpand\relax\noexpand\checkm@de%
{\s@csym\the\meqno\noexpand\f@rst{##1}}{\hbox{$\secsym\the\meqno##1$}})}
\writedef{#1\numbersign1\leftbracket#1{\numbersign1}}\global\advance\meqno by1}
\def\eqn#1#2{\DefWarn#1%
\xdef #1{(\noexpand\hyperref{}{equation}{\s@csym\the\meqno}%
{\secsym\the\meqno})}$$#2\eqno(\hyperdef\hypernoname{equation}%
{\s@csym\the\meqno}{\secsym\the\meqno})\eqlabeL#1$$%
\writedef{#1\leftbracket#1}\global\advance\meqno by1}
\def\xeqn{\expandafter\xe@n}\def\xe@n(#1){#1}
\def\xeqna#1{\expandafter\xe@n#1}
\def\eqns#1{(\e@ns #1{\hbox{}})}
\def\e@ns#1{\ifx\UNd@FiNeD#1\message{eqnlabel \string#1 is undefined.}%
\xdef#1{(?.?)}\fi{\let\hyperref=\relax\xdef\next{#1}}%
\ifx\next\em@rk\def\next{}\else%
\ifx\next#1\xeqn#1\else\def\n@xt{#1}\ifx\n@xt\next#1\else\xeqna#1\fi
\fi\let\next=\e@ns\fi\next}

\def\DefWarn#1{\ifx\UNd@FiNeD#1\else
\immediate\write16{*** WARNING: the label \string#1 is already defined ***}\fi}
%
%			 footnotes
\newskip\footskip\footskip14pt plus 1pt minus 1pt %sets footnote baselineskip
\def\footnotefont{\ninepoint}\def\f@t#1{\footnotefont #1\@foot}
\def\f@@t{\baselineskip\footskip\bgroup\footnotefont\aftergroup\@foot\let\next}
\setbox\strutbox=\hbox{\vrule height9.5pt depth4.5pt width0pt}
\global\newcount\ftno \global\ftno=0
\def\foot{\global\advance\ftno by1\def\foot@rg{\hyperref{}{footnote}%
{\the\ftno}{\the\ftno}\xdef\foot@rg{\noexpand\hyperdef\noexpand\hypernoname%
{footnote}{\the\ftno}{\the\ftno}}}\footnote{$^{\foot@rg}$}}
%
%say \footend to put footnotes at end
%will cause problems if \ref used inside \foot, instead use \nref before
\newwrite\ftfile
\def\footend{\def\foot{\global\advance\ftno by1\chardef\wfile=\ftfile
%%$^{\the\ftno}$\ifnum\ftno=1\immediate\openout\ftfile=\jobname.fts\fi%
\hyperref{}{footnote}{\the\ftno}{$^{\the\ftno}$}%
\ifnum\ftno=1\immediate\openout\ftfile=\jobname.fts\fi%
\immediate\write\ftfile{\noexpand\smallskip%
%%\noexpand\item{f\the\ftno:\ }\pctsign}\findarg}%
\noexpand\item{\noexpand\hyperdef\noexpand\hypernoname{footnote}
{\the\ftno}{f\the\ftno}:\ }\pctsign}\findarg}%
\def\footatend{\vfill\eject\immediate\closeout\ftfile{\parindent=20pt
\centerline{\bf Footnotes}\nobreak\bigskip\input \jobname.fts }}}
\def\footatend{}
%
%     \ref\label{text}
% generates a number, assigns it to \label, generates an entry.
% To list the refs on a separate page,  \listrefs
%
\global\newcount\refno \global\refno=1
\newwrite\rfile
\def\ref{[\hyperref{}{reference}{\the\refno}{\the\refno}]\nref}
\def\nref#1{\DefWarn#1%
\xdef#1{[\noexpand\hyperref{}{reference}{\the\refno}{\the\refno}]}%
\writedef{#1\leftbracket#1}%
\ifnum\refno=1\immediate\openout\rfile=\jobname.refs\fi
\chardef\wfile=\rfile\immediate\write\rfile{\noexpand\item{[\noexpand\hyperdef%
\noexpand\hypernoname{reference}{\the\refno}{\the\refno}]\ }%
\reflabeL{#1\hskip.31in}\pctsign}\global\advance\refno by1\findarg}
%	horrible hack to sidestep tex \write limitation
\def\findarg#1#{\begingroup\obeylines\newlinechar=`\^^M\pass@rg}
{\obeylines\gdef\pass@rg#1{\writ@line\relax #1^^M\hbox{}^^M}%
\gdef\writ@line#1^^M{\expandafter\toks0\expandafter{\striprel@x #1}%
\edef\next{\the\toks0}\ifx\next\em@rk\let\next=\endgroup\else\ifx\next\empty%
\else\immediate\write\wfile{\the\toks0}\fi\let\next=\writ@line\fi\next\relax}}
\def\striprel@x#1{} \def\em@rk{\hbox{}}
\def\lref{\begingroup\obeylines\lr@f}
\def\lr@f#1#2{\DefWarn#1\gdef#1{\let#1=\UNd@FiNeD\ref#1{#2}}\endgroup\unskip}

\def\addref#1{\immediate\write\rfile{\noexpand\item{}#1}} %now unnecessary
\def\listrefs{\footatend\vfill\supereject\immediate\closeout\rfile\writestoppt
\baselineskip=\footskip\centerline{{\bf References}}\bigskip{\parindent=20pt%
\frenchspacing\escapechar=` \input \jobname.refs\vfill\eject}\nonfrenchspacing}
\def\startrefs#1{\immediate\openout\rfile=\jobname.refs\refno=#1}
\def\xref{\expandafter\xr@f}\def\xr@f[#1]{#1}
\def\refs#1{\count255=1[\r@fs #1{\hbox{}}]}
\def\r@fs#1{\ifx\UNd@FiNeD#1\message{reflabel \string#1 is undefined.}%
\nref#1{need to supply reference \string#1.}\fi%
\vphantom{\hphantom{#1}}{\let\hyperref=\relax\xdef\next{#1}}%
\ifx\next\em@rk\def\next{}%
\else\ifx\next#1\ifodd\count255\relax\xref#1\count255=0\fi%
\else#1\count255=1\fi\let\next=\r@fs\fi\next}
%

%
% this is ugly, but moore insists
\newwrite\ffile\global\newcount\figno \global\figno=1
\def\fig{fig.~\hyperref{}{figure}{\the\figno}{\the\figno}\nfig}
\def\nfig#1{\DefWarn#1%
\xdef#1{fig.~\noexpand\hyperref{}{figure}{\the\figno}{\the\figno}}%
\writedef{#1\leftbracket fig.\noexpand~\xfig#1}%
\ifnum\figno=1\immediate\openout\ffile=\jobname.figs\fi\chardef\wfile=\ffile%
{\let\hyperref=\relax
\immediate\write\ffile{\noexpand\medskip\noexpand\item{Fig.\ %
\noexpand\hyperdef\noexpand\hypernoname{figure}{\the\figno}{\the\figno}. }
\reflabeL{#1\hskip.55in}\pctsign}}\global\advance\figno by1\findarg}
\def\listfigs{\vfill\eject\immediate\closeout\ffile{\parindent40pt
\baselineskip14pt\centerline{{\bf Figure Captions}}\nobreak\medskip
\escapechar=` \input \jobname.figs\vfill\eject}}
\def\xfig{\expandafter\xf@g}\def\xf@g fig.\penalty\@M\ {}
\def\figs#1{figs.~\f@gs #1{\hbox{}}}
\def\f@gs#1{{\let\hyperref=\relax\xdef\next{#1}}\ifx\next\em@rk\def\next{}\else
\ifx\next#1\xfig #1\else#1\fi\let\next=\f@gs\fi\next}
\def\figin{\epsfcheck\figin}\def\figins{\epsfcheck\figins}
\def\epsfcheck{\ifx\epsfbox\UNd@FiNeD
\message{(NO epsf.tex, FIGURES WILL BE IGNORED)}
\gdef\figin##1{\vskip2in}\gdef\figins##1{\hskip.5in}% blank space instead
\else\message{(FIGURES WILL BE INCLUDED)}%
\gdef\figin##1{##1}\gdef\figins##1{##1}\fi}
\def\DefWarn#1{}
\def\figinsert{\goodbreak\midinsert}
\def\ifig#1#2#3{\DefWarn#1\xdef#1{Fig.~\noexpand\hyperref{}{figure}%
{\the\figno}{\the\figno}}\writedef{#1\leftbracket fig.\noexpand~\xfig#1}%
\figinsert\figin{\centerline{#3}}\medskip\centerline{\vbox{\baselineskip12pt
\advance\hsize by -1truein\noindent\wrlabeL{#1=#1}\footnotefont%
{\bf Fig.~\hyperdef\hypernoname{figure}{\the\figno}{\the\figno}:} #2}}
\bigskip\endinsert\global\advance\figno by1}
\newwrite\lfile
{\escapechar-1\xdef\pctsign{\string\%}\xdef\leftbracket{\string\{}
\xdef\rightbracket{\string\}}\xdef\numbersign{\string\#}}
\def\writedefs{\immediate\openout\lfile=\jobname.defs \def\writedef##1{%
{\let\hyperref=\relax\let\hyperdef=\relax\let\hypernoname=\relax
 \immediate\write\lfile{\string\def\string##1\rightbracket}}}}%
\def\writestop{\def\writestoppt{\immediate\write\lfile{\string\pageno
 \the\pageno\string\startrefs\leftbracket\the\refno\rightbracket
 \string\def\string\secsym\leftbracket\secsym\rightbracket
 \string\secno\the\secno\string\meqno\the\meqno}\immediate\closeout\lfile}}
\def\writestoppt{}\def\writedef#1{}
\def\seclab#1{\DefWarn#1%
\xdef #1{\noexpand\hyperref{}{section}{\the\secno}{\the\secno}}%
\writedef{#1\leftbracket#1}\wrlabeL{#1=#1}}
\def\subseclab#1{\DefWarn#1%
\xdef #1{\noexpand\hyperref{}{subsection}{\secn@m.\the\subsecno}%
{\secn@m.\the\subsecno}}\writedef{#1\leftbracket#1}\wrlabeL{#1=#1}}
\def\applab#1{\DefWarn#1%
\xdef #1{\noexpand\hyperref{}{appendix}{\secn@m}{\secn@m}}%
\writedef{#1\leftbracket#1}\wrlabeL{#1=#1}}
\newwrite\tfile \def\writetoca#1{}
\def\leaderfill{\leaders\hbox to 1em{\hss.\hss}\hfill}
%	use this to write file with table of contents
\def\writetoc{\immediate\openout\tfile=\jobname.toc
   \def\writetoca##1{{\edef\next{\write\tfile{\noindent ##1
   \string\leaderfill {\string\hyperref{}{page}{\noexpand\number\pageno}%
                       {\noexpand\number\pageno}} \par}}\next}}}
%       and this lists table of contents on second pass
\newread\ch@ckfile
\def\listtoc{\immediate\closeout\tfile\immediate\openin\ch@ckfile=\jobname.toc
\ifeof\ch@ckfile\message{no file \jobname.toc, no table of contents this pass}%
\else\closein\ch@ckfile\centerline{\bf Contents}\nobreak\medskip%
{\baselineskip=18.5pt  \footnotefont
\parskip=2pt\catcode`\@=12\input\jobname.toc
\catcode`\@=12\bigbreak\bigskip}\fi}
\catcode`\@=12 % at signs are no longer letters
%
%	Unpleasantness in calling in abstract and title fonts
\edef\tfontsize{\ifx\answ\bigans scaled\magstep3\else scaled\magstep4\fi}
\font\titlerm=cmr10 \tfontsize \font\titlerms=cmr7 \tfontsize
\font\titlermss=cmr5 \tfontsize \font\titlei=cmmi10 \tfontsize
\font\titleis=cmmi7 \tfontsize \font\titleiss=cmmi5 \tfontsize
\font\titlesy=cmsy10 \tfontsize \font\titlesys=cmsy7 \tfontsize
\font\titlesyss=cmsy5 \tfontsize \font\titleit=cmti10 \tfontsize
\skewchar\titlei='177 \skewchar\titleis='177 \skewchar\titleiss='177
\skewchar\titlesy='60 \skewchar\titlesys='60 \skewchar\titlesyss='60
\def\titlefont{\def\rm{\fam0\titlerm}% switch to title font
\textfont0=\titlerm \scriptfont0=\titlerms \scriptscriptfont0=\titlermss
\textfont1=\titlei \scriptfont1=\titleis \scriptscriptfont1=\titleiss
\textfont2=\titlesy \scriptfont2=\titlesys \scriptscriptfont2=\titlesyss
\textfont\itfam=\titleit \def\it{\fam\itfam\titleit}\rm}
 \ifx\answ\bigans\else scaled\magstep1\fi
\ifx\answ\bigans\def\abstractfont{\tenpoint}\else
\font\absit=cmti10 scaled \magstep1
\font\abssl=cmsl10 scaled \magstep1
\font\absrm=cmr10 scaled\magstep1 \font\absrms=cmr7 scaled\magstep1
\font\absrmss=cmr5 scaled\magstep1 \font\absi=cmmi10 scaled\magstep1
\font\absis=cmmi7 scaled\magstep1 \font\absiss=cmmi5 scaled\magstep1
\font\abssy=cmsy10 scaled\magstep1 \font\abssys=cmsy7 scaled\magstep1
\font\abssyss=cmsy5 scaled\magstep1 \font\absbf=cmbx10 scaled\magstep1
\skewchar\absi='177 \skewchar\absis='177 \skewchar\absiss='177
\skewchar\abssy='60 \skewchar\abssys='60 \skewchar\abssyss='60
\def\abstractfont{\def\rm{\fam0\absrm}% switch to abstract font
\textfont0=\absrm \scriptfont0=\absrms \scriptscriptfont0=\absrmss
\textfont1=\absi \scriptfont1=\absis \scriptscriptfont1=\absiss
\textfont2=\abssy \scriptfont2=\abssys \scriptscriptfont2=\abssyss
\textfont\itfam=\absit \def\it{\fam\itfam\absit}\def\footnotefont{\tenpoint}%
\textfont\slfam=\abssl \def\sl{\fam\slfam\abssl}%
\textfont\bffam=\absbf \def\bf{\fam\bffam\absbf}\rm}\fi
\def\tenpoint{\def\rm{\fam0\tenrm}% switch back to 10-point type
\textfont0=\tenrm \scriptfont0=\sevenrm \scriptscriptfont0=\fiverm
\textfont1=\teni  \scriptfont1=\seveni  \scriptscriptfont1=\fivei
\textfont2=\tensy \scriptfont2=\sevensy \scriptscriptfont2=\fivesy
\textfont\itfam=\tenit \def\it{\fam\itfam\tenit}\def\footnotefont{\ninepoint}%
\textfont\bffam=\tenbf \def\bf{\fam\bffam\tenbf}\def\sl{\fam\slfam\tensl}\rm}
\font\ninerm=cmr9 \font\sixrm=cmr6 \font\ninei=cmmi9 \font\sixi=cmmi6
\font\ninesy=cmsy9 \font\sixsy=cmsy6 \font\ninebf=cmbx9
\font\nineit=cmti9 \font\ninesl=cmsl9 \skewchar\ninei='177
\skewchar\sixi='177 \skewchar\ninesy='60 \skewchar\sixsy='60
\def\ninepoint{\def\rm{\fam0\ninerm}% switch to footnote font
\textfont0=\ninerm \scriptfont0=\sixrm \scriptscriptfont0=\fiverm
\textfont1=\ninei \scriptfont1=\sixi \scriptscriptfont1=\fivei
\textfont2=\ninesy \scriptfont2=\sixsy \scriptscriptfont2=\fivesy
\textfont\itfam=\ninei \def\it{\fam\itfam\nineit}\def\sl{\fam\slfam\ninesl}%
\textfont\bffam=\ninebf \def\bf{\fam\bffam\ninebf}\rm}
%
%---------------------------------------------------------------------
%
\def\noblackbox{\overfullrule=0pt}
\hyphenation{anom-aly anom-alies coun-ter-term coun-ter-terms}
\def\inv{^{\raise.15ex\hbox{${\scriptscriptstyle -}$}\kern-.05em 1}}

\def\Dsl{\,\raise.15ex\hbox{/}\mkern-13.5mu D} %this one can be subscripted
\def\dsl{\raise.15ex\hbox{/}\kern-.57em\partial}

 %pound sterling
\def\lspace{\ifx\answ\bigans{}\else\qquad\fi}
\def\lbspace{\ifx\answ\bigans{}\else\hskip-.2in\fi} % $$\lbspace...$$
\def\boxeqn#1{\vcenter{\vbox{\hrule\hbox{\vrule\kern3pt\vbox{\kern3pt
	\hbox{${\displaystyle #1}$}\kern3pt}\kern3pt\vrule}\hrule}}}
\def\mbox#1#2{\vcenter{\hrule \hbox{\vrule height#2in
		\kern#1in \vrule} \hrule}}  %e.g. \mbox{.1}{.1}
%	matters of taste
%\def\tilde{\widetilde} \def\bar{\overline} \def\hat{\widehat}
%
% some sample definitions
  %     curly letters

\def\vev#1{\langle #1 \rangle}

\def\darr#1{\raise1.5ex\hbox{$\leftrightarrow$}\mkern-16.5mu #1}
 %pound sterling

 %puts a small half in a displayed eqn
\def\roughly#1{\raise.3ex\hbox{$#1$\kern-.75em\lower1ex\hbox{$\sim$}}}

%%%%%%%%%%%%%%%%%%%%%%%%%%%%%%%%%%%%%%%%%%%%%%%%%%%%%%%%%%%%%%%%%%%%%
%%%%%%%%%%%%%%%   Subsubsection  %%%%%%%%%%%%%%%%%%%%%%%%%%%%%%%%%%%%
%%%%%%%%%%%%%%%%%%%%%%%%%%%%%%%%%%%%%%%%%%%%%%%%%%%%%%%%%%%%%%%%%%%%%
\global\newcount\subsubsecno \global\subsubsecno=0
\def\subsubsec#1{\global\advance\subsubsecno by1%
{\toks0{#1}\message{(\the\secno\the\subsecno\the\subsubsecno. \the\toks0)}}%
\ifnum\lastpenalty>9000\else\bigbreak\fi
\noindent{\it\hyperdef\hypernoname{subsubsection}{\the\secno.\the\subsecno\the\subsubsecno}%
{\the\secno.\the\subsecno.\the\subsubsecno.} #1}
%%% Add Subsubsections to Index
%% \writetoca{\string\quad{\string\hyperref{}{subsubsection}{\the\secno\the\subsecno\the
%%\subsubsecno}{\baselineskip=9pt\it\the\secno.\the\subsecno.\the\subsubsecno.}}
%% {\baselineskip=9pt\it\ #1}}
\par\nobreak\medskip\nobreak}
%%%%%%%%%%%%%%%%%%%%%%%%%%%%%%%%%%%%%%%%%%%%%%%%%%%%%%%%%%%%%%%%%%%%%
%%%%%%%%%%%%%%%%%%%%%%%%%%%%%%%%%%%%%%%%%%%%%%%%%%%%%%%%%%%%%%%%%%%
%%%%%% BOX
%%%%%%%%%%%%%%%%%%%%%%%%%%%%%%%%%%%%%%%%%%
\def\boxit#1{\vbox{\hrule\hbox{\vrule\kern8pt
\vbox{\hbox{\kern8pt}\hbox{\vbox{#1}}\hbox{\kern8pt}}
\kern8pt\vrule}\hrule}}
\def\mathboxit#1{\vbox{\hrule\hbox{\vrule\kern8pt\vbox{\kern8pt
\hbox{$\displaystyle #1$}\kern8pt}\kern8pt\vrule}\hrule}}
%%%%%%%%%%%%%%%%%%%%%%%%%%%%%%%%%%%%%%%%%%%%%%%%%%%%%%%%%%%%%%%%%%%
%%%%%%%%%%%%%%%%%%%%%%%%%%%%%%%%%%%%%%%%%%%%%%%%%%%%%%%%%%%%%%%%
%%%%%   Dirac-Slash
%%%%%%%%%%%%%%%%%%%%%%%%%%%%%%%%%%%%%%%%%%%%%%%%%%%%%%%%%%%%%%%%
\def\slashchar#1{\setbox0=\hbox{$#1$}           % set a box for #1
   \dimen0=\wd0                                 % and get its size
   \setbox1=\hbox{/} \dimen1=\wd1               % get size of /
   \ifdim\dimen0>\dimen1                        % #1 is bigger
      \rlap{\hbox to \dimen0{\hfil/\hfil}}      % so center / in box
      #1                                        % and print #1
   \else                                        % / is bigger
      \rlap{\hbox to \dimen1{\hfil$#1$\hfil}}   % so center #1
      /                                         % and print /
   \fi}
%%%%%%%%%%%%%%%%%%%%%%%%%%%%%%%%%%%%%%%%%%%%%%%%%%%%%%%%%%%%%%%%%
%%%%%%%%%%%%%%%%%%%%%%%%%%%%%%%%%%%%%%%%%%%%%%%%%%%%%%%%%%%
%  To produce a box for a Dalembertian (adapted from p. 320 of TeXbook):
\def\sqr#1#2{{\vcenter{\vbox{\hrule height.#2pt
         \hbox{\vrule width.#2pt height#1pt \kern#1pt
            \vrule width.#2pt}
         \hrule height.#2pt}}}}
\def\square{\mathop{\mathchoice\sqr56\sqr56\sqr{3.75}4\sqr34}\nolimits}
%%%%%%%%%%%%%%%%%%%%%%%%%%%%%%%%%%%%%%%%%%%%%%%%%%%%%%%%%%%

%%%%%%%%%%%%%%%%%%%%%%%%%%%%%%%%%%%%%%%%
%%%%%  load AMS-fonts cf. TeX book p.158  ** can be downloaded from http://arxiv.org/macros/
\input amssym.def
\input amssym.tex
%%% \eqq : $\frak A$  $\Bbb  A$
%%%\input amstex.tex
%%%  e.g.: $\Cal A$
%%%\let\footnote\plainfootnote
%%%%%%%%%%%%%%%%%%%%%%%%%%%%%%%%%%%%%%%%%%%%%%%%%%%%%%%%%%%%%%%%%%%%%
\noblackbox
%%\draftmode  
%%%%%%%%%%%%%%%% Lineskip
%%%%%%%%%%%%%%%%%%%%%%%%%%%%%%%%%%%%%%%
\baselineskip=14.5pt
%%%%%%%%%%%% Local definitions %%%%%%%%%%%%%%%%%%%%%%%%%%%%%%%%%%%%

\def\ds#1{{\displaystyle{#1}}}
\def\comment#1{{}}

\def\ap{\alpha'}

\def\ie{{\it i.e.\ }}

\def\al{\alpha}

\def\si{\sigma}

\def\bet{\beta}

\def\fs{{\goth s}}
\def\Ar{{\rm A}}\def\Br{{\rm B}}

%%%%%%%%%%%%%%%%%%%%%%%%%%%%%%%%%%%%%%%%%%%%%%%%%%%%%%%%%%%%%%%%%
%%%%% Referencing  %%%%%%%%%%%%%%%%%%%%%%%%%%%%%%%%%%%%%%%%%%%%%%
%%%%%%%%%%%%%%%%%%%%%%%%%%%%%%%%%%%%%%%%%%%%%%%%%%%%%%%%%%%%%%%%%
\newif\ifnref

\def\doubref#1#2{\refs{{#1},{#2} }}

\nreffalse
%%%%%%%%%%%%%%%%%%%%%%%%%%%%%%%%%%%%%%%%%%%%%%%%%%%%%%%%%%%%%%%%%%
%%%%%%%%%%%%%%%%%   Stuff for Figures  %%%%%%%%%%%%%%%%%%%%%%%%%%%
%%%%%%%%%%%%%%%%%%%%%%%%%%%%%%%%%%%%%%%%%%%%%%%%%%%%%%%%%%%%%%%%%%

\input epsf

\def\figin{\epsfcheck\figin}\def\figins{\epsfcheck\figins}
\def\epsfcheck{\ifx\epsfbox\UnDeFiNeD
\message{(NO epsf.tex, FIGURES WILL BE IGNORED)}
\gdef\figin##1{\vskip2in}\gdef\figins##1{\hskip.5in}% blank space instead
\else\message{(FIGURES WILL BE INCLUDED)}%
\gdef\figin##1{##1}\gdef\figins##1{##1}\fi}
\def\DefWarn#1{}
\def\figinsert{\goodbreak\midinsert}  % instead \topinsert
\def\ifig#1#2#3{\DefWarn#1\def#1{Fig.~\the\figno}
\writedef{#1\leftbracket fig.\noexpand~\the\figno}%
\figinsert\figin{\centerline{#3}}\medskip\centerline{\vbox{\baselineskip12pt
\advance\hsize by -1truein\noindent\footnotefont\centerline{{\bf
Fig.~\the\figno}\ \sl #2}}}
\bigskip\endinsert\global\advance\figno by1}

%%%%%%%%  Second line in Figure caption
\def\iifig#1#2#3#4{\DefWarn#1\xdef#1{Fig.~\the\figno}
\writedef{#1\leftbracket fig.\noexpand~\the\figno}%
\figinsert\figin{\centerline{#4}}\medskip\centerline{\vbox{\baselineskip12pt
\advance\hsize by -1truein\noindent\footnotefont\centerline{{\bf
Fig.~\the\figno}\ \ \sl #2}}}\smallskip\centerline{\vbox{\baselineskip12pt
\advance\hsize by -1truein\noindent\footnotefont\centerline{\ \ \ \sl #3}}}
\bigskip\endinsert\global\advance\figno by1}

%%%%%%%%%%%%%%%%%%%%%%%%%%%%%%%%%%%%%%%%%%%%%%%%%%%%%%%%%%%%%%%%%%%%%
%%%%%%%%%%%%%%%   Standard alltime definitions   %%%%%%%%%%%%%%%%%%%%
%%%%%%%%%%%%%%%%%%%%%%%%%%%%%%%%%%%%%%%%%%%%%%%%%%%%%%%%%%%%%%%%%%%%%

\def\appA{A}
\def\appB{B}

\def\tilde{\widetilde}
\def\hatt{\widehat}
\def\h {{1\over 2}}

\def\ov {\overline}
\def\o {\over}
\def\fc#1#2{{#1 \o #2}}

\def\IC{{\bf C}}\def\IR{ {\bf R}}

\def\E {\hatt E}      % For Eisenstein E2
  % For Polylogarithm

\def\br{\hfill\break}

\def\det {{\rm det}}
\def\mod {{\rm mod}}
\def\lf {\left}
\def\ri {\right}
\def\ra {\rightarrow}

\def\p {\partial}

\def\Bc {{\cal B}}\def\Dc{{\cal D}}

\def\Cc {{\cal C}} \def\Oc {{\cal O}}
\def\Lc {{\cal L}} 
 \def\Ac {{\cal A}}
\def\Pc {{\cal P}} \def\Tc {{\cal T}}
 
\def\Ic {{\cal I}}

%%\def\Acb{{\pmb{\Cal A}}}

%%%%%%%%  My Shuffle Product

%%%%%%%%%%%%%%%%%%%%%%%%%%%%%%%%%%%%%%%%%%%%%%%%%%%%%%%%%%%%%%%%%%%

%%%%%%%%%%%%%%%%%%%%%%%%%%%%%%%%%%%%%%%%%%%%%%%%%%%%%%%%%%%%%%%%

%%%%%%%%%% Stephan's New Defs.

\def\lf{\left}
\def\ri{\right}

\def\ra{{\rightarrow}}

\def\ds{\displaystyle}
%\def\floor[1]{\lfloor#1\rfloor}
%\def\ceiling[1]{\lceil#1\rceil}

%%%%%%%%%%%%%%%%%%
\lref\SrisangyingcharoenLHX{
  P.~Srisangyingcharoen and P.~Mansfield,
``Plahte Diagrams for String Scattering Amplitudes,''
JHEP {\bf 2104}, 017 (2021).
[arXiv:2005.01712 [hep-th]].
%%CITATION = arXiv:2005.01712%%
}
\lref\MaUM{
  Q.~Ma, Y.J.~Du and Y.X.~Chen,
``On Primary Relations at Tree-level in String Theory and Field Theory,''
JHEP {\bf 1202}, 061 (2012).
[arXiv:1109.0685 [hep-th]].
%%CITATION = arXiv:1109.0685%%
}

\lref\StiebergerHQ{
  S.~Stieberger,
``Open \& Closed vs. Pure Open String Disk Amplitudes,''
[arXiv:0907.2211 [hep-th]].
%%CITATION = MPP-2008-01%%
}

\lref\CohenPV{
  A.G.~Cohen, G.W.~Moore, P.C.~Nelson and J.~Polchinski,
``Semi Off-shell String Amplitudes,''
Nucl.\ Phys.\ B {\bf 281}, 127 (1987).
%%CITATION = HUTP-86/A028%%
}

\lref\DHokerPDL{
  E.D'Hoker and D.H.~Phong,
  ``The Geometry of String Perturbation Theory,''
Rev.\ Mod.\ Phys.\  {\bf 60}, 917 (1988).
%%CITATION = PUPT-1039%%
}

\lref\NeveuIQ{
  A.~Neveu and J.~Scherk,
``Parameter-free regularization of one-loop unitary dual diagram,''
Phys.\ Rev.\ D {\bf 1}, 2355 (1970).
}

\lref\HsueRA{
  C.S.~Hsue, B.~Sakita and M.A.~Virasoro,
``Formulation of dual theory in terms of functional integrations,''
Phys.\ Rev.\ D {\bf 2}, 2857 (1970).
}

\lref\AbouelsaoodGD{
  A.~Abouelsaood, C.G.~Callan, Jr., C.R.~Nappi and S.A.~Yost,
 ``Open Strings in Background Gauge Fields,''
Nucl.\ Phys.\ B {\bf 280}, 599 (1987).
%%CITATION = Print-86-1189 (PRINCETON)%%
}

\lref\HoheneggerKQY{
  S.~Hohenegger and S.~Stieberger,
``Monodromy Relations in Higher-Loop String Amplitudes,''
Nucl.\ Phys.\ B {\bf 925}, 63 (2017).
[arXiv:1702.04963 [hep-th]].
%%CITATION = MPP-2017-001%%
}

\lref\PolchinskiFM{
  J.~Polchinski, S.~Chaudhuri and C.V.~Johnson,
``Notes on D-branes,''
[hep-th/9602052];\br
%%CITATION = hep-th/9602052%%
  J.~Polchinski,
``Tasi lectures on D-branes,''
[hep-th/9611050].
%%CITATION = hep-th/9611050%%
}

\lref\AntoniadisVW{
  I.~Antoniadis, C.~Bachas, C.~Fabre, H.~Partouche and T.R.~Taylor,
``Aspects of type I - type II - heterotic triality in four-dimensions,''
Nucl.\ Phys.\ B {\bf 489}, 160 (1997).
[hep-th/9608012].
%%CITATION = hep-th/9608012%%
}

\lref\WeinbergKQ{
  S.~Weinberg and E.~Witten,
``Limits on Massless Particles,''
Phys.\ Lett.\  {\bf 96B}, 59 (1980).
%%CITATION = HUTP-80/A056%%
}

\lref\MOS{W. Magnus, F. Oberhettinger, and R.P. Soni,
{\it Formulas and Theorems for the Special Functions of Mathematical Physics}, Springer 1966;\br
A. Erd\'elyi, W. Magnus, F. Oberhettinger, and F.G. Tricomi,
{\it Higher transcendental functions}, Volume II, McGraw-Hill Book Company 1953. }

\lref\StiebergerVYA{
  S.~Stieberger and T.R.~Taylor,
``Disk Scattering of Open and Closed Strings (I),''
Nucl.\ Phys.\ B {\bf 903}, 104 (2016).
[arXiv:1510.01774 [hep-th]].
%%CITATION = MPP-2015-184%%
}

\lref\GubserWT{
  S.S.~Gubser, A.~Hashimoto, I.R.~Klebanov and J.M.~Maldacena,
``Gravitational lensing by p-branes,''
Nucl.\ Phys.\ B {\bf 472}, 231 (1996).
[hep-th/9601057].
%%CITATION = hep-th/9601057%%
}

\lref\ForgerTU{
  K.~F\"orger and S.~Stieberger,
``String amplitudes and N=2, $d\!\!=\!\!4$ prepotential in heterotic $K_3 \times T^2$ compactifications,''
Nucl.\ Phys.\ B {\bf 514}, 135 (1998).
[hep-th/9709004].
%%CITATION = hep-th/9709004%%
}

\lref\HoheneggerKQY{
  S.~Hohenegger and S.~Stieberger,
``Monodromy Relations in Higher-Loop String Amplitudes,''
Nucl.\ Phys.\ B {\bf 925}, 63 (2017).
[arXiv:1702.04963 [hep-th]].
%%CITATION = MPP-2017-001%%
}

\lref\BroedelGBA{
  J.~Broedel and A.~Kaderli,
``Amplitude recursions with an extra marked point,''
[arXiv:1912.09927 [hep-th]].
%%CITATION = arXiv:1912.09927%%
}

\lref\TourkineBAK{
  P.~Tourkine and P.~Vanhove,
``Higher-loop amplitude monodromy relations in string and gauge theory,''
Phys.\ Rev.\ Lett.\  {\bf 117}, no. 21, 211601 (2016).
[arXiv:1608.01665 [hep-th]].
%%CITATION = arXiv:1608.01665%%
}
\lref\CasaliIHM{
  E.~Casali, S.~Mizera and P.~Tourkine,
``Monodromy relations from twisted homology,''
JHEP {\bf 1912}, 087 (2019).
[arXiv:1910.08514 [hep-th]].
%%CITATION = arXiv:1910.08514%%
}

\lref\CasaliKNC{
  E.~Casali, S.~Mizera and P.~Tourkine,
``Loop amplitudes monodromy relations and color-kinematics duality,''
JHEP {\bf 2021}, 048 (2020).
[arXiv:2005.05329 [hep-th]].
%%CITATION = arXiv:2005.05329%%
}

\lref\KawaiXQ{
  H.~Kawai, D.C.~Lewellen and S.H.H.~Tye,
``A Relation Between Tree Amplitudes of Closed and Open Strings,''
Nucl.\ Phys.\ B {\bf 269}, 1 (1986).
%%CITATION = CLNS-85/667%%
}

\lref\BernSV{
  Z.~Bern, L.J.~Dixon, M.~Perelstein and J.S.~Rozowsky,
``Multileg one loop gravity amplitudes from gauge theory,''
Nucl.\ Phys.\ B {\bf 546}, 423 (1999).
[hep-th/9811140].
%%CITATION = hep-th/9811140%%
}

\lref\BjerrumBohrHN{
  N.E.J.~Bjerrum-Bohr, P.H.~Damgaard, T.~Sondergaard and P.~Vanhove,
 ``The Momentum Kernel of Gauge and Gravity Theories,''
JHEP {\bf 1101}, 001 (2011).
[arXiv:1010.3933 [hep-th]].
%%CITATION = arXiv:1010.3933%%
}
%\NandanODY
\lref\NandanODY{
  D.~Nandan, J.~Plefka and G.~Travaglini,
``All rational one-loop Einstein-Yang-Mills amplitudes at four points,''
JHEP {\bf 1809}, 011 (2018).
[arXiv:1803.08497 [hep-th]].
%%CITATION = arXiv:1803.08497%%
}

\lref\StiebergerCEA{
  S.~Stieberger and T.R.~Taylor,
``Graviton as a Pair of Collinear Gauge Bosons,''
Phys.\ Lett.\ B {\bf 739}, 457 (2014).
[arXiv:1409.4771 [hep-th]];
%%CITATION = MPP-2014-343%%
``Graviton Amplitudes from Collinear Limits of Gauge Amplitudes,''
Phys.\ Lett.\ B {\bf 744}, 160 (2015).
[arXiv:1502.00655 [hep-th]].
%%CITATION = MPP-2015-001%%
``Subleading terms in the collinear limit of Yang--Mills amplitudes,''
Phys.\ Lett.\ B {\bf 750}, 587 (2015).
[arXiv:1508.01116 [hep-th]].
%%CITATION = MPP-2015-183%%
}

\lref\StiebergerLNG{
  S.~Stieberger and T.R.~Taylor,
``New relations for Einstein--Yang--Mills amplitudes,''
Nucl.\ Phys.\ B {\bf 913}, 151 (2016).
[arXiv:1606.09616 [hep-th]].
%%CITATION = MPP-2016-140%%
}

\lref\MizeraCQS{
  S.~Mizera,
``Combinatorics and Topology of Kawai-Lewellen-Tye Relations,''
JHEP {\bf 1708}, 097 (2017).
[arXiv:1706.08527 [hep-th]].
%%CITATION = arXiv:1706.08527%%
}

\lref\CachazoAOL{
  F.~Cachazo, S.~He and E.Y.~Yuan,
 ``One-Loop Corrections from Higher Dimensional Tree Amplitudes,''
JHEP {\bf 1608}, 008 (2016).
[arXiv:1512.05001 [hep-th]].
%%CITATION = arXiv:1512.05001%%
}

\lref\AntoniadisVW{
  I.~Antoniadis, C.~Bachas, C.~Fabre, H.~Partouche and T.R.~Taylor,
``Aspects of type I - type II - heterotic triality in four dimensions,''
Nucl.\ Phys.\ B {\bf 489}, 160 (1997).
[hep-th/9608012];\br
%%CITATION = hep-th/9608012%%
M.~Berg, M.~Haack, J.U.~Kang and S.~Sj\"ors,
``Towards the one-loop K\"ahler metric of Calabi-Yau orientifolds,''
JHEP {\bf 1412}, 077 (2014).
[arXiv:1407.0027 [hep-th]].
%%CITATION = LMU-ASC-43-14%%
}

\lref\BlumenhagenCI{
  R.~Blumenhagen, B.~K\"ors, D.~L\"ust and S.~Stieberger,
``Four-dimensional String Compactifications with D-Branes, Orientifolds and Fluxes,''
Phys.\ Rept.\  {\bf 445}, 1 (2007).
[hep-th/0610327].
%%CITATION = hep-th/0610327%%
}

%%%%%%%%%%%%%%%%%%%%%%%%%%%%%%%%%%%%%
%%%%%%%%%%%%%%%%%%%%%%%%%%%%%%%%%%%%%%%%%%%%%%%%%%%%%%%%%%%%%%%%%%%
%%%%%%%%%%%%%%%%%%%%%%%%%%%%%%%%%%%%%%%%%%%%%%%%%%%%%%%%%%%%%%%%%%%
\Title{\vbox{\rightline{MPP--2021--019}
}}
{\vbox{
\centerline{Open \& Closed vs. Pure Open String One-Loop Amplitudes}\vskip5mm
%\centerline{\it One-loop Einstein--Yang--Mills Amplitudes from pure gluon amplitudes}
}}
\medskip
\centerline{S. Stieberger}
\bigskip

\medskip
\centerline{\it Max--Planck--Institut  f\"ur Physik}
\centerline{\it Werner--Heisenberg--Institut, 80805 M\"unchen, Germany}
\medskip

\vskip15pt

\vskip15pt

\medskip
\bigskip\bigskip\bigskip
\centerline{\bf Abstract}
\vskip .2in
\noindent
We express one--loop string amplitudes involving both open and closed strings as sum over pure open string amplitudes. 
These findings generalize the analogous tree--level result to higher loops and extend the 
tree--level observation that in gravitational amplitudes a graviton can be traded for  
two gluons.
Our results are derived from analytic continuation of closed string world--sheet coordinates on the cylinder resulting in pairs of real open string coordinates located at the two cylinder boundaries subject to a one--loop kernel.
The latter depends on the loop momentum flowing between the two cylinder boundaries
and relates to intersection theory for twisted cycles.
Finally, contact  is made with one--loop open string monodromy relations. The latter contain a boundary term, which is related to non--physical contours on the cylinder. A physical interpretation of the latter in terms of a closed string insertion is given.

\noindent

\Date{}
\noindent
\goodbreak
\listtoc
\writetoc
\break
%%%%%%%%%%%%%%%%%%%%%%%%%%%%%%%%%%%%%%%%%%%%%%%%%%%%%%%%%%%%%%%%%%%%%%%%%%%%%%%
\newsec{Introduction}

Both gravity and gauge theories contain a local symmetry and mediate long--range forces at the classical level. Furthermore, they have a number of other well--known common features.
In particular, at the quantum level both forces are believed to be mediated through elementary bosons.
One may speculate that a massless graviton of spin--two can in some way be interpreted as a pair of two spin--one gluons.
In fact, it is a long standing question to what extent a graviton can be described by 
composite gluons. A priori there is a no--go theorem, which seems to negate this question \WeinbergKQ. The latter states, that in a relativistic quantum field theory  with a Lorentz--covariant energy--momentum tensor composite as well as elementary massless particles with spin higher than one are forbidden. This theorem holds for any known renormalizable field--theory, e.g. QCD and its proof relies on the construction of a conserved and 
Lorentz--covariant stress tensor. As a consequence the graviton cannot be a bound state of 
known elementary particles.
However, a closer look reveals that there are ways to circumvent this theorem like massive gravity, conformal field theory, or string theory  do not seem to fulfil all assumptions for this theorem. E.g. in string theory a Lorentz--covariant stress tensor cannot consistently be defined as a function of target space coordinates. Moreover, at the level of interactions unexpected relations between gravitational and gauge amplitudes hold, which signals that there is indeed  some gauge structure in quantum gravity. 

In fact, it is string theory where gauge--gravity connections become manifest -- at least in the most direct and obvious way.
In string theory a massless graviton state  appears as lowest closed string mode whilst a gluon is the lowest massless open string state. The closed string modes can be described by a tensor product of left--movers and right--movers with each one describing open strings. On the  string world--sheet, which describes the relevant perturbative string amplitude, the left-- and right movers are linked by monodromies, which are specified by  some kernel or intersection matrix.
It is this hybrid construction which entails the connection between gravitational and gauge amplitudes  in perturbative string theory. In fact, many gauge--gravity relations derive from properties of the underlying string world--sheet and give already rise to unexpected relations in field theory. 
One of the most well--known relation is the Kawai--Lewellen--Tye (KLT) relation~\KawaiXQ.
This relation which is derived from pure string world--sheet properties without any assumptions on the underlying string background (except that there exists an underlying conformal field theory description) expresses 
 closed string tree amplitudes as weighted sum over squares of open string tree amplitudes.  
 This identity holds to all orders in the inverse string tension $\ap$, with the latter  related to the string mass as follows $\ap\!\sim\! M_{\rm string}^{-2}$.
 In the field--theory limit $\ap\ra0$, this relation implies that at tree--level gravitational amplitudes can be expressed as sum over squares of gauge amplitudes.
Furthermore, in \StiebergerHQ\ it has been shown that any tree--level string amplitude involving 
$n_c$ closed strings and $n_o$ open strings can be written as linear combination of pure open string amplitudes involving only $2n_c+n_o$ open strings. 
These structures have been further exploited and extended leading to new relations for Einstein--Yang--Mills (EYM) amplitudes both at string and field--theory level \refs{\StiebergerCEA,\StiebergerVYA,\StiebergerLNG}. Cf.  also \SrisangyingcharoenLHX\ 
for some interesting complementary aspects. 
At any rate in the field--theory limit, these findings imply that at least at tree--level properties of gravitational amplitudes  are inherited from gauge amplitudes.
 Therefore, it is natural to ask for a one--loop generalization thereof.

In tackling any closed string amplitude calculation it useful to first consider the corresponding open string computation. At field--theory level this means that for graviton amplitudes one should recycle results from gauge amplitudes \BernSV.
In this work we study one--loop string amplitudes involving both open and closed strings.
Their interaction at one--loop is described by a cylinder world--sheet.
We show that one--loop cylinder amplitudes of $n_c$ closed and $n_o$ open strings can be written as pure open string one--loop amplitudes involving $2n_c+n_o$ open strings with each closed string replaced by two open strings.
The present work essentially aims at a one--loop generalization of \StiebergerHQ\ and \StiebergerCEA.

The organization of this work is as follows. In Section 2 we introduce generic one--loop cylinder amplitudes involving both open and closed strings. We also discuss various setups
resulting in different interactions of closed string left-- and right moving fields.
In Section 3 we solve the monodromy problem on the cylinder by deforming the closed string position integrations along a closed contour thereby disentangling holomorphic and anti--holomorphic coordinates and converting them into a pair of real coordinates describing open string positions located along the two cylinder boundaries. Finally, in Section 4 we relate these steps to the one--loop open string monodromy relations. The latter contain a boundary term, which is related to non--physical contours on the cylinder. A physical interpretation of the latter in terms of a closed string insertion is attributed.
In Section 5 we summarize our results and give some concluding remarks.

\newsec{One--loop cylinder amplitudes involving open and closed strings}

We consider the scattering of $n_o$ open and $n_c$ closed strings at one--loop.
The on--shell string amplitude $A_{n_o,n_c}^{(1)}$ is described by the topology of a world--sheet cylinder or annulus with two boundaries or a M\"obius strip with one boundary.
In the following we shall concentrate on the first case. The underlying string world--sheet is described by a cylinder $\Cc$ with  complex structure modulus $\tau=it$ (modular parameter), with $t\in(0,\infty)$. The complex coordinate $z=\sigma^1+i\sigma^2$ on the cylinder is parameterized by
$(\sigma^1,\sigma^2)\in [0,1]\times[0,\fc{t}{2}]$. The $n_o$ open strings are described through vertex operator insertions at the two boundaries boundaries $\sigma^2=0$ and $\sigma^2=\fc{t}{2}$, respectively.
On the other hand, the $n_c$  closed strings are inserted at points $z_r$ in the bulk. Hence, for the cylinder diagram under consideration $n_1$ of the  $n_o$ open string insertions $x_i,\;i=1,\ldots,n_1$ may be put on the first boundary and $n_2$ insertions $x_j',\;j=1,\ldots,n_2$ on the second boundary, with $n_o=n_1+n_2$.
The bosonic open string Green's function on the cylinder (referring to  Neumann boundary conditions for the open string ends) is given by
\eqn\BOSG{
G^{(1)}(z_1,z_2)=\ln\lf|\fc{\theta_1(z_1-z_2,\tau)}{\theta_1'(0,\tau)}\ri|-\fc{\pi}{\Im\tau}[\Im(z_2-z_1)]^2\ ,}
and $q=e^{2\pi i\tau}$ with the complex structure modulus $\tau\!=\!i\tau_2\!=\!it$. The second term of \BOSG\ renders 
the function to be single--valued on the torus under the shift $z\ra z+\tau$ \DHokerPDL.
Furthermore, the  scalar open string Green function on the cylinder describing two positions on two distinct boundaries is given by:
\eqn\BOSGT{
G_T^{(1)}(z_1,z_2)=\ln\lf|\fc{\theta_4(z_1-z_2,\tau)}{\theta_1'(0,\tau)}\ri|-\fc{\pi}{\Im\tau}[\Im(z_2-z_1)]^2\ .}

On surfaces with boundaries  there is an interaction between left-- and right--moving closed string fields. This effect can conveniently be described by the doubling trick \GubserWT. 
As a result the Green's function $G_B$ on the cylinder for the closed string bosonic fields can be written as
\eqn\BOSGD{
G_B(z_1,z_2)=G^{(1)}(z_1,z_2)\pm\ G^{(1)}(z_1,\ov z_2)\ ,}
with positive sign referring to Neumann  and  negative sign to Dirichlet boundary conditions, respectively.
The different boundary conditions for the fields can be specified by a matrix $D^{\mu\nu}$, with $\mu,\nu=0,\ldots,d-1$ with $d$ the space--time dimension, cf. also Appendix \appA. 
If we represent a generic plane wave closed string state by
\eqn\plane{
|q\rangle=\lim_{\rho,\bar \rho\ra0} :e^{iq_L^\mu X_\mu(\rho)}e^{iq_R^\mu X_\mu(\bar \rho)}:|0\rangle\ ,}
with complex plane coordinates $\rho,\bar\rho\in\IC$ and the total momentum $q_L+q_R$ the interaction between left-- and right--moving fields can conveniently be described  by correlators on the torus. 
In particular, for a specific D--brane setup with boundary matrix $D$ we choose the left-- and right--moving  momenta $q_L,q_R$ as
\eqn\Split{
q_{L}=\h q\ \ \ ,\ \ \ q_{R}=\h Dq\ ,}
with $q_{L}^2=q_{R}^2=0$  representing massless closed  states.
Then, the total (parallel) closed string momentum $q^\parallel\equiv q_L+q_R$ is given by 
\eqn\qparallel{
q^\parallel=\h(q+Dq)\ ,}
while the (orthogonal) closed string momentum $q^\perp\equiv q_L-q_R$ is determined by
\eqn\qperp{
q^\perp=\h(q-Dq)\ ,}
with $(q^\parallel)^2=2q_{L}q_{R}=\h qDq$ for the massless case $q^2=(q^\parallel)^2+(q^\perp)^2=0$, i.e. $(q^\perp)^2\!=\!-(q^\parallel)^2$. The notions parallel and orthogonal refer to a D$p$--brane describing a $p\!+\!1$--dimensional hyperplane in the $d$
dimensional space--time. Amplitudes involving open and closed strings describe excited D--branes emitting and absorbing bulk closed string states.

\subsec{Generic configuration}

To describe a non--planar cylinder configuration we introduce the  two orderings 
$\Ac\!=\!\{\alpha_1,\ldots,\alpha_{n_1}\},\ \Bc\!=\!\{\beta_1,\ldots,\beta_{n_2}\}$, with $n_1+n_2=n_o$. The elements  $\alpha_i,\beta_j\in \Pc_{n_o}$ of the two sets $\Ac$ and $\Bc$ with $\Ac\cup\Bc=\Pc_{n_o}$ and $\Pc_n=\{1,\ldots,n\}$ label $n_o$ open strings inserted along the two cylinder boundaries. Furthermore, we have a set $\Cc=\{q_{L,1},q_{R,1},\ldots,q_{L,n_c},q_{R,n_c}\}$ of left-- and right--moving closed string momenta referring to 
$n_c$ closed strings. Then, in $d$ space--time dimensions the color--ordered 
amplitude  of $n_o$ open and $n_c$ closed strings at one--loop assumes the generic form 
\eqn\Start{\eqalign{
A_{n_o,n_c}^{(1)}(\Ac|\Bc||\Cc)&=g_o^{n_o}g_c^{n_c}\ \delta^{(d)}\lf(\sum_{i=1}^{n_1}p_{\al_i}+\sum_{j=1}^{n_2}p_{\bet_j}+\sum_{k=1}^{n_c}q^\parallel_{k}\ri)\cr 
&\times\int d\tau_2\; V_{CKG}^{-1} \left(\!\int_{\Ic_\Ac}\prod_{i=1}^{n_1}dx_{\al_i}\int_{\Ic_\Bc}\prod_{j=1}^{n_2}dx_{\bet_j}'\!\right) E(\{x_i,x_j'\})\cr
&\times \left(\!\int_{\Cc}\prod_{s=1}^{n_c}d^2z_s\!\right) \ I(\{x_i,x_j',z_s,\bar z_s\})\ Q(\tau,\{x_i,x_j',z_s,\bar z_s\})\ ,}}
with the open and closed string coupling constants $g_o$ and $g_c$, respectively and the two domains of integrations  
\eqn\domain{\eqalign{
\Ic_\Ac&=\{x_{\al_i}\in \IR\ |\ 0<x_{\al_1}<\ldots<x_{\al_a}<1\}\ ,\cr
\Ic_\Bc&=\{x_{\bet_j}'\in \IR\ |\ 0<x'_{\bet_1}<\ldots<x'_{\bet_b}<1\}\ ,}}
subject to the orderings $\Ac,\Bc$ of open string vertex operator insertions  along the two cylinder boundaries, respectively.
Furthermore, we have
\eqnn\KN{
$$\eqalignno{
E(\{x_i,x_j'\})&=
\prod_{1\leq i<j\leq n_1}e^{2\ap p_{\al_i}p_{\al_j} G^{(1)}(x_{\al_j},x_{\al_i})}
\!\!\prod_{1\leq i<j\leq n_2}e^{2\ap p_{\bet_i}p_{\bet_j}G^{(1)}(x'_{\bet_j},x_{\bet_i}')}
\cr
&\times\prod_{1\leq i\leq n_1\atop 1\leq j\leq n_2}e^{2\ap p_{\al_i}p_{\bet_j}G_T^{(1)}(x_{\bet_ j}',x_{\al_i})},\cr
I(\{x_i,x_j',z_s,\bar z_s\})&=\prod_{r=1}^{n_c}e^{\h\ap (q^\parallel_r)^2 G^{(1)}_N(z_r,\ov z_r)}e^{\h\ap (q^\perp_r)^2 G^{(1)}_D(z_r,\ov z_r)} \cr
&\times\prod_{1\leq r<s\leq n_c}e^{\ap q_rq_s G^{(1)}(z_s,z_r)}\ 
e^{\ap q_rDq_s G^{(1)}(z_s,\ov z_r)}\cr
&\times\prod_{1\leq i\leq n_1\atop 1\leq r\leq n_c}e^{2\ap p_{\al_i}q_rG^{(1)}(z_r,x_{\al_i})}
\prod_{1\leq j\leq n_2\atop 1\leq r\leq n_c}e^{2\ap p_{\bet_j}q_rG_T^{(1)}(z_r,x'_{\bet_j})
}\ ,&\KN}$$}
accounting for the insertion of an exponential factor universal to all vertex operators.
Finally, we have the momentum conservation condition (parallel to the D--brane world volume):
\eqn\conserv{
\sum_{i=1}^{n_1}p_{\al_i}+\sum_{j=1}^{n_2}p_{\bet_j}+\sum_{r=1}^{n_c}q^\parallel_{r}=0\ .}
Above we have used the bosonic closed string Green's functions on the cylinder \eqn\GND{\eqalign{
G^{(1)}_N(z,\bar z)&=\ln\lf|\fc{\theta_1(z-\bar z,\tau)}{\theta_1'(0,\tau)}\ri|-\fc{4\pi}{\tau_2}\ \Im(z)^2\ ,\cr
G^{(1)}_D(z,\bar z)&=\ln\lf|\fc{\theta_1'(0,\tau)}{\theta_1(z-\bar z,\tau)}\ri|+\fc{4\pi}{\tau_2}\ \Im(z)^2\ ,}}
referring to Neumann and  Dirichlet boundary conditions along the 
boundary~$z\!=\!\bar z$, respectively. The modular function $Q$ of weight $w$, \ie $Q(-\fc{1}{\tau},\{\fc{x_i}{\tau},\fc{x_j'}{\tau},\fc{z_s}{\tau},\fc{\bar z_s}{\tau}\})=|\tau|^wQ(\tau,\{x_i,x_j',z_s,\bar z_s\})$  depends on the set of positions $\{x_i,x_j',z_s,\bar z_s\}$ and describes the remaining contractions of vertex operators including
ghost and matter contributions and the partition function. The simplest case represents  tachyon scattering in $d\!=\!26$ bosonic string theory, with $Q\!=\!\eta(\tau)^{-24}$.  
The volume of the conformal Killing group $V_{CKG}=\tau_2=t$
can be cancelled by fixing one real part of one of the open or closed vertex operator positions. 
The corresponding world--sheet cylinder with open and closed string insertions is depicted in the next figure Fig.~1.
\ifig\SigmaWorldsheet{String world--sheet cylinder with $n_1+n_2$ open and $n_c$ closed string insertions.}{\epsfxsize=0.5\hsize\epsfbox{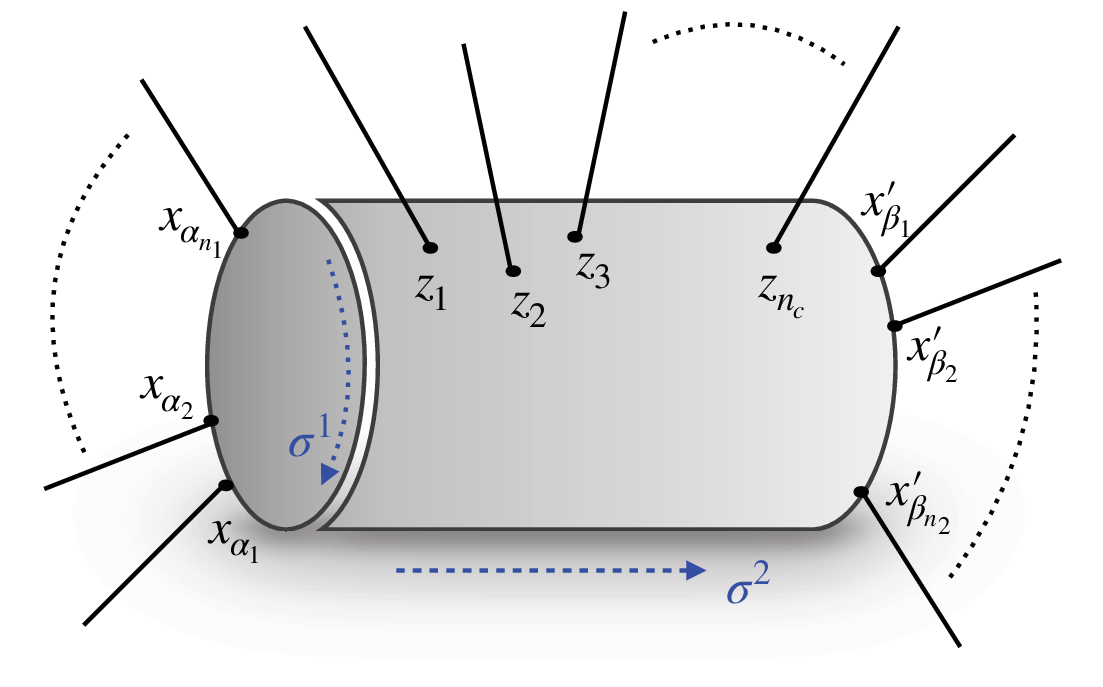}}
\noindent Since $x_i,x_j'\in \IR$ only the closed string positions are sensitive to the second term of the Green's function \BOSG. Explicitly, we have
\eqn\wehaveE{\eqalign{
E(\{x_i,x_j'\})&=\prod_{1\leq i<j\leq n_1}\lf|\fc{\theta_1(x_{\al_j}-x_{\al_i},\tau)}{\theta_1'(0,\tau)}\ri|^{2\ap p_{\al_i}p_{\al_j}}
\prod_{1\leq i<j\leq n_2}\lf|\fc{\theta_1(x_{\bet_j}'-x_{\bet_i}',\tau)}{\theta_1'(0,\tau)}\ri|^{2\ap p_{\bet_i}p_{\bet_j}}\cr
&\times\prod_{1\leq i\leq n_1\atop 1\leq j\leq n_2}\lf|\fc{\theta_4(x_{\bet_j}'-x_{\al_i},\tau)}{\theta_1'(0,\tau)}\ri|^{2\ap p_{\al_i}p_{\bet_j}}\ ,}}
and
\eqn\Generali{\eqalign{
I(\{x_i,x_j',z_s,\bar z_s\})&=
\Theta_L(\{x_i,x_j',z_s,\bar z_s\})\ 
\Theta_R(\{x_i,x_j',z_s,\bar z_s\})\ I_{nh}(\{z_s,\bar z_s\})\cr
&\times \prod_{r=1}^{n_c}\lf|\fc{\theta_1(z_r-\bar z_r,\tau)}{\theta_1'(0,\tau)}\ri|^{\ap (q^\parallel_r)^2}\ ,}}
with the functions
\eqnn\thetaLR{
$$\eqalignno{
\Theta_L(\{x_i,x_j',z_s,\bar z_s\})&=\!\!\!\prod_{1\leq r<s\leq n_c}\lf[\fc{\theta_1(z_s-z_r,\tau)}{\theta_1'(0,\tau)}\ri]^{2\ap q_{L,s}q_{L,r}} \lf[\fc{\theta_1(z_s-\bar z_r,\tau)}{\theta_1'(0,\tau)}\ri]^{2\ap q_{L,s}q_{R,r}}\cr
&\!\!\!\times\prod_{1\leq i\leq n_1\atop 1\leq r\leq n_c}\lf[\fc{\theta_1(z_r-x_{\al_i},\tau)}{\theta_1'(0,\tau)}\ri]^{2\ap p_{\al_i}q_{L,r}}\!\!
\prod_{1\leq j\leq n_2\atop 1\leq r\leq n_c}\lf[\fc{\theta_4(z_r-x_{\bet_j}',\tau)}{\theta_1'(0,\tau)}\ri]^{2\ap p_{\bet_j}q_{L,r}}\ ,\cr
\Theta_R(\{x_i,x_j',z_s,\bar z_s\})&=\!\!\! \prod_{1\leq r< s\leq n_c}\lf[\fc{\bar \theta_1(\bar z_s-\bar z_r,\bar \tau)}{\bar\theta_1'(0,\bar \tau)}\ri]^{2\ap q_{R,s}q_{R,r}}\ \lf[\fc{\bar\theta_1(\bar z_s- z_r,\bar\tau)}{\bar\theta_1'(0,\bar\tau)}\ri]^{2\ap q_{R,s}q_{L,r}}\cr
&\!\!\!\times \prod_{1\leq i\leq n_1\atop 1\leq r\leq n_c}\lf[\fc{\bar\theta_1(\bar z_r-x_{\al_i},\bar\tau)}{\bar\theta_1'(0,\bar\tau)}\ri]^{2\ap p_{\al_i}q_{R,r}}\!\!
\prod_{1\leq j\leq n_2\atop 1\leq r\leq n_c}\lf[\fc{\bar\theta_4(\bar z_r-x_{\bet_j}',\bar\tau)}{\bar\theta_1'(0,\bar\tau)}\ri]^{2\ap p_{\bet_j}q_{R,r}}&\thetaLR}$$}
and the non--holomorphic factor only depending on the closed string positions $z_r$:
\eqnn\nonholo{
$$\eqalignno{
I_{nh}(\{z_s,\bar z_s\})\ &=e^{-\fc{4\pi\ap}{\tau_2}\sum\limits_{r=1}^{n_c}(q^\parallel_r)^2\Im(z_r)^2}\ e^{-\fc{\pi\ap}{\tau_2}\sum\limits_{r,s=1\atop r<s}^{n_c}q_{r}q_{s}\Im(z_r-z_s)^2}\ e^{-\fc{\pi\ap}{\tau_2}\sum\limits_{r,s=1\atop r<s}^{n_c}q_{r}Dq_{s}\Im(z_r-\bar z_s)^2}\cr
&\times e^{-\fc{\pi\ap}{\tau_2}\sum\limits_{r=1}^{n_c}\sum\limits_{i=1}^{n_1}[q_{r}p_{\al_i}\Im(z_r)^2+q_{r}p_{\al_i}\Im(\bar z_r)^2]}  e^{-\fc{\pi\ap}{\tau_2}\sum\limits_{r=1}^{n_c}\sum\limits_{j=1}^{n_2}[q_{r}p_{\bet_j}\Im(z_r)^2+q_{r}p_{\bet_j}\Im(\bar z_r)^2]}\cr
&=
%e^{-\fc{\pi\ap}{2\tau_2}\sum\limits_{r=1}^{n_c}(q_r+Dq_r)^2\Im(z_r)^2}\ 
e^{\fc{\pi\ap}{2\tau_2}\sum\limits_{r,s=1}^{n_c}(q_r-Dq_r)^t(q_s-Dq_s)\Im(z_r)\Im(z_s)}\ .&\nonholo}$$}
Note, that above we have used the relations $p_iq_r=p_iDq_r$ following from $p_iq_r=p_iq^\parallel_r$. Furthermore, with \Split\ the non--holomorphic factor \nonholo\ may also be expressed in 
terms of generic left-- and right--moving momenta:
\eqn\nonholoo{
I_{nh}(\{z_s,\bar z_s\})=e^{-\fc{4\pi\ap}{\tau_2}\sum\limits_{r=1}^{n_c}(q_{L,r}^tq_{L,r}+q_{R,r}^tq_{R,r})\Im(z_r)^2}\ e^{\fc{2\pi\ap}{\tau_2}\sum\limits_{r,s=1}^{n_c}(q_{L,r}-q_{R,r})^t(q_{L,s}-q_{R,s})\Im(z_r)\Im(z_s)}\ .}

%%%%%%%%%%%%%%%%%%%%%%%%%%%%%%%%%%%%%
\subsec{Sample configurations}

\subsubsec{Collinear left-- and right--moving closed string momenta}

\noindent
Instead of the splitting \Split\  we may distribute the closed string momentum $q$ into left-- and right--moving (collinear) momenta as  \StiebergerVYA\
\eqn\split{
q_{L}=\rho q\ \ ,\ \ q_{R}=(1-\rho) q\ ,}
with $0\leq \rho\leq 1$ and the total momentum 
\eqn\total{
q^\parallel\equiv q_L+q_R=q\ ,} 
with $q^2=0$ and $q_{L}^2=q_{R}^2=0$ for the massless case. 
Of course, the simplest choice is $\rho_s=\h$ corresponding to
$q_{L,s}=q_{R,s}=\fc{q_s}{2}$.
Obviously, in this case we have $\Theta_R=\ov\Theta_R$.
However, in the following we shall discuss the case $\rho_s\neq 0$. The parameterization is 
 most suited for studying  collinear limits of closed string momenta \StiebergerVYA.

\subsubsec{Generic D--brane setup}

\noindent
For generic D$p$--branes the non--holomorphic factor \nonholo\ assumes the general form
\eqn\Nonholo{
I_{nh}(\{z_s,\bar z_s\})=
e^{\fc{2\pi\ap}{\tau_2}\sum\limits_{r,s=1}^{n_c}(q_r^\perp)^tq_s^\perp\Im(z_r)\Im(z_s)}=
e^{-\fc{\pi\ap}{2\tau_2}\lf(\sum\limits_{r=1}^{n_c}q_r^\perp(z_r-\bar z_r)\ri)^2}\ ,}
which in turn by introducing the loop momentum $\ell$ can be converted into
\eqn\looprepr{(\fc{\ap \tau_2}{2})^{-d/2}\   I_{nh}(\{z_s,\bar z_s\})=\int_{-\infty}^\infty d^{d}\ell\ \Lc(\{z_s,\bar z_s\};\ell)\ ,}
with the loop momentum dependent factor depending on the closed string positions $z_r$:
\eqn\KleineWolfsschlucht{
\Lc(\{z_s,\bar z_s\};\ell)=\exp\lf\{-\h\pi\ap \tau_2\ell^2-\pi i\ap\ell\sum_{r=1}^{n_c}
q^\perp_r(z_r-\bar z_r)\ri\}\ .}
The loop momentum $\ell$ has been introduced for the case of closed strings on the torus to disentangle the integrand into holomorphic and anti--holomorphic parts \DHokerPDL.
For generic D$p$--brane scattering  with \looprepr\ now the integrand \Generali\ can be written as:
\eqn\LoopE{
(\fc{\ap \tau_2}{2})^{-d/2}\ I(\{x_i,x_j',z_s,\bar z_s\})=\int_{-\infty}^\infty d^d\ell\ I(\{x_i,x_j',z_s,\bar z_s\};\ell)\ ,}
with\foot{Note that compared to \Generali\ in the following definition of $I$ we have dropped a total factor of $|\theta_1(0,\tau)|^{\ap\lf(\sum\limits_{r=1}^{n_c}q_r^\parallel\ri)^2}$,  which will be taken into account when combining it with \KN\ into \Start.}: 
\eqn\GrosseWolfsschlucht{\eqalign{
I(\{x_i,x_j',z_s,\bar z_s\};\ell)&:=\Lc(\{z_s,\bar z_s\};\ell)
\prod_{r,s=1}^{n_c}\lf|\theta_1(z_s-z_r,\tau)\ri|^{\h\ap q_{s}q_{r}} \lf|\theta_1(z_s-\bar z_r,\tau)\ri|^{\h\ap q_{s}Dq_{r}}\cr
&\times\prod_{1\leq i\leq n_1\atop 1\leq r\leq n_c}\lf|\theta_1(z_r-x_{\al_i},\tau)\ri|^{2\ap p_{\al_i}q_{r}}
\prod_{1\leq j\leq n_2\atop 1\leq r\leq n_c}\lf|\theta_4(z_r-x_{\bet_j}',\tau)\ri|^{2\ap p_{\bet_j}q_{r}}\ .}}

\subsubsec{Pure Neumann boundary conditions}

\noindent
For the case $D=+{\bf 1}$ describing e.g. the world--volume scattering on a space--time filling D$9$--brane the factor \Nonholo\ becomes trivial:
\eqn\nonholoeasy{
I_{nh}(\{z_s,\bar z_s\})=1\ .}
Note, that by systematically applying $T$--duality one can obtain other configurations 
with $D\neq -{\bf 1}$ describing generic D$p$--branes with $p+1$ Neumann directions  and the remaining $d\!-\!p\!-\!1$ directions with Dirichlet boundary conditions \refs{\PolchinskiFM}.

\subsubsec{Pure Dirichlet boundary conditions}

\noindent
On the other hand, the case $D=-{\bf 1}$ describes e.g. a D$(-1)$--brane configuration or D--instanton background. 
According to \qparallel\ for $D=-{\bf 1}$ we have $q_r^\parallel=0$ and 
$q_r=q_r^\perp$, with $q_r^2=(q_r^\perp)^2=0$. Then all scalar products between open $p_i$ and closed string momenta $q_r$ vanish, i.e. $p_iq_{L,r}=p_iq_{R,r}=0$ and the momentum conservation \conserv\ is supplemented  
by the bulk condition $\sum\limits_{r=1}^{n_c}q_r^\perp=0$. As a consequence \Generali\ becomes
a pure closed string object
\eqnn\EllmauerTor{
$$\eqalignno{
I(\{z_s,&\bar z_s\})\equiv I_{nh}(\{z_s,\bar z_s\})\ \prod_{r=1}^{n_c}\theta_1(z_r-\bar z_r,\tau)^{-\h\ap (q_r^\perp)^2}&\EllmauerTor\cr 
&\times\prod_{r<s}^{n_c}\theta_1(z_s-z_r,\tau)^{\fc{\ap}{2} q_r q_s} 
\theta_1(\bar z_s-\bar z_r,\tau)^{\fc{\ap}{2} q_r q_s}\theta_1(z_s-\bar z_r,\tau)^{\fc{\ap}{2} q_s Dq_r} 
\theta_1(\bar z_s- z_r,\tau)^{\fc{\ap}{2} q_r Dq_s},}$$}
with the non--holomorphic expression \nonholo.
On the other hand, the open string part \wehaveE\ decouples from the closed string part \EllmauerTor.
Then we have:
\eqn\WochenbrunnerAlm{\eqalign{
(\fc{\ap \tau_2}{2})^{-d/2}\   I(\{z_s,\bar z_s\})&=\int_{-\infty}^\infty d^{d}\ell\ \exp\lf\{-\h\pi\ap \tau_2\ell^2+2\pi \ap\ell\sum_{r=1}^{n_c} q_r\Im(z_r)\ri\}\cr
&\times \prod_{r<s}^{n_c}\lf(\fc{\theta_1(z_s-z_r,\tau)}
{\theta_1( z_s-\bar z_r,\tau)}\ri)^{\fc{\ap}{2} q_r q_s}\ \lf(\fc{\theta_1(\bar z_s-\bar z_r,\tau)}
{\theta_1( \bar z_s- z_r,\tau)}\ri)^{\fc{\ap}{2} q_r q_s}\ .}}
Eventually, without open strings inserting \WochenbrunnerAlm\ into \Start\ gives rise to the $n$--point amplitude  $A_{0,n_c}^{(1)}$ involving $n=n_c$ closed string
\eqn\HintereGoingerHalt{\eqalign{
(\fc{\ap \tau_2}{2})^{-d/2}\ A_{0,n}^{(1)}(q_1,\ldots,q_c)&=g_c^n\ \int_{-\infty}^\infty d^{d}\ell\ V_{CKG}^{-1}\ \Tc(q_1,\ldots,q_{n};-\ell,\ell)\ ,}}
with
\eqnn\vordereGoingerHalt{
$$\eqalignno{\Tc(q_1,\ldots,q_{n};-\ell,\ell)&:=\delta^{(d)}\lf(\sum_{r=1}^nq_r\ri)\int d\tau_2 \left(\!\int_{\Cc}\prod_{s=1}^{n}d^2z_s\!\right)\ Q(\tau,\{z_s,\bar z_s\})\cr
&\times \exp\lf\{-\h\pi\ap \tau_2\ell^2+2\pi \ap\ell\sum_{r=1}^{n}
q_r\Im(z_r)\ri\}&\vordereGoingerHalt\cr 
&\times \prod_{r<s}^{n}\lf(\fc{\theta_1(z_s-z_r,\tau)}
{\theta_1( z_s-\bar z_r,\tau)}\ri)^{\fc{\ap}{2} q_r q_s}\ \lf(\fc{\theta_1(\bar z_s-\bar z_r,\tau)}
{\theta_1( \bar z_s- z_r,\tau)}\ri)^{\fc{\ap}{2} q_r q_s}\ .}$$}
Due to the decoupling of open and closed string adding open strings is simply accomplished by  the additional  factor \wehaveE. With the additional (closed string) data 
\eqn\posiVG{
z_{n+1}=-\fc{\tau}{2}\ \ \ ,\ \ \ q_{n+1}=\ell\ ,}
the following identity holds:
\eqn\identVG{
\exp\lf\{+2\pi \ap\ell\sum_{r=1}^{n} q_r\Im(z_r)\ri\}=\prod_{r=1}^{n}\lf(\fc{\theta_1(z_{n+1}-z_r,\tau)}
{\theta_1( z_{n+1}-\bar z_r,\tau)}\ri)^{\fc{\ap}{2} q_r q_{n+1}}\ \lf(\fc{\theta_1(\bar z_{n+1}-\bar z_r,\tau)}
{\theta_1( \bar z_{n+1}- z_r,\tau)}\ri)^{\fc{\ap}{2} q_r q_{n+1}}\ .}
Eventually, with  \posiVG, \identVG\ and
\eqn\charge{
q_0=-\ell\ ,}
we can generalize \vordereGoingerHalt\ to:
\eqn\VordereGoingerHalt{\eqalign{
\Tc(q_1,\ldots,q_{n};q_0,q_{n+1})&=\delta^{(d)}\lf(q_0+q_{n+1}+\sum_{r=1}^nq_r\ri)\int d\tau_2\;  \left(\!\int_{\Cc}\prod_{s=1}^{n}d^2z_s\!\right) Q(\tau,\{z_s,\bar z_s\})\cr
&\times  e^{-\h\pi\ap \tau_2q_{n+1}^2}\ \prod_{r,s=1\atop r\neq s}^{n+1}\lf|\fc{\theta_1(z_s-z_r,\tau)}{\theta_1( z_s-\bar z_r,\tau)}\ri|^{\fc{\ap}{2} q_r q_s}\ .}}
This expression $\Tc(q_1,\ldots,q_{n};q_0,q_{n+1})$ is the off--shell closed string amplitude derived in \CohenPV\ (with $\fc{\ap}{2}=(4\pi T)^{-1}$, $\tau=2i\lambda$, the tube length $\lambda$ and  $Q\!=\!\eta(\tau)^{-24}$).
It describes the (semi) off--shell scattering of $n$ massless closed strings $q_i^2=0$ ($i\!=\!1,\ldots,n$) and two off--shell string states $q_i^2=\ell^2\neq 0$ ($i\!=\!0,n+1$), cf. Fig.~2. 
\ifig\Region{Scattering of two off--shell string states and $n$ massless closed strings.}{\epsfxsize=0.55\hsize\epsfbox{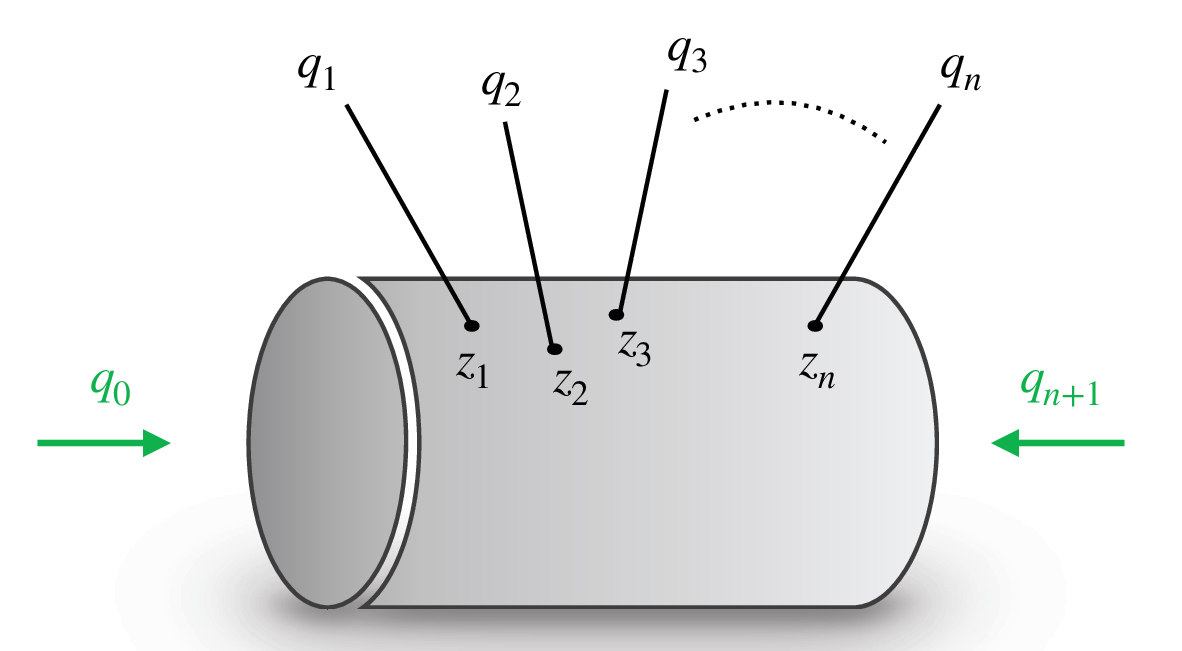}}
\noindent Thus, the loop momentum dependent factor 
\looprepr\ has an interesting interpretation within off--shell string scattering. 
Integrating over the loop momentum $\ell$ converts the off--shell amplitude \VordereGoingerHalt\ into the on--shell $n$--point amplitude \HintereGoingerHalt.
Performing in \vordereGoingerHalt\ the integration  over the closed string positions $z_s$ will be accomplished in the next section.

%%%%%%%%%%%%%%%%%%%%%%%%%%%%%%%%%%%%%
\newsec{Open and closed string amplitudes as pure open string amplitudes}

In this subsection we perform the necessary steps to convert integrations over complex closed string positions  on the cylinder $\Cc$ into pairs  of real integrations over open string coordinates. We shall perform an analytic continuation of the closed string vertex operator positions  $z_t\in \Cc$ on the cylinder (with $t\!=\!1,\ldots,n_c$) and consider some closed contour integrals on the cylinder.
As a result for each  complex coordinate $z_t$ we obtain a pair of two real coordinates $\xi_t,\eta_t$  describing two real open string positions  along one of the two cylinder  boundaries.

\subsec{Amplitudes with one closed string and $n_o$ open strings}

In order to explain our methods for simplicity we first focus on the case $n_c\!=\!1$.
We consider a planar one--loop cylinder amplitude with canonical open string ordering $\Ac\!=\!\{1,\ldots,n_1\}$, with  any number $n_o=n_1$ of open strings.  
Let us discuss the dependence of the integrand \GrosseWolfsschlucht\ on the closed string coordinate $z$. Applying (B.13) the latter~becomes
\eqn\relevanti{\eqalign{
I(z,\bar z;\ell)&:=I(\{x_i,x_j',z,\bar z\};\ell)\equiv\exp\lf\{-\h\pi\ap \tau_2\ell^2-\pi i\ap\ell
q^\perp(z-\bar z)\ri\}\ |\theta_1(z-\bar z,\tau)|^{\ap (q^\parallel)^2}\cr
&\times \prod_{i=1}^{n_o}\theta_1(z-x_i,\tau)^{\ap  qp_i}\ \theta_1(\bar z-x_i,\tau)^{\ap qp_i}\ ,}}
including the non--holomorphic factor  \KleineWolfsschlucht\ for $n_c\!=\!1$. The 
combination of theta--functions in the integrand 
\relevanti\ is single--valued in $z$. After introducing the paramterization
\eqn\sig{
z=\sigma^1+i\sigma^2\ \ \ ,\ \ \ \sigma^1\in(0,1)\ ,\ \sigma^2\in(0,\fc{t}{2})\ ,}
we investigate the holomorphic $\sigma^2$--dependence of
\eqn\relevant{
I(\si^1,\si^2;\ell):= \Lc(\sigma^2;\ell)\ \theta_1(2i\sigma^2,\tau)^{\ap (q^\parallel)^2}
\ \prod_{i=1}^{n_o}\theta_1(\sigma^1+i\sigma^2-x_i,\tau)^{\ap  qp_i}\ \theta_1(\sigma^1-i\sigma^2-x_i,\tau)^{\ap qp_i},}
with
\eqn\Laschet{
\Lc(\sigma^2;\ell)=\exp\lf\{-\h\pi\ap \tau_2\ell^2+2\pi \ap\ell q^\perp\si^2\ri\}\ ,}
in the complex $\sigma^2$--plane and consider the closed cycle (polygon) $C$ defined by the 
four edges $C_1,C_1',C_2$ and~$C_2'$
\eqn\edges{
C=C_1\cup C_2' \cup C_1'\cup C_2\ ,}
depicted in Fig.~3. 
\ifig\SigmaWorldsheet{A closed contour $C$ in the complex $\sigma^2$--plane and branch points.}{\epsfxsize=0.55\hsize\epsfbox{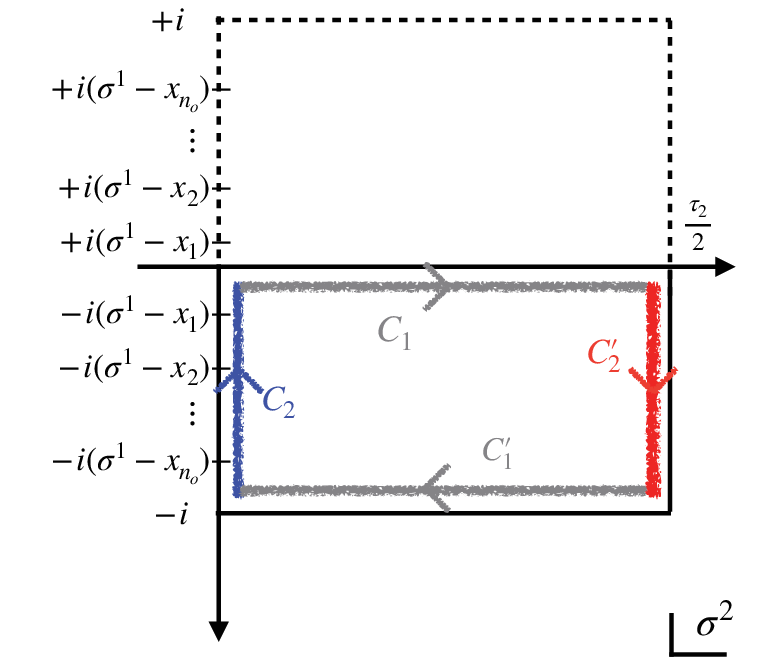}}
\noindent  Since $x_i\in (0,1)$ the function \relevant\ as a holomorphic function in $\sigma^2$  has $n_o$ pairs of  branch points at $\sigma^2=\pm i(\sigma^1-x_i)$ along the imaginary axis $\sigma^2=0$ ($z=\bar z$) and one branch point at $\sigma^2=0$ where the arguments of the theta--functions become zero. The corresponding branching phases are $e^{\pm\pi i \ap qp_i}$ and $e^{\pi i \ap (q^\parallel)^2}$, respectively.  These monodromy phases need to be accommodated  when passing these branch cuts.  Their effect will be taken into account below. 
The holomorphic  integrand along $C_1'$ differs from that along $C_1$ by a phase
\eqn\differ{
I(\sigma^1,\sigma^2-i;\ell)=\vartheta\ I(\sigma^1,\sigma^2;\ell))\ ,}
given by:
\eqn\phase{
\vartheta:=e^{-2\pi i\ap q^\perp \ell}\ .}
The last manipulation follows from \conserv\ and (B.10). On the other hand, moving from $C_2$ to $C_2'$ 
can be furnished by a shift of the loop momentum $\ell$
\eqn\shiff{
\ell\longrightarrow \ell-q^\perp\ ,}
leading to
\eqn\KlooAscher{
I(\sigma^1,\sigma^2+\fc{\tau_2}{2};\ell)=e^{-\pi i\ap (q^\perp)^2}\ I'(\sigma^1,\sigma^2;\ell-q^\perp)\ ,}
with:
\eqn\Streinalm{\eqalign{
I'(\sigma^1,\sigma^2;\ell)&=\Lc(\sigma^2;\ell)\ \theta_1(2i\sigma^2,\tau)^{\ap (q^\parallel)^2}\cr
&\times \prod_{i=1}^{n_o}\theta_4(\sigma^1+i\sigma^2-x_i,\tau)^{\ap  qp_i}\ \theta_4(\sigma^1-i\sigma^2-x_i,\tau)^{\ap qp_i}\ .}}

According to \Start, eventually we need to integrate the coordinate $z$  over the full cylinder $\Cc$:
\eqn\need{
\int_{\Cc}d^2z\ I(z,\bar z;\ell)=-2i\int_0^1d\sigma^1\int_0^{t/2}d\sigma^2\ I(\sigma^1,\sigma^2;\ell)\ .}
Thus in \need\ we are supposed to integrate $\sigma^2$ along the (grey) edge $C_1$ in Fig.~3.
We seek to convert this integration to an integration along the imaginary axes $C_2,C_2'$ by considering the closed cycle \edges\ and using Cauchy's integral theorem stating, that\foot{Note, that the sum of residua of the poles of an elliptic function on the torus vanishes. 
Therefore, on a torus such a contour integral with $I$ an elliptic function vanishes trivially.}:
\eqn\cauchy{
\oint_{C}d\sigma^2\  \hat I(\sigma^1,\sigma^2;\ell)=0\ .}
The integrand $\hat I$ differs from $I$  by choosing a correct branch  when moving in the complex $\sigma^2$--plane and passing the branch cuts discussed above.
By using Cauchy's theorem \cauchy\ and taking into account the branch cuts when moving along the paths $C_2$ and $C_2'$ thanks to the relations \differ\ and \KlooAscher\ the $\sigma_2$--integral along $C_1$ can be expressed as
\eqn\Contouri{
d(\ell,q^\perp)\ \int_0^{\fc{t}{2}}d\sigma^2\ I(\sigma^1,\sigma^2;\ell)=-i\int_0^1d\tilde\sigma^2\ 
\lf[\ \hat I(\sigma^1,-i\tilde\sigma^2;\ell)-\hat I'(\sigma^1,-i\tilde\sigma^2;\ell-q^\perp)\ \ri]\ ,}
with the symbol:
\eqn\kernetr{
d(\ell,q^\perp)=1-e^{-2\pi i \ap \ell q^\perp}\ .}
In \Contouri\ the prime denotes the respective non--planar configuration \Streinalm\ at $C_2'$. On the other hand, the hat indicates that the integrands \relevant\ and \Streinalm\ have to be supplemented by appropriate  phase factors taking into account the branch points along $C_2$ and $C_2'$, respectively. Above for $C_2$ we have introduced the parameterization $\tilde\sigma^2:=i\sigma^2$ subject to $\sigma^2\in i(-1,0)$. Similarly  for   $C_2'$  we have $\sigma^2\in \fc{t}{2}+i(0,-1)$. Along $C_2$ we can introduce the following new real coordinates 
\eqn\newcoords{\eqalign{
\xi&=\sigma^1+\tilde\sigma^2\ ,\cr
\eta&=\sigma^1-\tilde\sigma^2\ ,}}
with $\det\lf(\fc{\p(\xi,\eta)}{\p(\sigma^1,\tilde\sigma^2)}\ri)=-2$.
With this parameterization the integrands in \Contouri\ can be specified as 
some function depending on $\xi$ and $\eta$ as
\eqn\wehave{\eqalign{
\hat I(\sigma^1,\sigma^2;\ell)&=\Pi_{C_2}(\{x_i\},\xi,\eta)\ I_{C_2}(\xi,\eta;\ell)\ ,\ \ \ \sigma^2\in C_2\ ,\cr
\hat I'(\sigma^1,\sigma^2;\ell)&=\Pi_{C_2'}(\{x_i\},\xi,\eta)\ I_{C_2'}(\xi,\eta;\ell)\ ,\ \ \ \sigma^2\in C_2'\ ,}}
with
\eqnn\integrands{
$$\eqalignno{\
I_{C_2}(\xi,\eta;\ell)&:=\Lc(\ell;\xi,\eta)\ |\theta_1(\xi-\eta)|^{\ap (q^\parallel)^2}\cr
&\times\prod_{i=1}^{n_o}|\theta_1(\xi-x_i)|^{\ap q p_i}\ 
|\theta_1(\eta-x_i)|^{\ap qp_i}\ ,\cr 
I_{C_2'}(\xi,\eta;\ell)&:=\Lc(\ell;\xi,\eta)\ |\theta_1(\xi-\eta)|^{\ap (q^\parallel)^2}\cr
&\times\prod_{i=1}^{n_o}\theta_4(\xi-x_i)^{\ap q p_i}\ \theta_4(\eta-x_i)^{\ap qp_i}\ ,&\integrands}$$}
with the loop momentum dependent factor: 
\eqn\Loopdepp{
\Lc(\ell;\xi,\eta)=\exp\lf\{-\h\pi\ap \tau_2\ell^2-i\pi\ap \ell q^\perp(\xi-\eta)\ri\} \ .}
The phases $\Pi_{C_2},\Pi_{C_2}'$ will be given below.
Now,  all the theta--functions of \integrands\  entering the integrand \Contouri\ are analytic and single--valued w.r.t. $\xi,\eta\in\IR$.

Due to \newcoords\ the region $(\sigma^1,\tilde\sigma^2)\in[0,1]^2$ is mapped to a rhombus 
accounting for the $\xi,\eta$--dependence of \integrands, cf. Fig.~4. 
\ifig\Region{Rhombus divided into the four domains $\Dc_i$}{\epsfxsize=0.5\hsize\epsfbox{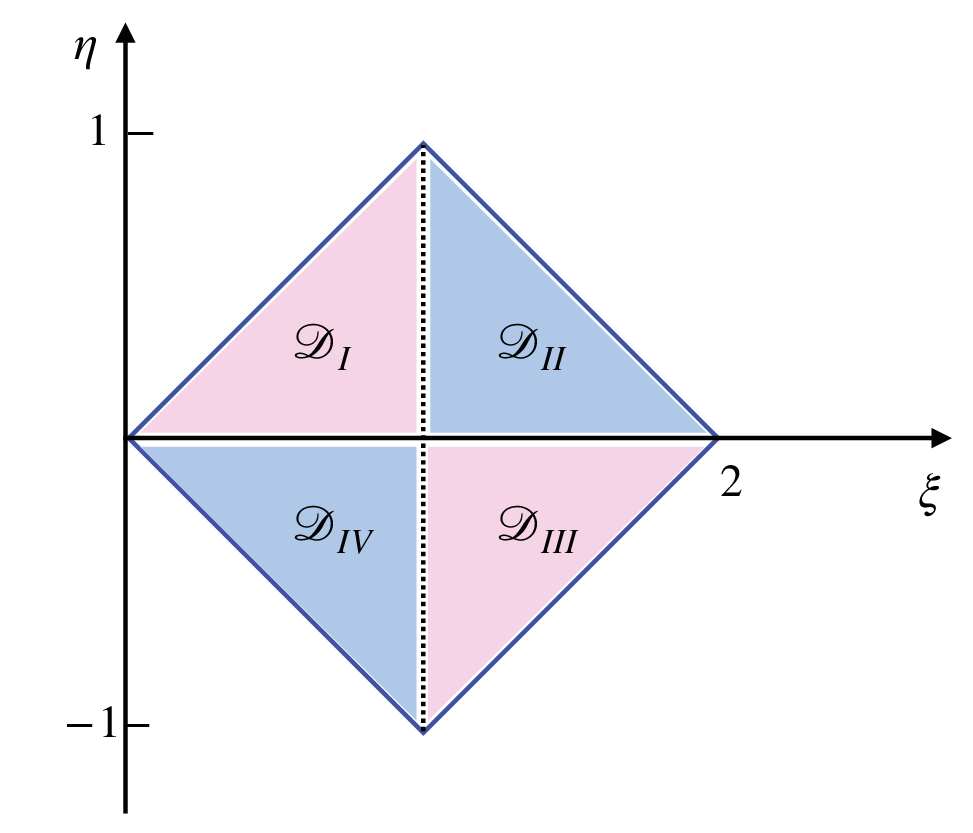}}
\noindent This diamond $\Dc=\Dc_{I}\cup\Dc_{II}\cup\Dc_{III}\cup\Dc_{IV}$ can be divided into the following four triangular domains
\eqn\domains{\eqalign{
\Dc_{I}&=\{(\xi,\eta)\ |\ 0<\xi<1\ \wedge\ 0\leq\eta\leq\xi\} \ ,\cr
\Dc_{II}&=\{(\xi,\eta)\ |\ 1<\xi<2\ \wedge\ 0\leq\eta\leq2-\xi\} \ ,\cr
\Dc_{III}&=\{(\xi,\eta)\ |\ 1<\xi<2\ \wedge\ \xi-2\leq\eta\leq0\}\ ,\cr
\Dc_{IV}&=\{(\xi,\eta)\ |\ 0<\xi<1\ \wedge\ -\xi\leq\eta\leq0\}  \ ,}}
depicted in Fig.~4.
In these four domains we can study the systems \integrands\ yet without specifying the phases $\Pi_{C_2},\Pi_{C_2}'$, which eventually reinstate the correct branch of the integrands \integrands\ when $\xi,\eta$ cross the open string points $x_i$. 
  After a careful inspection of the dependence of the local system \integrands\ on the regions \domains\ and performing changes of integration variables subject to  (B.10) we find that at $C_2$ by shifts in $\xi$ and $\eta$ the two blue triangles 
$\Dc_{II}$ and $\Dc_{IV}$ combine into the unit square $[0,1]^2$:
\eqn\Together{
\int_{\Dc_{II}}\hat I(\xi,\eta;\ell)+\int_{\Dc_{IV}}\hat I(\xi,\eta;\ell)=e^{-\pi i\ap\ell q^\perp}\int_0^1d\xi   \int_{0}^1 d\eta\ \hat I(\xi,\eta;\ell)\ .}
Similarly, for the red triangles $\Dc_{I}$ and $\Dc_{III}$ we find:
\eqn\Togetheri{\eqalign{
\int_{\Dc_{I}}\hat I(\xi,\eta;\ell)&=  \int_0^1d\xi    \int_0^{\xi} d\eta\ \hat I(\xi,\eta;\ell)\ ,\cr
\int_{\Dc_{III}}\hat I(\xi,\eta;\ell)&= e^{-2\pi i\ap\ell q^\perp}  
\int_0^1d\xi   \int_{\xi}^1 d\eta\ \hat I(\xi,\eta;\ell)\ .}}
Similar manipulations can be made at $C_2'$ by taking into account  the effect of the shift $\ell\ra\ell-q^\perp$. 
Eventually, with \Contouri, \Together\ and \Togetheri\ the complex integral \need\ over the complex closed string position $z$ can  be combined to the unit square $[0,1]^2$:
\eqn\Baeckeralm{\eqalign{
2i\ \sin\lf(\fc{\pi\ap\ell q^\perp}{2}\ri)\int_{\Cc}d^2z\ &I(z,\bar z;\ell)=e^{\h\pi i\ap\ell q^\perp} \int_0^1d\xi\int_0^{\xi}d\eta\ 
\lf[ \hat I(\xi,\eta;\ell)-\hat I'(\xi,\eta;\ell-q^\perp)\ri]\cr
&+ e^{-\h\pi i\ap\ell q^\perp}\  \int_0^1d\xi\int_{\xi}^1d\eta\ 
\lf[ \hat I(\xi,\eta;\ell)-\hat I'(\xi,\eta;\ell-q^\perp)\ri]\cr
&=\int_0^1d\xi\int_0^1d\eta\ 
e^{\h\pi i\ap\ell q^\perp\;{\rm sgn}(\xi-\eta)}\lf[ \hat I(\xi,\eta;\ell)-\hat I'(\xi,\eta;\ell-q^\perp)\ri].}}
Above we have used the trigonometric identity:
$$e^{\h\pi i\ap\ell q^\perp}\ \fc{1-e^{-2\pi i\ap\ell q^\perp}}{1+e^{-\pi i\ap\ell q^\perp}}=2i\ \sin\lf(\fc{\pi\ap\ell q^\perp}{2}\ri)\ .$$
We are now able to specify the phase factors $\Pi_{C_2},\Pi_{C_2'}$  accounting for the correct branch when moving in the unit interval $(\xi,\eta)\in[0,1]^2$ and crossing some of the branch cuts. For a generic configuration of $n_o$ open  string vertex positions $x_i$ the phase factors $\Pi_{C_2},\Pi_{C_2'}$ account for the correct branch of the integrand \need\ along $C_2$ and $C_2'$, respectively
\eqn\phases{\eqalign{
\Pi_{C_2}(\{x_i\},\xi,\eta)&=e^{-\pi i\ap (q^\perp)^2}\ e^{\pi i\ap(q^\parallel)^2\theta(\eta-\xi)}\ \prod_{i=1}^{n_o}e^{\pi i\ap qp_i\{1-\theta[(\xi-x_i)(\eta-x_i)]\}}\ ,\cr
\Pi_{C_2'}(\{x_i\},\xi,\eta)&=e^{-\pi i\ap (q^\perp)^2}\  e^{\pi i\ap(q^\parallel)^2\theta(\eta-\xi)}\ ,}}
with the step function $\theta$. Eventually, with \Contouri, the complex integral \need\ over the closed string vertex position $z$ becomes:
\eqn\final{\eqalign{
\int_{\Cc}d^2z\ I(z,\bar z;\ell)=\int_0^1d\xi\int_0^1d\eta\ \fs(\ell,q^\perp)& \ 
\lf[\ \Pi_{C_2}(\{x_i\},\xi,\eta)\ I_{C_2}(\xi,\eta;\ell)-\ri.\cr
&\lf.- \Pi_{C_2'}(\{x_i\},\xi,\eta)\ I_{C_2'}(\xi,\eta;\ell-q^\perp)\ \ri]\ ,}}
with the kernel:
\eqn\kernell{
\fs(\ell,q^\perp)^{-1}=2i\ \sin\lf(\fc{\pi\ap\ell q^\perp}{2}\ri)\ e^{-\h\pi i\ap\ell q^\perp\;{\rm sgn}(\xi-\eta)}\ .}

Eq. \final\ enables us to express the complex cylinder integral \need\ in terms of two terms  accounting for the contributions from $C_2$ and $C_2'$, respectively.
These two contributions, which are given by real
(iterated) integrals at the two boundaries of the cylinder correspond to pure open string one--loop subamplitudes. 
Thus, we have accomplished to convert one complex closed string coordinate $z$  into a pair of two real open string positions $\xi,\eta$ by means of considering a closed 
contour \edges\ in the complex $\sigma^2$--plane and applying Cauchy's theorem \cauchy, which results in the relation \final. This result enables us to express the complex cylinder integral \need\ over $z$ in terms two real  integrations along the cylinder boundaries. This manipulation  is illustrated in Fig.~5. 
\iifig\Region{Deforming the double cover coordinates $\pm\Im(z)$ of the cylinder}{to its boundary coordinate $\Re(z)$ resulting in $\xi,\eta$.}{\epsfxsize=0.6\hsize\epsfbox{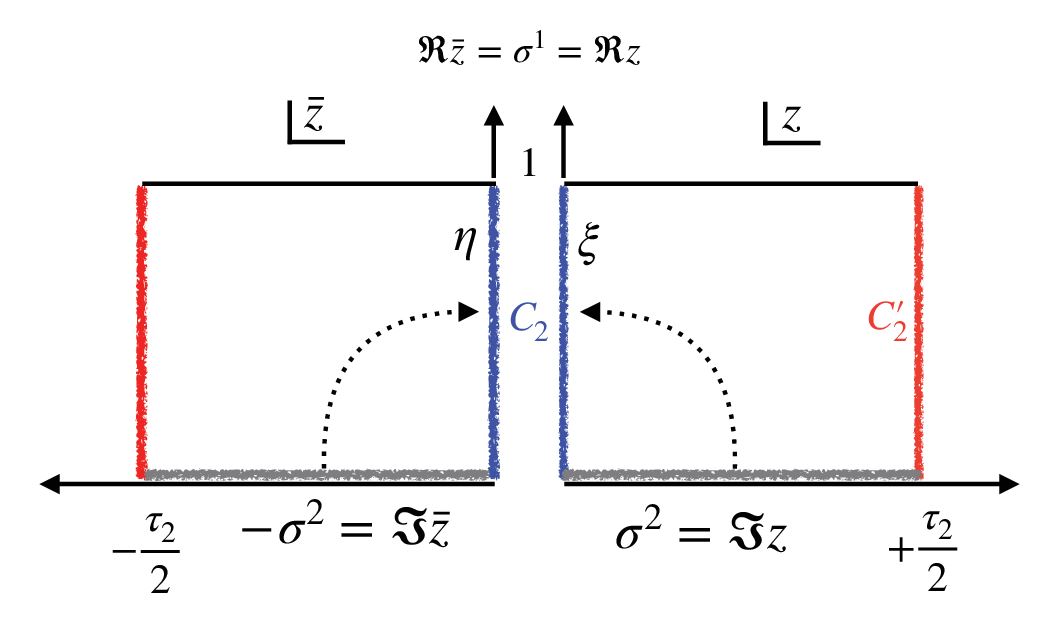}}
\noindent 
After splitting the complex cylinder coordinate as \sig\ the $\sigma^2\!=\!\Im(z)$--integration 
is deformed to be aligned along $\sigma^1\!=\!\Re(z)$ yielding the real integration variable 
$\xi$. Simultaneously, the $\Im(\bar z)$--integration 
is deformed to be aligned along $\Re(z)$ yielding the real integration variable $\eta$.  
In the double cover the real integrations $\xi,\eta$ describe (iterated) integrals along the two boundaries edges $C_2$ and $C_2'$ of two cylinders, cf. also Fig.~5.
As a consequence  the dependence on the complex closed string coordinate $z$ has been converted into contributions from a pair of open strings. 
On the cylinder that pair of open strings is located at either one of its boundaries.
The resulting non--planar cylinder configurations of $n_o\!+\!2$ open string positions are drawn in Fig.~6. 
\ifig\Region{Non--planar open string cylinder configurations corresponding to $C_2,C_2'$.}{\epsfxsize=0.95\hsize\epsfbox{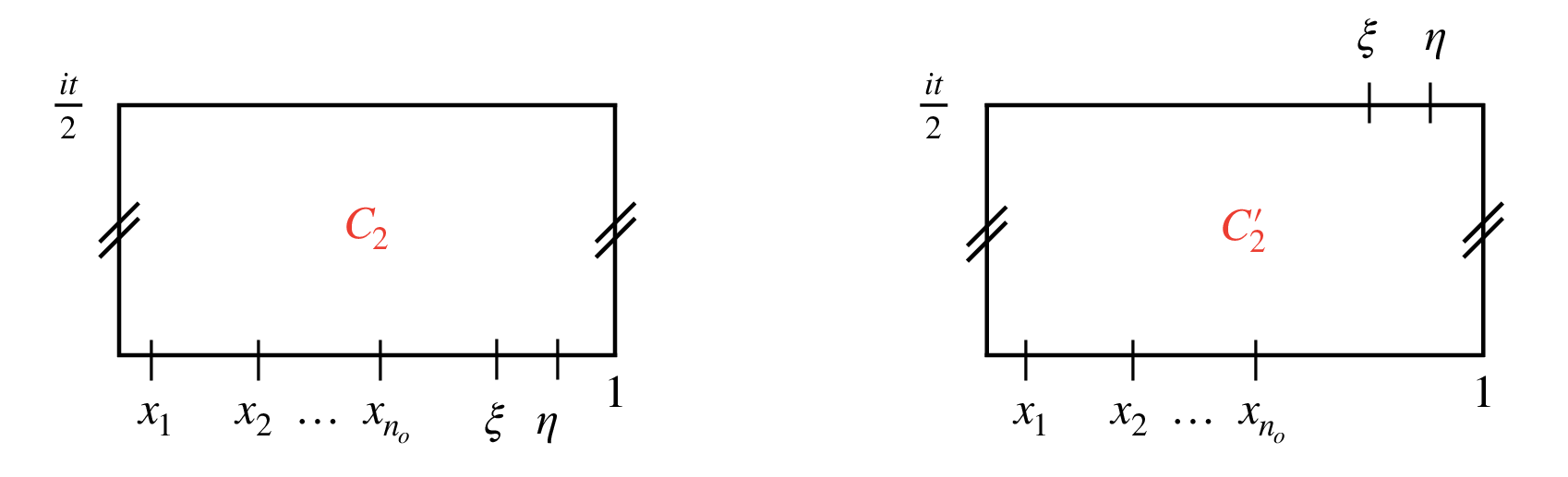}}
\noindent 
More precisely, the configuration referring to $C_2$ gives rise to a planar subamplitude with all $n_o$ open string positions and $\xi,\eta$ located at one cylinder boundary. On the other hand, the configuration referring to $C_2'$ amounts to a non--planar configuration with the pair $\xi,\eta$ located at the opposite boundary.
To get the full amplitude \Start\ we have to combine \final\ with the open string part \wehaveE\ subject to the planar case with $n_c=1$ and integrate over the loop momentum $\ell$.
%% For our planar case with $n_c=1$ the latter becomes:
%% \eqn\WehaveE{\
%% E(\{x_i,x_j\})\equiv
%%% \prod_{1\leq i<j\leq n_o}\lf|\theta_1(x_{j}-x_{i},\tau)\ri|^{2\ap p_ip_j}\ .}
In total we obtain for the full amplitude \Start\ 
\eqnn\traithen{
$$\eqalignno{
\hskip0.5cm A_{n_o,1}^{(1)}(\{1,&\ldots,n_o\}||\{q_L,q_R\})=g_o^{n_o}g_c\ \delta^{(d)}\lf(\sum_{i=1}^{n_o}p_i+q_L+q_R\ri)\int d\tau_2 \cr 
&\times \int_{-\infty}^\infty d^d\ell\ V_{CKG}^{-1} \left(\!\int_{\Ic}\prod_{i=1}^{n_o}dx_i\ri) \prod_{1\leq i<j\leq n_o}\lf|\theta_1(x_{j}-x_{i},\tau)\ri|^{2\ap p_ip_j}\cr
&\times\int_0^1d\xi\int_0^1d\eta\ |\theta_1(\xi-\eta)|^{\ap (q^\parallel)^2}\ 
\Lc(\ell;\xi,\eta)\ \ \hat Q(\tau,\{x_i,\xi,\eta\})&\traithen \cr
&\times \lf\{\ \fs(\ell,q^\perp)\ \Pi_{C_2}(\{x_i\},\xi,\eta)\ \prod_{i=1}^{n_o}|\theta_1(\xi-x_i)|^{\ap q p_i}\ 
|\theta_1(\eta-x_i)|^{\ap qp_i}\ri.\cr
&-\lf. \fs(\ell+q^\perp,q^\perp)\ \Pi_{C_2'}(\{x_i\},\xi,\eta)\ \prod_{i=1}^{n_o}\theta_4(\xi-x_i)^{\ap q p_i}\ \theta_4(\eta-x_i)^{\ap qp_i}\ \ri\},}$$}
with the expressions \integrands, \Loopdepp\ and \phases.
The function $\hat Q$ differs from $Q$ by the universal zero mode factor 
generated from the loop momentum integral \looprepr, i.e. $\hat Q=\lf(\fc{\ap\tau_2}{2}\ri)^{d/2}Q$.

Note, that in \Loopdepp\ the loop momentum $\ell$ can be split into its orthogonal $\ell^\perp$ and transversal $\ell^\parallel$ parts as:
\eqn\splitloop{
\ell=\ell^\parallel+\ell^\perp\ .}
Then, in \traithen\ the integral over $\ell^\parallel$ decouples and gives for a D$p$--brane its zero mode factor:
\eqn\decoup{
\int_{-\infty}^\infty d^{p+1}\ell^\parallel\ e^{-\h\pi\ap\tau_2(\ell^\parallel)^2}=\lf(\fc{\ap\tau_2}{2}\ri)^{-\fc{p+1}{2}}\ .}
On the other hand, for the orthogonal loop momentum $\ell^\perp$ dependent part we proceed as follows. We apply the relation \identVG, which now reads ($\tau=i\tau_2$):
\eqnn\identVGi{
$$\eqalignno{
\exp\lf\{-i\pi\ap \ell q^\perp(\xi-\eta)\ri\}
&=\lf(\fc{\theta_1(\xi+\fc{\tau}{2},\tau)}{\theta_1(\xi-\fc{\tau}{2},\tau)}\fc{\theta_1(\eta-\fc{\tau}{2},\tau)}{\theta_1(\eta+\fc{\tau}{2},\tau)}\ri)^{\h\ap \ell q^\perp}\cr
&=\lf(\fc{\theta_4(\xi+\fc{\tau}{2},\tau)}{\theta_4(\xi-\fc{\tau}{2},\tau)}\fc{\theta_4(\eta-\fc{\tau}{2},\tau)}{\theta_4(\eta+\fc{\tau}{2},\tau)}\ri)^{\h\ap \ell q^\perp}\hskip-0.5cm.&\identVGi}$$}
The above combinations of theta--function will be accommodated to describe interactions of two (auxiliary) open string states inserted at
\eqn\posORTH{
x_{0}=-\fc{\tau}{2}\ \ \ ,\ \ \ x_{0}'=+\fc{\tau}{2}\ ,}
with the momenta:
\eqn\chargeORTH{
p_{0}=\h\ell^\perp\ \ \ ,\ \ \ p_{0}'=\h D\ell^\perp\ .}
Note, that these definitions are to be compared to the closed string case \Split\ encompassing \posiVG\ and \charge. They give rise  to two additional open string insertions at \posORTH, which only interact with the transverse part of the closed string momentum $q^\perp$.
We have $\ell q^\perp=\ell^\perp q$ and $l^\perp q^\parallel=0$ from which we derive 
$\ell^\perp q=-\ell^\perp Dq$. Moreover, we have $p_0+p_0'=\h(\ell^\perp+D\ell^\perp)=0$.

For a given ordered configuration of the vertex operator positions $\xi,\eta\in(0,1)$ relative to $x_i$ along one of the cylinder boundaries (cf.~Fig.~6) all the phase factors entering \phases\ and \kernell\ are constants depending only on the external momenta. 
Therefore, we may represent \traithen\ as a sum over open string non--planar subamplitudes.
For this purpose let us introduce the one--loop non--planar open string subamplitude involving $N\!=\!n_1\!+\!n_2$ open strings which follows from  \Start\ by dropping 
the closed string part
\eqnn\Open{
$$\eqalignno{
A^{(1)}_N(\Ac|\Bc)&=g_o^N\delta^{(d)}\lf(k_{0}+k_{N+1}+\sum_{i=1}^{N}k_i\ri)\ V_{CKG}^{-1} \ \int d\tau_2\left(\!\int_{\Ic_\Ac}\prod_{i=1}^{n_1}dx_{\al_i}\int_{\Ic_\Bc}\prod_{j=1}^{n_2}dx_{\bet_j}\!\right)\cr
&\times \prod_{1\leq i<j\leq n_1}\lf|\theta_1(x_{\al_j}-x_{\al_i},\tau)\ri|^{2\ap k_{\al_i}k_{\al_j}}
\prod_{0\leq i<j\leq n_2+1\atop(i,j)\neq (0,n_2+1)}\lf|\theta_1(x_{\bet_j}-x_{\bet_i},\tau)\ri|^{2\ap k_{\bet_i}k_{\bet_j}}\cr
&\times\prod_{1\leq i\leq n_1\atop 0\leq j\leq n_2+1}\lf|\theta_4(x_{\bet_j}-x_{\al_i},\tau)\ri|^{2\ap k_{\al_i}k_{\bet_j}}\ \hat Q(\tau,\{x_i\})\ ,&\Open }$$}
with some non--planar configurations $\Ac\!=\!\{\alpha_1,\ldots,\alpha_{n_1}\}, \Bc\!=\!\{\beta_0,\beta_1,\ldots,\beta_{n_2},\beta_{n_2+1}\}$ and the integration domain \domain. Supplementary, the amplitude \Open\ is 
extended by the two additional open string insertions \posORTH\ with momenta \chargeORTH\ 
at the second cylinder boundary. More precisely, we have
$x_{\beta_0}=-\fc{\tau}{2},\ x_{\beta_{n_2+1}}=+\fc{\tau}{2}$ with momenta $k_{\bet_0}=p_0$ and $k_{\bet_{n_2+1}}=p_0'$, respectively.
Our open string configuration in \traithen\ can be described by
the following $N+2$ (with $N=n_o+2$) open string momenta entering the open string amplitudes \Open
\eqn\mautt{\eqalign{
k_r&=p_r\ ,\ \ \ r=1,\ldots,N-2\ ,\cr
k_{N-1}&=q_L=\h q\ ,\ \ \ k_{N}=q_R=\h Dq\ ,\cr
k_0&=\h\ell^\perp\ ,\ \ \ \ \ \ k_{N+1}=\h D\ell^\perp\ ,}}
while the $N+2$ open string insertions $x_s$ along the two boundaries of the cylinder:
\eqn\Mautt{\eqalign{
x_r&\ ,\ \ \ r=1,\ldots,N-2\ ,\cr
x_{N-1}&=\xi\ ,\ \ \ x_N=\eta\ ,\cr
x_0&=-\fc{\tau}{2}\ ,\ \ \ \ \ \ x_{N+1}=+\fc{\tau}{2}\ .}}
In the following we shall collect all phases arising from \phases\ and \kernell\ and exploit their  net effect in \traithen.  
Those phases  are familiar from the tree--level KLT relations \KawaiXQ\ or \StiebergerVYA\ 
involving the KLT kernel \doubref\BernSV\BjerrumBohrHN. The closed string coordinates  $\xi,\eta\in(0,1)$ give rise to an ordering along one cylinder boundary with regard to the open string positions $x_i$. 
For either ordered configuration of $\xi$ and $\eta$ along the two boundaries of the cylinder (cf. Fig.~6), respectively  the  result \traithen\  represents a sum over  one--loop $N$--point open string cylinder amplitudes \Open, with  the additional two open string insertions \posORTH. 
%%   For two sets  $\Ar,\Br$ each referring to a set of ordered points $x_{a_i},x_{b_j}$, 
%%  respectively  along one cylinder boundary we introduce the kernel 
%% \eqn\Bohrkern{
%%  S(\Ar|\Br):=\prod_{i=1}^{a}\prod_{j=1}^{b}e^{2\pi i\ap p_{a_i}p_{b_j} 
%%   \theta(x_{a_i}-%%  x_{b_j})}\ ,}
%%%  with the cardinalities $a=|\Ar|$ and $b=|\Br|$. 
Eventually, with $g_c=g_o^2$ we can express \traithen\ in the following way
\eqnn\Traithen{
$$\eqalignno{
A_{n_o,1}^{(1)}(\{1,\ldots,n_o\}&||\{k_{N-1},k_N\})=(2i)^{-1}\lf(\fc{\ap\tau_2}{2}\ri)^{-\fc{p+1}{2}}\ \int_{-\infty}^\infty d^{d-p-1}\ell^\perp\ e^{-\h\pi\ap\tau_2(\ell^\perp)^2}\cr &\times\lf\{\ \sin\lf(2\pi\ap k_{N-1}k_0\ri)^{-1}\sum_{\Ac \in OP(\Pc_{n_o},\Oc)} S(\Ac)S(\Oc)\ A_N^{(1)}(\Ac)\ri.&\Traithen\cr 
& \lf.+\sin\lf(2\pi\ap k_{N-1}(k_0-k_N)\ri)^{-1}\sum_{\Ac=\Pc_{n_o} } S(\Oc)\ 
A_N^{(1)}(\Ac|\Oc)\ \ri\}}$$} 
as sum over the $N$--point  non--planar open string subamplitudes $A^{(1)}_N$ defined in \Open, with $N=n_o+2$ and some kernels $S$ accounting for all phases. Above we have introduced the two sets $\Pc_{n_o}=\{1,\ldots,n_o\}$ and $\Oc=\{N-1,N\}$
referring to open string orderings and $OP$ denotes the ordered products of two sets.

\subsec{Amplitudes with $n_c$ closed strings and $n_o$ open strings}

In the previous subsection we have exhibited in full detail our method for the case of one closed string $n_c\!=\!1$. Here we shall generalize these steps to an arbitrary number $n_c$ of closed strings. Again, we consider the canonical open string orderings $\Ac=\{1,\ldots,n_1\}$ and $\Bc=\{n_1+1,\ldots,n_2\}$. In what follows we shall use $p_i:=p_{\al_i},\ i=1,\ldots,n_1$ and $p_j'=p_{\bet_j},\ j=1,\ldots,n_2$, with  $n_o=n_1+n_2$.

The combination of theta--functions
in the integrand \GrosseWolfsschlucht\ is single--valued in the complex closed string coordinates $z_r$. Following \sig\ for each  $z_r$ let us write 
\eqn\Sig{
z_r=\sigma_r^1+i\sigma_r^2\ \ \ ,\ \ \ \sigma^1_r\in(0,1)\ ,\ \sigma^2_r\in(0,\fc{t}{2})\ \ \ ,\ \ \ r=1,\ldots,n_c\ ,}
and discuss the holomorphic dependence of the integrand \GrosseWolfsschlucht\ on the closed string coordinates $\sigma_r^1,\sigma^2_r$. Therefore, we introduce the function 
\eqnn\Relevant{
$$\eqalignno{
I(\{\si^1_s,\si^2_s\};\ell)&=\Lc(\{\si^2_s\};\ell)\ \prod_{r=1}^{n_c}\theta_1(2i\si_r^2)^{\ap(q_r^\parallel)^2}\prod_{r<s}^{n_c}
\Theta(\sigma^1_r,\sigma^2_r;\sigma^1_s,\sigma^2_s)\cr
&\times\prod_{r=1}^{n_c} \Theta_1(\sigma^1_r,\sigma^2_r)\ \prod_{r=1}^{n_c} \Theta_4(\sigma^1_r,\sigma^2_r)\ ,&\Relevant}$$}
\eqn\abki{\eqalign{
\Theta(\sigma^1_r,\sigma^2_r;\sigma^1_s,\sigma^2_s)&=\theta_1(\si^1_s+i\si^2_s-\si^1_r-i\si_r^2,\tau)^{\h\ap q_{s} q_{r}}\ \theta_1(\si^1_s-i\si^2_s-\si^1_r+i\si_r^2,\tau)^{\h\ap q_{s} q_{r}}\cr
&\times\theta_1(\si^1_s+i\si^2_s-\si^1_r+i\si_r^2,\tau)^{\h\ap q_{r} Dq_{s}}\ \theta_1(\si^1_s-i\si^2_s-\si^1_r-i\si_r^2,\tau)^{\h\ap q_{r} Dq_{s}}\ ,}}
with the property $\Theta(\sigma^1_r,\sigma^2_r;\sigma^1_s,\sigma^2_s)=
\Theta(\sigma^1_s,\sigma^2_s;\sigma^1_r,\sigma^2_r)$, 
\eqn\abkii{\eqalign{
\Theta_1(\sigma^1_r,\sigma^2_r)&=\prod_{i=1}^{n_1}\theta_1(\si^1_r+i\si^2_r-x_i,\tau)^{\ap q_{r} p_i}\ \theta_1(\si^1_r-i\si^2_r-x_i,\tau)^{\ap q_{r}p_i}\ ,\cr
\Theta_4(\sigma^1_r,\sigma^2_r)&=\prod_{j=1}^{n_2}\theta_4(\si^1_r+i\si^2_r-x_j',\tau)^{\ap q_{r} p_j'}\ \theta_4(\si^1_r-i\si^2_r-x_j',\tau)^{\ap q_{r} p_j'}\ ,}}
and the non--holomorphic factor
\eqn\Nonholoo{
\Lc(\{\si^2_s\};\ell)=\exp\lf\{-\h\pi\ap \tau_2\ell^2+2\pi \ap\ell\sum_{r=1}^{n_c}
q^\perp_r\si^2_r\ri\}}
following from   \KleineWolfsschlucht.
In \Start\ the integration of all closed string coordinates $z_r$ over the full cylinder~$\Cc$ becomes
\eqn\Need{
\lf(\prod_{r=1}^{n_c}\int_{\Cc}d^2z_r\ri)\ I(\{z_s,\bar z_s\};\ell)=(-2i)^{n_c}\ \lf(\prod_{r=1}^{n_c}\int_0^1d\sigma^1_r\int_0^{t/2}d\sigma^2_r\ri)\ I(\{\sigma^1_s,\sigma^2_s\};\ell)\ ,}
with the integrand \Relevant\ to be supplemented by the open string part \KN\ later.

The expressions \Relevant\ and \Nonholoo\ are holomorphic in all the coordinates $\sigma_r^2,\ r=1,\ldots,n_c$.
In the previous subsection we have exhibited how one complex closed string coordinate $z$ is 
converted into a pair of two real open string positions $\xi,\eta$ by means of exploiting  a closed 
contour \edges\ in the complex $\sigma^2$--plane and applying Cauchy's theorem \cauchy\  resulting in the relation \final. 
We want to repeat the steps from the previous subsection and perform an analytic continuation of all closed string coordinates $\sigma_t^2,\ t=1,\ldots,n_c$, 
defined in \Sig. We shall discuss the dependence of the integrands \Relevant\ and \Nonholoo\ in  the complex $\sigma^2_t$--planes and consider\foot{This can be achieved by discussing forgetful fibrations of the moduli space of the cylinder $\Cc$ with $n_c$ punctures,  i.e. integrating one puncture $z_t$ at a time while keeping all other $n_c\!-\!1$ punctures fix. Then forgot this point and repeat these steps for $n_c\!-\!2$ punctures. This map removes one puncture at each step.} in each complex $\sigma^2_t$--plane closed cycles (polygons) $C$ defined by the four edges \edges. We simultaneously deform all of the $\sigma^2_t$--integration contours from the real axis to the pure imaginary axes $C_2,C_2'$, i.e. trade in a $\sigma_r^2$ integration by means of deforming its integration region $C_1$ to $C_2,C_2'$. This way we obtain real coordinates $\xi_t,\eta_t$ given by 
\eqn\Newcoords{\eqalign{
\xi_t&=\sigma^1_t+\tilde\sigma^2_t\ ,\cr
\eta_t&=\sigma^1_t-\tilde\sigma^2_t\ ,}}
with $\tilde \si^2_t=i\sigma^2_t,\ t=1,\ldots,n_c$ and $\det\lf(\fc{\p(\xi_t,\eta_t)}{\p(\sigma^1_t,\tilde\sigma^2_t)}\ri)=-2$.
In terms of  the real coordinates \Newcoords\
  at $\sigma^2_t\in C_2$ \Relevant\ becomes the holomorphic function
\eqnn\Relevantt{
$$\eqalignno{
&\hskip0.75cm I (\{\xi_s,\eta_s\};\ell)\equiv\Lc(\{\xi_s,\eta_s\};\ell)\ \prod_{r=1}^{n_c}\theta_1(\xi_r-\eta_r)^{\ap(q_r^\parallel)^2}&\Relevantt\cr 
&\hskip0.75cm\times\prod_{r<s}^{n_c}\theta_1(\xi_s-\xi_r,\tau)^{\fc{\ap}{2} q_{s} q_{r}}\; \theta_1(\eta_s-\eta_r,\tau)^{\fc{\ap}{2} q_{s} q_{r}}\;\theta_1(\xi_s-\eta_r,\tau)^{\fc{\ap}{2} q_{r} Dq_{s}}\; \theta_1(\eta_s-\xi_r,\tau)^{\fc{\ap}{2} q_{r} Dq_{s}}\cr
&\times\prod_{r=1}^{n_c}\prod_{i=1}^{n_1}\theta_1(\xi_r-x_i,\tau)^{\ap q_{r} p_i}\; \theta_1(\eta_r-x_i,\tau)^{\ap q_{r}p_i}\prod_{r=1}^{n_c}\prod_{j=1}^{n_2} \theta_4(\xi_r-x_j',\tau)^{\ap q_{r} p_j'}\;\theta_4(\eta_r-x_j',\tau)^{\ap q_{r} p_j'},}$$}
with:
\eqn\LoopF{
\Lc(\{\xi_s,\eta_s\};\ell)=\exp\lf\{-\h\pi\ap \tau_2\ell^2-\pi i\ap\ell
\sum_{r=1}^{n_c}q_r^\perp(\xi_r-\eta_r)\ri\}\ .}

For each coordinate $\sigma^2_t$ we consider the same closed cycle \edges\ in the complex $\sigma^2_t$--plane.
However, in the multi--dimensional complex  space $\sigma^2\in\IC^{n_c}$ there are now many branch points, all of them lying along the imaginary axes given by $\sigma^2_s-\sigma^2_r=\pm i \sigma^1_{sr}$ and $\sigma^2_s+\sigma^2_r=\pm i \sigma^1_{sr}$. These conveniently divide themselves into pairs (given by $z_s-z_r=0$ and $\bar z_s-\bar z_r=0$ as well as  $z_s-\bar z_r=0$ and $\bar z_s- z_r=0$, respectively).
Furthermore,  since $x_i,x_j'\in (0,1)$ the function \Relevant\ as a holomorphic function in $\sigma^2_{r}$  has $2n_1$ branch points at $\sigma^2_{r}=\pm i(\sigma^1_{r}-x_i)$ along the imaginary axis $\sigma^2_{r}=0$ and $2n_2$ branch points at
$\sigma_{r}^2=\pm i(\sigma^1_r-x_i')+\fc{t}{2}$ along $\sigma^2_r=\fc{\tau_2}{2}$. Finally, there are also $n_c$ branch points at the origin $\sigma^2_r=0$.
These branch cut structures may be examined essentially in the same way as in the previous subsection.
Again, by applying  Cauchy's integral theorem we shall convert the integration $C_1$ of $\sigma^2_t$ along the real $\sigma_t^2$--axis  to an integration along the two  axes given by $C_2:\ \sigma_t^2=0$ and $C_2':\ \sigma^2_t=\fc{\tau_2}{2}$, respectively.

Let us consider a generic closed string position $z_t$ with $t\in \{1,\ldots,n_c\}$. 
Eventually, we need to integrate the $z_t$ coordinate over the full cylinder  $\Cc$:
\eqn\needii{
\int_{\Cc}d^2z_t\ I(z_t,\bar z_t;\ell)=-2i\int_0^1d\sigma^1_t\int_0^{\tau_2/2}d\sigma^2_t\ I(\sigma^1_t,\sigma^2_t;\ell)\ .}
As above we shall explore  the holomorphic $\sigma^2_t$--dependence of $I(\{\sigma^1_s,\sigma^2_s\};\ell)$ along the edges \edges\ in the complex $\sigma^2_t$--plane. The latter are depicted in the figure  Fig.~3 with $\sigma^2$ replaced by $\sigma^2_t$. 
Thus in \needii\ we are supposed to integrate $\sigma_t^2$ along the (grey) edge $C_1$ depicted in Fig.~3.
We seek to convert this integration to an integration along the imaginary axis $C_2,C_2'$ by considering the closed cycle  $C$  and using Cauchy's integral theorem stating, that:
\eqn\Cauchy{
\oint_{C}d\sigma_t^2\  \hat I(\sigma^1_t,\sigma^2_t;\ell)=0\ .}
The integrand $\hat I$ differs from $I$  by choosing a correct branch  when moving in the complex $\sigma_t^2$--plane and passing the branch cuts discussed above.
Moving $\sigma^2_t$ from $C_2$ to $C_2'$ gives rise to a different (non--planar) 
configuration of theta--functions than \Relevant. To describe this effect let us define
\eqnn\Streinalmi{
$$\eqalignno{
I_{\al,\bet}(\{\si^1_s,\si^2_s\};\ell)&=\Lc(\{\si^2_s\};\ell)\ \prod_{r=1}^{n_c}\theta_1(2i\si_r^2)^{\ap(q_r^\parallel)^2}&\Streinalmi\cr 
&\times\prod_{r,s\in\al\atop r<s}\Theta(\sigma^1_r,\sigma^2_r;\sigma^1_s,\sigma^2_s)\ 
\prod_{r,s\in\beta\atop r<s}\Theta(\sigma^1_r,\sigma^2_r;\sigma^1_s,\sigma^2_s)\ 
\prod_{r\in\al\atop s\in\bet}\Theta_T(\sigma^1_r,\sigma^2_r;\sigma^1_s,\sigma^2_s)\cr
&\times\prod_{r\in\al}\Theta_1(\sigma^1_r,\sigma^2_r)\ \prod_{r\in\al}\Theta_4(\sigma^1_r,\sigma^2_r)\ \prod_{r\in\bet}\Theta_{1T}(\sigma^1_r,\sigma^2_r)\ \prod_{r\in\bet}\Theta_{4T}(\sigma^1_r,\sigma^2_r)\ ,}$$}
with $\alpha\cup\beta=\{1,\ldots,n_c\}$, 
\eqn\Abki{\eqalign{
\Theta_T(\sigma_r^1,\si^2_r;\sigma_s^1,\si_s^2)&=\theta_4(\si^1_s+i\si^2_s-\si^1_r-i\si_r^2,\tau)^{\h\ap q_{s} q_{r}}\ \theta_4(\si^1_s-i\si^2_s-\si^1_r+i\si_r^2,\tau)^{\h\ap q_{s} q_{r}}\cr
&\times\theta_4(\si^1_s+i\si^2_s-\si^1_r+i\si_r^2,\tau)^{\h\ap q_{r} Dq_{s}}\ \theta_4(\si^1_s-i\si^2_s-\si^1_r-i\si_r^2,\tau)^{\h\ap q_{r} Dq_{s}}\ ,}}
with the property $\Theta_T(\sigma^1_r,\sigma^2_r;\sigma^1_s,\sigma^2_s)=
\Theta_T(\sigma^1_s,\sigma^2_s;\sigma^1_r,\sigma^2_r)$, and:
\eqn\abkiii{\eqalign{
\Theta_{1T}(\sigma^1_r,\sigma^2_r)&=\prod_{i=1}^{n_1}\theta_4(\si^1_r+i\si^2_r-x_i,\tau)^{\ap q_{r} p_i}\ \theta_4(\si^1_r-i\si^2_r-x_i,\tau)^{\ap q_{r}p_i}\ ,\cr
\Theta_{4T}(\sigma^1_r,\sigma^2_r)&=\prod_{j=1}^{n_2}\theta_1(\si^1_r+i\si^2_r-x_j',\tau)^{\ap q_{r} p_j'}\ \theta_1(\si^1_r-i\si^2_r-x_j',\tau)^{\ap q_{r} p_j'}\ .}}
The integrands \Relevant\ and \Streinalmi\ along $C_1'$ differ from that along $C_1$ by a phase factor
\eqn\Differ{
I_{\al,\bet}(\sigma^1_t,\sigma^2_t-i;\ell)=\vartheta_t\ I_{\al,\bet}(\sigma^1_t,\sigma^2_t;\ell)\ ,}
given by
\eqn\phasep{
\vartheta_t:=e^{-2\pi i\ap q^\perp_t\ell}\ ,}
following from  \Nonholoo, i.e.: 
$\Lc(\sigma^1_t,\sigma^2_t-i;\ell)=\vartheta_t\Lc(\sigma^1_t,\sigma^2_t;\ell)$.
Furthermore, moving $\sigma^2_t$ from $C_2$ to $C_2'$ involves a shift in the loop momentum $\ell$
\eqn\shiff{
\ell\longrightarrow \ell-q_t^\perp\ ,}
leading to
\eqn\KlooAscheri{
I_{\al,\bet}(\sigma^1_t,\sigma^2_t+\fc{\tau_2}{2};\ell)=e^{-\pi i\ap (q_t^\perp)^2}\ I_{\al',\bet'}(\sigma^1_t,\sigma^2_t;\ell-q^\perp_t)\ ,}
with $\alpha'=\alpha\slash\{t\}$ and $\beta'=\beta\cup\{t\}$.

In the following in \Need\  for each $\sigma_t^2$ integration we consider the contour \edges\ in the complex $\sigma_t^2$--plane, cf. Fig.~3 with $\sigma^2$ replaced by $\sigma^2_t$.
We trade the $\sigma_t^2$--integration along the edges $C_1,C_1'$ for integrations along  the edges $C_2,C_2'$. Taking the 
coordinate $\sigma_t^2$ and displaying only the dependence on $\sigma_t^1,\sigma^2_t$  we can write for \Relevant\ or \Streinalmi
\eqn\CONTOUR{
d(l,q_t^\perp)\ \int_{C_1}d\sigma_t^2\ I_{\al,\bet}(\sigma_t^1,\si^2_t;\ell)
=-\int_{C_2}d\sigma_t^2\ \hat I_{\al,\bet}(\sigma_t^1,\si_t^2;\ell)-\int_{C_2'}d\sigma_t^2\
\hat I_{\al',\bet'}(\sigma_t^1,\si_t^2;\ell-q^\perp_t)\ ,}
with the symbol:
\eqn\kernelI{
d(l,q_t^\perp)=1-e^{-2\pi i\ap l q^\perp_t}\ .}
The prime denotes the respective non--planar configuration at $C_2'$ following from \KlooAscheri. On the other hand, the hat indicates that the integrand \Relevant\ has to be supplemented by appropriate  phase factors taking into account the branch points along $C_2$ and $C_2'$, respectively. 
We need to repeat the steps from above and perform an analytic continuation of all closed string coordinates $\sigma_t^2,\;t=1,\ldots,n$, defined in \Sig.
We establish for each coordinate $\sigma_t^2$ an equation of the sort \CONTOUR.
This way, the $n_c$ closed strings are traded for
$2n_c$ open strings together with a transverse loop momentum flow $\ell^\perp$ between the two cylinder boundaries. While omitting the open strings this is illustrated in Fig.~7. 
\iifig\Region{Trading $n_c$ closed strings into configurations of $2n_c$ non--planar open strings}{subject to a transverse loop momentum flow $\ell^\perp$ between the two cylinder boundaries.}{\epsfxsize=0.9\hsize\epsfbox{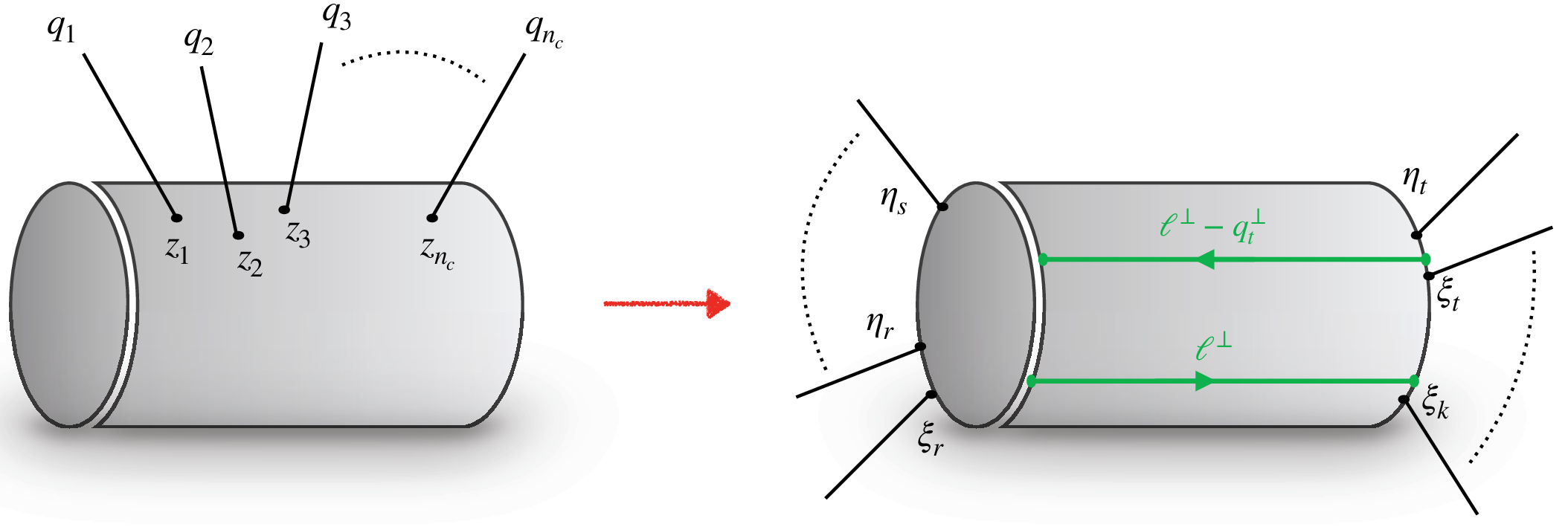}}
\noindent 
Including open strings is straightforward but tedious and does not change the generic picture. In this case there are more open strings attached to the two cylinder boundaries whose branch cut structure is taken into account by additional phase factors at $C_2,C_2'$.
In the following let us detail our method by one example and  concentrate on 
the case $n_o\!=\!0$.
In the sequel we investigate the  setup $D\!=\!-{\bf 1}$ introduced in Subsection 2.2.4. For this case we first deal with the expression \vordereGoingerHalt\ representing  the integrand before the loop momentum $\ell$ integration and then perform its integration to get the amplitude \HintereGoingerHalt.

\subsec{Cylinder amplitudes with $n_c$ closed strings}

After introducing the parameterisation \Sig\ the integrand of \WochenbrunnerAlm\ becomes ($n_c=n$)
\eqn\GoingerHalt{
I(\{\sigma^1_p,\sigma^2_p\};\ell)= \Lc(\{\sigma_p^2\};\ell)\ \prod_{r<s}^{n}\ \Theta(\sigma_r^1,\si^2_r;\sigma_s^1,\si_s^2)\ ,}
with the loop momentum factor
\eqn\Loopfactor{
\Lc(\{\sigma_p^2\};\ell):=\exp\lf\{-\h\pi\ap \tau_2\ell^2+2\pi \ap\ell\sum_{r=1}^{n}
q_r\sigma_r^2\ri\}\ ,}
and
\eqn\abk{
\Theta(\sigma_r^1,\si^2_r;\sigma_s^1,\si_s^2)=\lf(\fc{\theta_1(\sigma^1_s+i\sigma^2_s-\sigma^1_r-i\sigma^2_r,\tau)}
{\theta_1( \sigma^1_s+i\sigma^2_s-\sigma^1_r+i\sigma^2_r,\tau)}\ \fc{\theta_1(\sigma^1_s-i\sigma^2_s-\sigma^1_r+i\sigma^2_r,\tau)}
{\theta_1( \sigma^1_s-i\sigma^2_s-\sigma^1_r-i\sigma^2_r,\tau)}\ri)^{\h\ap q_r q_s}\ ,}
with the property $\Theta(\sigma_r^1,\si^2_r;\sigma_s^1,\si_s^2)=\Theta(\sigma_s^1,\si^2_s;\sigma_r^1,\si_r^2)$ and following from \abki.
In what follows similarly to \Streinalmi\ we shall also pay attention to object 
\eqn\Object{\eqalign{
I_{\alpha,\beta}(\{\sigma^1_p,\sigma^2_p\};\ell)&=\Lc(\{\sigma_p^2\};\ell)\ 
\prod_{r,s\in\alpha\atop r<s}^{n}\ \Theta(\sigma_r^1,\si^2_r;\sigma_s^1,\si_s^2)\ \prod_{r,s\in\beta\atop r<s}^{n}\ \Theta(\sigma_r^1,\si^2_r;\sigma_s^1,\si_s^2)\cr
&\times\prod_{r\in\alpha\atop s\in\beta}^{n}\Theta_T(\sigma_r^1,\si^2_r;\sigma_s^1,\si_s^2)\ ,}}
with  $\alpha\cup\beta=\{1,\ldots,n\}$,
\eqn\Abk{
\Theta_T(\sigma_r^1,\si^2_r;\sigma_s^1,\si_s^2)=\lf(\fc{\theta_4(\sigma^1_s+i\sigma^2_s-\sigma^1_r-i\sigma^2_r,\tau)}
{\theta_4( \sigma^1_s+i\sigma^2_s-\sigma^1_r+i\sigma^2_r,\tau)}\ \fc{\theta_4(\sigma^1_s-i\sigma^2_s-\sigma^1_r+i\sigma^2_r,\tau)}
{\theta_4( \sigma^1_s-i\sigma^2_s-\sigma^1_r-i\sigma^2_r,\tau)}\ri)^{\h\ap q_r q_s}\ ,}
and $\Theta_T(\sigma_r^1,\si^2_r;\sigma_s^1,\si_s^2)=\Theta_T(\sigma_s^1,\si^2_s;\sigma_r^1,\si_r^2)$  and following from \Abki.

For the moment let us single out one particular coordinate $\sigma^2_t$ with $t\in\{1,\ldots,n\}$ and 
investigate the $\sigma_t^2$--dependence of \GoingerHalt.
More precisely we shall be concerned with the local system
\eqn\Local{\eqalign{
I(\{\sigma^1_p,\sigma^2_p\};\ell)\hookrightarrow  I(\sigma^1_t,\sigma^2_t;\ell)\ ,}}
which is holomorphic in the coordinate $\sigma_t^2$. Despite the expression \Local\ is the same as \Relevant, in this subsection we only display its  dependence on $\sigma^1_t$ and $\sigma^2_t$ and consider the remaining coordinates as spectators. We want to discuss the dependence of \Local\ in  the complex $\sigma^2_t$--plane and consider the closed cycle (polygon) defined by \edges.
First, we determine  in \Differ\  the loop--momentum dependent phase
\eqn\Phase{
\vartheta_t=e^{-2\pi i \ap\ell q_t}\ ,}
i.e. we have $I(\sigma^1_t,\sigma^2_t-i;\ell)=\vartheta_t\; 
I(\sigma^1_t,\sigma^2_t;\ell)$.
On the other hand, shifting $\sigma_t^2$ by a half--period $\fc{\tau_2}{2}$ involves a shift
in the loop momentum $\ell$ 
\eqn\shiftloop{
\ell\longrightarrow \ell-q_t\ ,}
leading to
\eqn\WochenbrunnerAlm{
I_{\alpha,\beta}(\sigma_t^1,\sigma_t^2+\fc{\tau_2}{2};\ell)=I_{\alpha',\beta'}(\sigma_t^1,\sigma_t^2;\ell-q_t)\ ,}
with $\alpha'=\alpha\slash\{t\}$ and $\beta'=\beta\cup\{t\}$.
In the following in \Need\  for each $\sigma_t^2$ integration we consider the contour \edges\ in the complex $\sigma_t^2$--plane, cf. Fig.~3, with $\sigma^2$ replaced by $\sigma^2_t$.
We trade the $\sigma_t^2$--integration along the edges $C_1,C_1'$ for integrations along  the edges $C_2,C_2'$. Similarly to \CONTOUR\ starting with the 
coordinate $\sigma_t^2$  we can write for the local system (displaying only the dependence on $\sigma_t^1,\sigma^2_t$) 
\eqn\Contour{
d(l,q_t)\ \int_{C_1}d\sigma_t^2\ I(\sigma_t^1,\si^2_t;\ell)
=-i\int_{0}^1d\tilde\sigma_t^2\ \lf[\ \hat I(\sigma_t^1,-i\tilde\si_t^2;\ell)-
\hat I'(\sigma_t^1,-i\tilde\si_t^2;\ell-q_t)\ \ri]\ ,}
with the symbol:
\eqn\kerneli{
d(l,q_t)=1-e^{-2\pi i\ap l q_t}\ .}
The prime denotes the respective non--planar configuration at $C_2'$ following from \WochenbrunnerAlm. On the other hand, the hat indicates that the integrand \Local\ has to be supplemented by appropriate  phase factors taking into account the branch points along $C_2$ and $C_2'$, respectively. Above for $C_2$ we have introduced the parameterization $\tilde\sigma^2_t:=i\sigma^2_t$ subject to $\sigma^2_t\in i(-1,0)$. Similarly  for   $C_2'$  we have $\sigma^2_t\in \fc{\tau_2}{2}+i(0,-1)$. 
The next step is to introduce the coordinates \Newcoords. 
With this parameterization the integrand of the r.h.s. of \Contour\ can be specified as 
some function depending on $\xi_t$ and $\eta_t$ as:
\eqn\integrands{
\hat I(\sigma^1_t,-i\tilde\sigma^2_t;\ell)\equiv \hat I(\xi_t,\eta_t;\ell)\ .}
Actually, the precise form of the latter with the correct phases will be displayed in the sequel.
At any rate, the combination of theta--functions contributing to  \integrands\  and entering the integrand \Contour\ is analytic and single--valued w.r.t. $\xi_t,\eta_t$.

Again, due to \newcoords\ the region $(\sigma^1_t,\tilde\sigma^2_t)\in[0,1]^2$ is mapped to a rhombus accounting for the $\xi_t,\eta_t$--dependence of \integrands. 
This diamond can be divided into the four triangular domains
\domains, depicted in Fig.~4.
In these four domains we can study the local system
\eqn\local{
\hat I\sim  \exp\lf\{-\h\pi\ap \tau_2\ell^2-\pi i\ap\ell\sum_{r=1}^{n}
q_r(\xi_r-\eta_r)\ri\}\ 
\prod_{s\neq t}^n\lf(\fc{\theta_1(\xi_s-\xi_t,\tau)}{\theta_1(\xi_s-\eta_t,\tau)}\fc{\theta_1(\eta_s-\eta_t,\tau)}
{\theta_1(\eta_s-\xi_t,\tau)}\ri)^{\h\ap q_t q_s}\Pi,}
with the phase factor $\Pi$ accounting for the correct branch when moving in the $(\xi_r,\eta_r)$--plane and crossing some branch cuts.  After a careful inspection of the dependence of the local system \local\ on the regions \domains\ and performing changes  of integration variables  we find that by shifts in $\xi_r$ and $\eta_r$ the two blue triangles $\Dc_{II}$ and $\Dc_{IV}$ can be  combined to give the unit square $[0,1]^2$:
\eqn\together{
\int_{\Dc_{II}}\hat I(\xi_t,\eta_t;\ell)+\int_{\Dc_{IV}}\hat I(\xi_t,\eta_t;\ell)=e^{-\pi i\ap\ell q_t}\ \int_0^1d\xi_t\int_0^1d\eta_t\ \hat I(\xi_t,\eta_t;\ell)\ .}
Similarly for the red triangles $\Dc_{I}$ and $\Dc_{III}$ we find:
\eqn\togetheri{\eqalign{
\int_{\Dc_{I}}\hat I(\xi_t,\eta_t;\ell)&= \int_0^1d\xi_t    \int_0^{\xi_t} d\eta_t\ 
\hat I(\xi_t,\eta_t;\ell)\ ,\cr
\int_{\Dc_{III}}\hat I(\xi_t,\eta_t;\ell)&=  e^{-2\pi i\ap\ell q_t}\ \int_0^1d\xi_t   \int_{\xi_t}^1 d\eta_t\ \hat I(\xi_t,\eta_t;\ell)\ .}}
The same relations \together\ and \togetheri\ apply for  $C_2'$ where our local system $\hat I'$ is constituted from \Object.
Eventually, with \Contour, \together\ and \togetheri\ the complex integral \need\ over the closed string vertex position $z_t$ becomes:
\eqnn\final{
$$\eqalignno{
2i\ \sin\lf(\fc{\pi\ap\ell q_t}{2}\ri)&\int_{\Cc}d^2z_t\ I(z_t,\bar z_t;\ell)&\final\cr
&=e^{\h\pi i\ap\ell q_t}\ \int_0^1d\xi_t\int_0^{\xi_t}d\eta_t\ 
\lf[ \hat I(\xi_t,\eta_t;\ell)-\hat I'(\xi_t,\eta_t;\ell-q_t)\ri]\cr
&+ e^{-\h\pi i\ap\ell q_t}\  \int_0^1d\xi_t\int_{\xi_t}^1d\eta_t\ 
\lf[ \hat I(\xi_t,\eta_t;\ell)-\hat I'(\xi_t,\eta_t;\ell-q_t)\ri]\cr
&=\int_0^1d\xi_t\int_0^1d\eta_t\ 
e^{\h\pi i\ap\ell q_t\;{\rm sgn}(\xi_t-\eta_t)}\lf[ \hat I(\xi_t,\eta_t;\ell)-\hat I'(\xi_t,\eta_t;\ell-q_t)\ri]\ .}$$}
Above we have used similar manipulations which lead to \Baeckeralm.
The expression \Object\  is holomorphic in all the coordinates $\sigma_r^2,\ r=1,\ldots,n$.
Above  we have exhibited how one complex closed string coordinate $z_t$ is 
converted into a pair of two real open string positions $\xi_t,\eta_t$ by means of considering a closed 
contour in the complex $\sigma^2_t$--plane and applying Cauchy's theorem \Cauchy. This leads to   \Contour\ giving rise to the relation \final. This result enables us to express the complex cylinder integral \need\ over $z_t$ in terms two real  integrations $\xi_t,\eta_t$ along the cylinder boundaries. This manipulation  is illustrated in Fig.~5 for one single closed string coordinate.

We want to repeat the steps from above and perform an analytic continuation of all closed string coordinates $\sigma_r^2,\ t=1,\ldots,n$, 
defined in \sig. We shall discuss the dependence of the integrand \Object\  in  the complex $\sigma^2_r$--planes and consider in each complex $\sigma^2_r$--plane closed cycles (polygons) $C$ defined by the four edges  \edges. We simultaneously deform all of the $\sigma^2_r$--integration contours from the real axis to the pure imaginary axes $C_2,C_2'$. This way we obtain real coordinates $\xi_r,\eta_r$ given by \newcoords. In other words, \final\ allows to trade in a $\sigma_r^2$--integration by means of deforming its integration region $C_1$ to $C_2,C_2'$.
In the multi--dimensional complex  space $\sigma^2\in\IC^{n}$ there are now many branch points, all of them lying along the imaginary axes given by $\sigma^2_s-\sigma^2_r=\pm i \sigma^1_{sr}$ and $\sigma^2_s+\sigma^2_r=\pm i \sigma^1_{sr}$. These conveniently divide themselves into pairs given by $z_s-z_r=0$ and $\bar z_s-\bar z_r=0$ as well as  $z_s-\bar z_r=0$ and $\bar z_s- z_r=0$, respectively. Essentially they may be examined independently in the same way as anticipated in the previous subsection.
For $\sigma^2_r\in C_2,C_2'$ with $\tilde\sigma^2_r=i\sigma^2_r$ and \Newcoords\ we now define the single--valued function
\eqnn\Gaudeamushutte{
$$\eqalignno{
I_{\alpha,\beta}(\{\xi_r,\eta_r\};\ell)&:= \exp\lf\{-\h\pi\ap \tau_2\ell^2-\pi i\ap\ell\sum_{r=1}^{n}
q_r(\xi_r-\eta_r)\ri\}&\Gaudeamushutte \cr
&\times \prod_{r,s\in\al\atop\cup r,s,\in\beta}\prod_{r<s}\ 
\Theta(\xi_r,\eta_r,\xi_s,\eta_s)\ \Pi(\xi_r,\xi_s,\eta_r,\eta_s)\ \prod_{s\in \al\atop t\in\beta}\ \Theta_T(\xi_t,\eta_t,\xi_s,\eta_s)\ ,}$$}
generalizing the local system \local\ and with the phases
\eqn\Phases{
\Pi(\xi_r,\xi_s,\eta_r,\eta_s)=
e^{\h\pi i \ap q_rq_s[1-\theta(\xi_s-\xi_r)\theta(\eta_s-\eta_r)]}\  e^{\h\pi i \ap q_rq_s[1-\theta(\xi_s-\eta_r)\theta(\eta_s-\xi_r)]}\ ,}
rendering the correct branch after deforming the contour $C_1$ to $C_2,C_2'$.
Above we have defined the anlog of \abk\ and \Abk:
\eqn\Analogs{\eqalign{
\Theta(\xi_r,\eta_r,\xi_s,\eta_s)&=\lf(\fc{\theta_1(\xi_s-\xi_r,\tau)}
{\theta_1(\xi_s-\eta_r,\tau)}\fc{\theta_1(\eta_s-\eta_r,\tau)}
{\theta_1(\eta_s-\xi_r,\tau)}\ri)^{\h\ap q_r q_s}\ ,\cr
\Theta_T(\xi_t,\eta_t,\xi_s,\eta_s)&=\lf(\fc{\theta_4(\xi_s-\xi_t,\tau)}
{\theta_4(\xi_s-\eta_t,\tau)}\fc{\theta_4(\eta_s-\eta_t,\tau)}
{\theta_4(\eta_s-\xi_t,\tau)}\ri)^{\h\ap q_t q_s}\ ,}}
respectively.
In other words, the phase factor $\Pi$ makes sure, that we stay in the correct branch when the coordinates $\xi_t,\eta_t$ are varied. 
Similarly as in the  one coordinate case giving rise to  \final\  now with \Contour\ the complex integral over the closed string vertex position $z_t$ becomes
\eqn\Gruttenhuette{\eqalign{
\int_{\Cc}d^2z_t\ I_{\al,\bet}(\{z_s,\bar z_s\};\ell)&=\int_0^1d\xi_t\int_0^1d\eta_t\ \fs(\ell,q_t)\cr
&\times\lf[\ \hat I_{\al,\bet}(\{\xi_s,\eta_s\};\ell)-\hat I_{\al',\bet'}(\{\xi_s,\eta_s\};\ell-q_t)\ \ri]\ ,}}
with $\alpha'=\alpha\slash\{t\}$ and $\beta'=\beta\cup\{t\}$ following from \WochenbrunnerAlm\  and with the kernel:
\eqn\kernel{
\fs(\ell,q_t)^{-1}=2i\ \sin\lf(\fc{\pi\ap\ell q_t}{2}\ri)\ e^{-\h\pi i\ap\ell q_t\;{\rm sgn}(\xi_t-\eta_t)}\ .}

To convert all $n$ complex world--sheet cylinder integrals into pairs of real integrals
we  successively  apply the formula \Gruttenhuette\ for all complex coordinates. Following  \Contour\ each time we trade in a $\sigma_t^2$ coordinate by means of deforming its integration region $C_1$ to $C_2,C_2'$.
We shall end up with $2n$ real integrations $\xi_r,\eta_r$ along the boundaries $C_2,C_2'$.
To convert all $n$ complex world--sheet cylinder integrals into pairs of real integrals
we start with a closed string position (e.g. $z_t$ with $t\!=\!1$) and apply \Gruttenhuette.
After successively  applying the formula \Gruttenhuette\ for all complex coordinates 
we end up with $2^n$ terms ($2^{n-1}$ after taking into account the conformal Killing factor $V_{CKG}$) each comprising $2n$ ($2n-2$ after cancelling the conformal Killing factor $V_{CKG}$) real integrations $\xi_r,\eta_r$ along the boundaries $C_2,C_2'$ of the cylinder.
In total, after converting all $n$ complex cylinder integrals \Need\ we end up at
\eqnn\Ellmau{
$$\eqalignno{
\lf(\prod_{t=1}^n\int_{\Cc}d^2z_t\ri) I(\{z_s,\bar z_s\}&;\ell)=\sum_{m=0}^n\sum_{b_m\subset \Pc_n}(-1)^m\prod_{i=1}^n \fs\lf(q_i,\ell-\sum_{j=1\atop \beta_j<i}^m q_{\beta_j}\ri)&\Ellmau\cr 
&\times\lf(\prod_{t=1}^n \int_0^1d\xi_t\int_0^1d\eta_t\ri)I_{a_{n-m},b_m}\lf(\{\xi_r,\eta_r\};\ell-\sum_{j=1}^mq_{\beta_j}\ri)\ ,}$$}
with $b_m$  an ordered subset of $\Pc_n:=\{1,\ldots,n\}$
with cardinality $m$ containing $m$ elements $\{\beta_1,\ldots,\beta_m\}$ from $\Pc_n$ and $a_{n-m}=\{\alpha_1,\ldots,\alpha_{n-m}\}$ the corresponding  ordered complement such that $a_{n-m}\cup b_m=\Pc_n$.

Eventually, after putting everything together we obtain the amplitude \vordereGoingerHalt\ with all complex integrations converted into real integrations
\eqnn\Gaudeamushuette{
$$\eqalignno{
\Tc(q_1,\ldots,q_n;-\ell,\ell)&=\delta^{(d)}\lf(\sum_{i=1}^n q_i\ri)  
\int d\tau_2&\Gaudeamushuette\cr
&\times \sum_{m=0}^n\sum_{b_m\subset \Pc_n}\lf(\prod_{r=1}^{n}\int_0^1d\xi_r\int_0^1d\eta_r \ri)\;  \Lc_{b_m}\lf(\ell-\sum_{j=1}^mq_{\beta_j};\{\xi_s,\eta_s\}\ri) \cr
&\times \prod_{r,s\in a_{n-m}\atop\cup r,s,\in b_m}\prod_{r<s}\ 
\Theta(\xi_r,\eta_r,\xi_s,\eta_s)\ \Pi(\xi_r,\xi_s,\eta_r,\eta_s)\cr
&\times\prod_{s\in a_{n-m}\atop t\in b_m}\Theta_T(\xi_t,\eta_t,\xi_s,\eta_s) \ Q(\tau,\{\xi_s,\eta_s\})\ ,}$$}
with the loop momentum dependent factor
\eqn\Loopdep{
\Lc_{b_m}(\ell;\{\xi_s,\eta_s\})=\exp\lf\{-\h\pi\ap \tau_2\ell^2-i\pi\ap \ell\sum\limits_{r=1}^{n}q_r(\xi_r-\eta_r)\ri\}\;(-1)^m\prod_{t=1}^n\fs\lf(q_t,\ell+\sum^m_{j=1\atop \beta_j\geq t}q_{\beta_j}\ri),}
referring to a given ordered subset $b_m\subset P_n$ of cardinality $m$.
Thus, in total the off--shell  closed string cylinder amplitude \VordereGoingerHalt\ involving $n$ on--shell closed strings and two off--shell closed string states reads
\eqnn\FINALVGH{
$$\eqalignno{
\Tc(q_1,\ldots,q_n;q_0,q_{n+1})&=\delta^{(d)}\lf(q_0+q_{n+1}+\sum_{i=1}^n q_i\ri)  
\int d\tau_2 &\FINALVGH\cr
&\times \sum_{m=0}^n\sum_{b_m\subset \Pc_n}\lf(\prod_{r=1}^{n}\int_0^1d\xi_r\int_0^1d\eta_r \ri)\;  \Lc_{b_m}\lf(q_{n+1}-\sum_{j=1}^mq_{\beta_j};\{\xi_s,\eta_s\}\ri) \cr
&\times \prod_{r,s\in a_{n-m}\atop\cup r,s,\in b_m}\prod_{r<s}\ 
\Theta(\xi_r,\eta_r,\xi_s,\eta_s)\ \Pi(\xi_r,\xi_s,\eta_r,\eta_s)\cr
&\times\prod_{s\in a_{n-m}\atop t\in b_m}\ \Theta_T(\xi_t,\eta_t,\xi_s,\eta_s)\ Q(\tau,\{\xi_s,\eta_s\})\ ,}$$}
with the loop momentum dependent factor \Loopdep
\eqn\LoopDep{\eqalign{
\Lc_{b_m}(q_{n+1};\{\xi_s,\eta_s\})&=e^{-\h\pi\ap \tau_2q_{n+1}^2}\ \prod_{r=1}^n \lf(\fc{\theta_1(\xi_r+\fc{\tau}{2},\tau)}{\theta_1(\xi_r-\fc{\tau}{2},\tau)}\fc{\theta_1(\eta_r-\fc{\tau}{2},\tau)}{\theta_1(\eta_r+\fc{\tau}{2},\tau)}\ri)^{\h\ap q_rq_{n+1}}\cr
&\times(-1)^m\ \prod_{t=1}^n\fs\lf(q_t,q_{n+1}+\sum^m_{j=1\atop \beta_j\geq t}q_{\beta_j}\ri)\ .}}
Above we have used the relations \identVG\ or \identVGi, which now read ($\tau=i\tau_2$):
\eqnn\IdentVG{
$$\eqalignno{
\exp\lf\{-i\pi\ap q_{n+1}\sum\limits_{r=1}^{n}q_r(\xi_r-\eta_r)\ri\}
&=\prod_{r=1}^n \lf(\fc{\theta_1(\xi_r+\fc{\tau}{2},\tau)}{\theta_1(\xi_r-\fc{\tau}{2},\tau)}\fc{\theta_1(\eta_r-\fc{\tau}{2},\tau)}{\theta_1(\eta_r+\fc{\tau}{2},\tau)}\ri)^{\h\ap q_rq_{n+1}}\cr
&=\prod_{r=1}^n
\lf(\fc{\theta_4(\xi_r+\fc{\tau}{2},\tau)}{\theta_4(\xi_r-\fc{\tau}{2},\tau)}\fc{\theta_4(\eta_r-\fc{\tau}{2},\tau)}{\theta_4(\eta_r+\fc{\tau}{2},\tau)}\ri)^{\h\ap q_rq_{n+1}}\hskip-0.5cm.&\IdentVG}$$}

Finally, integrating \Gaudeamushuette\ over the loop momentum $\ell$ and 
performing the shift in the loop momentum  $\ell$ by 
$\sum\limits_{j=1}^m q_{\beta_j}$ we obtain the on--shell one--loop amplitude \HintereGoingerHalt,
which thanks to \IdentVG\ we may cast into the compact form
\eqnn\Scheffau{
$$\eqalignno{
A^{(1)}_{0,n}(q_1,\ldots,q_n)&=g_c^n\ \delta^{(d)}\lf(\sum_{i=1}^n q_i\ri)  
\int d\tau_2\ V_{CKG}^{-1}\int_{-\infty}^\infty d^d\ell\ e^{-\h\pi\ap \tau_2\ell^2}&\Scheffau\cr
&\times \sum_{m=0}^n\sum_{b_m\subset \Pc_n} (-1)^m\ \lf\{\prod_{r=1}^n\int_0^1d\xi_r\int_0^1d\eta_r\ \fs\lf(q_r,\ell+\sum^m_{j=1\atop \beta_j\geq r}q_{\beta_j}\ri)\ri\}\cr
&\times \prod_{r,s\in a_{n-m}\atop\cup r,s\in b_m\cup\{n+1\}}\prod_{r<s}\ 
\Theta(\xi_r,\eta_r,\xi_s,\eta_s)\ \Pi(\xi_r,\xi_s,\eta_r,\eta_s)\cr
&\times\prod_{s\in a_{n-m}\atop t\in b_m\cup\{n+1\}}\ \Theta_T(\xi_t,\eta_t,\xi_s,\eta_s)\ \hat Q(\tau,\{\xi_s,\eta_s\})\ ,}$$}
subject to \posiVG, \ie $\ell=q_{n+1}$, $\xi_{n+1}\!=\!-\fc{\tau}{2}, \eta_{n+1}\!=\!\fc{\tau}{2}$ and
$\Pi(\xi_r,\xi_{n+1},\eta_r,\eta_{n+1})\!:=\!1$.

For a given configuration of coordinates $\xi_r,\eta_r\in(0,1)$ within the unit interval parameterizing  one of the cylinder boundaries both the phase factor \Phases\ and the phase in \kernel\ are constants.
Therefore, subject to a phase the result \Scheffau\ can be written as a sum over pure open string one--loop  cylinder amplitudes \Open\ involving $N\!=\!2n$ open strings supplemented by the two additional
insertions \posORTH.
More precisely, the open string momenta $k_s$ entering the open string amplitudes \Open\   are specified by
\eqn\maut{\eqalign{
k_r&=\h\ q_r\ ,\cr
k_{n+r}&=-\h\ q_r\ \ \ ,\ \ \ r=1,\ldots,n\ ,\cr
k_0&=\ell\ \ \ ,\ \ \ k_{2n+1}=-\ell\ ,}}
and $N\!+\!2=2n+2$ open string insertions $x_s$ along the two cylinder boundaries:
\eqn\Maut{\eqalign{
x_r&=\xi_r\ ,\cr
x_{n+r}&=\eta_r\ \ \ ,\ \ \ r=1,\ldots,n\ ,\cr
x_0&=-\fc{\tau}{2}\ \ \ ,\ \ \ x_{2n+1}=+\fc{\tau}{2}\ .}}
In \Scheffau\ for any configuration $b_m\subset\Pc_n$ the closed string coordinates $\xi_r,\eta_r\in(0,1)$  give rise to some orderings $\Ar\in S_{2n-2m}$ and $\Br\in S_{2m}$ along the two cylinder boundaries, respectively. Here, $\Ar$ denotes a permutation of the set consisting of $n-m$ positions $x_r\!=\!\xi_r$ and their partners $x_{n+r}\!=\!\eta_r$. Similarly, $\Br$ denotes a permutation of the set consisting of the remaining $m$ positions $x_s\!=\!\xi_s$ and their partners $x_{n+s}\!=\!\eta_s$. Given orderings $\Ar,\Br$ completely specify all phase factors appearing in \Scheffau. After exploiting their net effect we may introduce 
a total kernel $S(\Ar,\Br)$ accounting for both the phase factors \Phases\ and 
the phase in \kernel. With these preliminaries we may represent \Scheffau\ as a sum over one--loop non--planar subamplitudes \Open\ involving $N\!=\!2n$ open strings and  extended by the additional two open string insertions $x_0$ and $x_{N+1}$ given in \Maut. 
Eventually, with $g_c=g_o^2$  we can express \Scheffau\ in the following way
\eqnn\Habdank{
$$\eqalignno{
A^{(1)}_{n}(q_1,\ldots,q_n)&=(2i)^{-n}\ \delta^{(d)}\lf(\sum_{i=1}^n q_i\ri)  \ \int d\tau_2
\int_{-\infty}^\infty d^d\ell\ e^{-\h\pi\ap \tau_2\ell^2}\cr
&\times\sum_{m=0}^n\sum_{b_m\subset \Pc_n}  (-1)^m\ \prod_{t=1}^n\sin\lf(k_t,\ell+\sum^m_{j=1\atop \beta_j\geq t}q_{\beta_j}\ri)^{-1}&\Habdank\cr
&\times \sum_{\Ar\in S_{2n-2m}\atop \Br\in S_{2m} }
 S(\Ar,\Br)\  A^{(1)}_{2n}(\Ar|\Br)\ ,}$$}
To conclude, in this subsection we have accomplished to perform the $n$ complex cylinder integrals of \vordereGoingerHalt\ by converting them into $2n$ real integrals
along the two boundaries of the cylinder, cf. also Fig.~7.

To familiarize  with our one--loop formula \Scheffau\ here we shall discuss one example. We consider the one--loop four closed string amplitude. For this case Eq. \Scheffau\ comprises the following sixteen terms 
\eqnn\scheffau{
$$\eqalignno{
(\fc{\ap\tau_2}{2})^{-d/2} \ &A_{0,4}^{(1)}(q_1,q_2,q_3,q_4)=g_c^4\ \delta^{(d)}\lf(q_1+q_2+q_3+q_4\ri)  \int_0^\infty d\tau_2&\scheffau\cr
&\times\int_{-\infty}^\infty d^d\ell\ e^{-\h\pi\ap \tau_2\ell^2}
\ V_{\rm CKG}^{-1}\lf(\prod_{r=1}^{4}\int_0^1\!\!d\xi_r\int_0^1 d\eta_r \ri)\cr 
&\times\Big\{\ \fs(q_1,\ell)\;\fs(q_2,\ell)\;\fs(q_3,\ell)\;\fs(q_4,\ell)\ 
[a_{z_5}(1,2,3,4;)-a_{z_5}(2,3,4;1)]\cr
& -\fs(q_1,\ell+q_2)\;\fs(q_2,\ell)\;\fs(q_3,\ell)\;\fs(q_4,\ell)\ 
[a_{z_5}(1,3,4;2)-a_{z_5}(3,4;1,2)\cr
&-\fs(q_1,\ell+q_3)\;\fs(q_2,\ell+q_3)\;\fs(q_3,\ell);\fs(q_4,\ell)\ 
[a_{z_5}(1,2,4;3)-a_{z_5}(2,4;1,3)]  \cr
&-\fs(q_1,\ell+q_4)\;\fs(q_2,\ell+q_4)\;\fs(q_3,\ell+q_4)\;\fs(q_4,\ell)\ 
[a_{z_5}(1,2,3;4)-a_{z_5}(2,3;1,4)]\cr
& +\fs(q_1,\ell-q_2)\;\fs(q_2,\ell-q_1)\;\fs(q_3,\ell+q_4);\fs(q_4,\ell)\ 
[a_{z_5}(1,2;3,4)-a_{z_5}(2;1,3,4)]\cr
& +\fs(q_1,\ell-q_3)\;\fs(q_2,\ell+q_4)\;\fs(q_3,\ell+q_4);\fs(q_4,\ell)\ 
[a_{z_5}(1,3;2,4)-a_{z_5}(3;1,2,4)]\cr
& +\fs(q_1,\ell-q_4)\;\fs(q_2,\ell+q_3)\;\fs(q_3,\ell);\fs(q_4,\ell)\ 
[a_{z_5}(1,4;2,3)-a_{z_5}(4;1,2,3)]\cr
& +\fs(q_1,\ell)\;\fs(q_2,\ell-q_1)\;\fs(q_3,\ell+q_4);\fs(q_4,\ell)\ 
[a_{z_5}(;1,2,3,4)-a_{z_5}(1;2,3,4)]\ \Big\}\ ,}$$}
with the (unintegrated) open string (ten--point) subamplitudes
\eqn\subOPEN{
a_{z_5}(i,j;k,l)=\prod_{r,s\in\{i,j\}\atop \cup r,s\in\{k,l,5\}}\prod_{r<s}\Theta(\xi_r,\eta_r,\xi_s,\eta_s)\ 
\Pi(\xi_r,\xi_s,\eta_r,\eta_s)
\prod_{s\in \{i,j\}\atop t\in \{k,l,5\} }  \Theta_T(\xi_t,\eta_t,\xi_s,\eta_s)\ ,}
involving the expressions \Analogs.
The point $z_5=-\fc{\tau}{2}$ denotes the additional marked point \posiVG\ referring to the two open string positions \Maut, i.e.
$\xi_5=-\fc{\tau}{2},\eta_5=\fc{\tau}{2}$.
We can further reduce the terms in \scheffau\ by  grouping terms with identical 
configurations at either boundaries, e.g. $a_{z_5}(;1,2,3,4)=a_{z_5}(1,2,3,4;)$ leading to:
\eqnn\scheffaul{
$$\eqalignno{
\hskip1cm(\fc{\ap\tau_2}{2})^{-d/2} \ A_{0,4}^{(1)}(q_1,&q_2,q_3,q_4)=g_c^4\ \delta^{(d)}\lf(q_1+q_2+q_3+q_4\ri)  \int_0^\infty d\tau_2\cr
&\times\int_{-\infty}^\infty d^d\ell\ e^{-\h\pi\ap \tau_2\ell^2}
\ V_{\rm CKG}^{-1}\lf(\prod_{r=1}^{4}\int_0^1\!\!d\xi_r\int_0^1 d\eta_r \ri)\cr 
&\hskip-3cm\times\Big\{\ \fs(q_1,\ell);\fs(q_4,\ell)\;[\fs(q_2,\ell)\;\fs(q_3,\ell)+\fs(q_2,\ell-q_1)\;\fs(q_3,\ell+q_4)]\cr
&\times [a_{z_5}(1,2,3,4;)-a_{z_5}(2,3,4;1)]\cr
&\hskip-3.25cm -\fs(q_4,\ell)\;[\fs(q_1,\ell+q_2)\;\fs(q_2,\ell)\;\fs(q_3,\ell)+\fs(q_1,\ell-q_2)\;\fs(q_2,\ell-q_1)\;\fs(q_3,\ell+q_4)]\cr 
&\times[a_{z_5}(1,3,4;2)-a_{z_5}(3,4;1,2)\cr
&\hskip-3.5cm-\fs(q_4,\ell)[\fs(q_1,\ell+q_3)\;\fs(q_2,\ell+q_3)\;\fs(q_3,\ell)+\fs(q_1,\ell-q_3)\;\fs(q_2,\ell+q_4)\;\fs(q_3,\ell+q_4)]\cr 
&\times[a_{z_5}(1,2,4;3)-a_{z_5}(2,4;1,3)]  \cr
&\hskip-3.5cm-\fs(q_4,\ell)\;[\fs(q_1,\ell+q_4)\;\fs(q_2,\ell+q_4)\;\fs(q_3,\ell+q_4)+\fs(q_1,\ell-q_4)\;\fs(q_2,\ell+q_3)\;\fs(q_3,\ell)] \cr 
&\times[a_{z_5}(1,2,3;4)-a_{z_5}(2,3;1,4)]\ \Big\}\ .&\scheffaul}$$}

%%%%%%%%%%%%%%%%%%%%%%%%%%%%%%%%
Finally, if $q_i=q_i^\perp$ for $d^\perp$ dimensions   we can simply add to the result \Scheffau\ open strings with momenta $p_i,p_j'$ lying in the remaining $d^\parallel$ dimensions. 
Then, due to the decoupling of open and closed strings adding open strings is simply accomplished by  including the additional  factor \wehaveE\ and integrating over the open string positions. There are no additional monodromies when crossing the open string positions:
\eqnn\schefffau{
$$\eqalignno{
A^{(1)}_{n_o,n_c}(\Ac&|\Bc||\{q_1,\ldots,q_n\})=\delta^{(d^\parallel)}\lf(\sum_{i=1}^{n_1}p_i+\sum_{j=1}^{n_2}p_j'\ri)\ \delta^{(d^\perp)}\lf(\sum_{i=1}^n q_i\ri) \int d\tau_2 \cr
&\times \int_{-\infty}^\infty d^d\ell\ e^{-\h\pi\ap \tau_2\ell^2}\ V_{CKG}^{-1} \ \left(\!\int_{\Ic_1}\prod_{i=1}^{n_1}dx_i\int_{\Ic_1'}\prod_{j=1}^{n_2}dx_j'\!\right) E(\{x_i,x_j'\})\cr
&\times  \sum_{m=0}^n\sum_{b_m\subset \Pc_n} (-1)^m\ \lf\{\prod_{r=1}^n\int_0^1d\xi_r\int_0^1d\eta_r \ \fs\lf(q_r,\ell+\sum^m_{j=1\atop \beta_j\geq r}q_{\beta_j}\ri)\ri\}\cr
&\times \prod_{r,s\in a_{n-m}\atop\cup r,s\in b_m\cup\{n+1\}}\prod_{r<s}\Theta(\xi_r,\eta_r,\xi_s,\eta_s)\ \Pi(\xi_r,\xi_s,\eta_r,\eta_s)\cr
&\times\prod_{s\in a_{n-m}\atop t\in b_m\cup\{n+1\}}\ \Theta_T(\xi_t,\eta_t,\xi_s,\eta_s)\ \hat Q(\tau,\{\xi_s,\eta_s\})\ .&\schefffau}$$}

\newsec{Open string one--loop monodromy relations and closed string insertion}

In this section we present a link between open string one--loop monodromy relations and 
cylinder amplitudes with one closed string insertion.

\subsec{Amplitude with one closed string and $n_o$ non--planar open strings}

In the following for simplicity we first focus on the case $n_c\!=\!1$ with collinear momenta \split,  with canonical open string orderings $\Ac=\{1,\ldots,n_1\}$ and $\Bc=\{n_1+1,\ldots,n_2\}$. In what follows we shall use $p_i:=p_{\al_i},\ i=1,\ldots,n_1$ and $p_j'=p_{\bet_j},\ j=1,\ldots,n_2$, with  $n_o=n_1+n_2$.  Let us discuss the dependence of the integrand \Generali\ on the closed string coordinate $z$.
As a consequence of \conserv\ for the non--holomorphic factor \nonholoo\ we have $I_{nh}=1$ and with (B.13) the integrand 
$I(z,\bar z):=I(\{x_i,x_j',z,\bar z\})$ becomes 
\eqn\relevant{
I(z,\bar z)\equiv\prod_{i=1}^{n_1}\theta_1(z-x_i,\tau)^{2\ap q_1 p_i}\ \theta_1(\bar z-x_i,\tau)^{2\ap q_2p_i}\ 
\prod_{j=1}^{n_2}\theta_4(z-x_j',\tau)^{2\ap q_1 p_j'}\ \theta_4(\bar z-x_j',\tau)^{2\ap q_2 p_j'} \ ,}
with $q:=q_1=q_{L,1}+q_{R,1}$ and $q_1:=q_{L,1}$ and $q_2:=q_{R,2}$.

In the following we may repeat some of the steps from Subsection 3.1. 
After introducing the parameterization \sig\ 
we investigate the $\sigma^2$--dependence of $I$ in the complex $\sigma^2$--plane and consider the closed cycle (polygon) defined by the 
four edges $C_1,C_1',C_2$ and~$C_2'$ \edges, depicted in Fig.~8. 
Since $x_i,x_j'\in (0,1)$ the function \relevant\ as a holomorphic function in $\sigma^2$  has $n_1$ pairs of  branch points at $\sigma^2=\pm i(\sigma^1-x_i)$ along the imaginary axis $\sigma^2=0$ ($z=\bar z$)  where the arguments of the theta--functions $\theta_1$ become zero  with branching phases structure $e^{2\pi i \ap q_1p_i}$ and $e^{2\pi i \ap q_2p_i}$, respectively. Furthermore, we have pairs of $n_2$ branch points at
$\sigma^2=\pm i(\sigma^1-x_i')+\fc{t}{2}$ along $\sigma^2=\fc{t}{2}$, where the arguments of the theta--functions $\theta_4$ become zero ($\mod\ \tau/2$)  with branching phases structure $e^{2\pi i \ap q_1p_j'}$ and $e^{2\pi i \ap q_2p_j'}$, respectively. These phases need to be introduced when passing those branch cuts. Their effect will be taken into account below.
The holomorphic  integrand along $C_1'$ differs from that along $C_1$ by a phase
\eqn\differ{
I(\sigma^1,\sigma^2-i)=\varphi\ I(\sigma^1,\sigma^2)\ ,}
given by:
\eqn\phase{
\varphi:=e^{-2\pi i\ap\sum\limits_{i=1}^{n_1}(q_1-q_2)p_i}
=        e^{2\pi i\ap(q_1-q_2)\sum\limits_{j=1}^{n_2}p_j'}\ .}
The last manipulation follows from  momentum conservation \conserv.
For $n_2\geq1$  and $q_1\neq q_2$ we have $\varphi\neq1$.
\ifig\SigmaWorldsheetI{A closed contour $C$ in the complex $\sigma^2$--plane and branch points.}{\epsfxsize=0.55\hsize\epsfbox{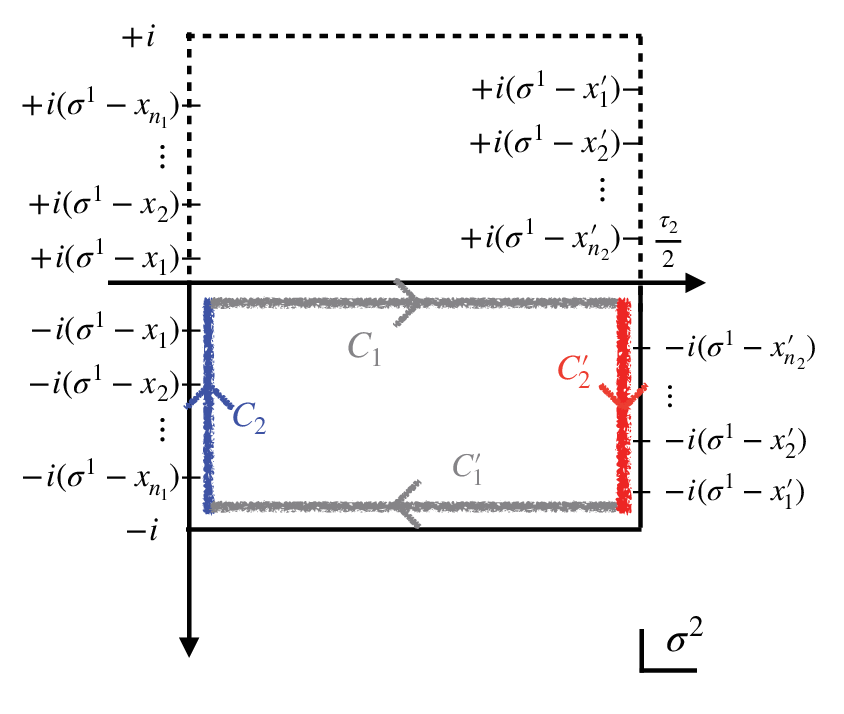}}
\noindent

Eventually, we need to integrate the coordinate $z$  over the full cylinder $\Cc$ as \need. Thus in \need\ we are supposed to integrate $\sigma^2$ along the (grey) edge $C_1$ in Fig.~8.
We seek to convert this integration to an integration along the imaginary axis $C_2,C_2'$ by considering the closed cycle \edges\ and using Cauchy's integral theorem \cauchy.
Again, the integrand $\hat I$ differs from $I$  by choosing a correct branch  when moving in the complex $\sigma^2$--plane and passing the branch cuts discussed above.
By using Cauchy's theorem \cauchy\ and taking into account the branch cuts when moving along the paths $C_2$ and $C_2'$ thanks to the relation \differ\ the $\sigma_2$--integral along $C_1$ can be expressed as:
\eqn\Contour{
(1-\varphi)\ \int_0^{\fc{t}{2}}d\sigma^2\ I(\sigma^1,\sigma^2)=-i\int_0^1d\tilde\sigma^2\ 
\lf[\ \hat I(\sigma^1,-i\tilde\sigma^2)-\hat I'(\sigma^1,-i\tilde\sigma^2)\ \ri]\ .}
The prime denotes the respective non--planar configuration at $C_2'$. On the other hand, the hat indicates that the integrand \relevant\ has to be supplemented by appropriate  phase factors taking into account the branch points along $C_2$ and $C_2'$, respectively. Above for $C_2$ we have introduced the parameterization $\tilde\sigma^2:=i\sigma^2$ subject to $\sigma^2\in i(-1,0)$. Similarly  for   $C_2'$  we have $\sigma^2\in \fc{t}{2}+i(0,-1)$. Along $C_2$ we can introduce the new real coordinates \newcoords\
with $\det\lf(\fc{\p(\xi,\eta)}{\p(\sigma^1,\tilde\sigma^2)}\ri)=-2$.
With this parameterization the integrands in \Contour\ can be specified as 
some function depending on $\xi$ and $\eta$ as
\eqn\wehave{\eqalign{
\hat I(\sigma^1,\sigma^2)&=\Pi_{C_2}(x,\xi,\eta)\ I_{C_2}(\xi,\eta)\ \ \ ,\ \ \ \sigma^2\in C_2\ ,\cr
\hat I'(\sigma^1,\sigma^2)&=\Pi_{C_2'}(x',\xi,\eta)\ I_{C_2'}(\xi,\eta)\ \ \ ,\ \ \sigma^2\in C_2'\ ,}}
with:
\eqn\integrandsi{\eqalign{
I_{C_2}(\xi,\eta)&:=\prod_{i=1}^{n_1}|\theta_1(\xi-x_i)|^{2\ap q_1 p_i}\ |\theta_1(\eta-x_i)|^{2\ap q_2p_i}\ \prod_{j=1}^{n_2}\theta_4(\xi-x_j')^{2\ap q_1 p_j'}\ \theta_4(\eta-x_j')^{2\ap q_2 p_j'},\cr 
I_{C_2'}(\xi,\eta)&:= 
\prod_{i=1}^{n_1}\theta_4(\xi-x_i)^{2\ap q_1 p_i}\ \theta_4(\eta-x_i)^{2\ap q_2p_i}\ 
e^{-2\pi i\ap (\xi-x_i)p_i q_1}\ e^{2\pi i\ap(\eta-x_i)p_iq_2}\cr
&\ \ \ \ \times\prod_{j=1}^{n_2}|\theta_1(\xi-x_j')|^{2\ap q_1 p_j'}\ |\theta_1(\eta-x_j')|^{2\ap q_2 p_j'}\ e^{-2\pi i\ap (\xi-x_j')p_j' q_1}\ e^{2\pi i\ap (\eta-x_j')p_j'q_2}\ .}}
The phases $\Pi_{C_2},\Pi_{C_2'}$ will be given below. They impose the correct monodromy phases at the branch points, which now are located at $\xi=x_i$ and $\eta=x_i$.
Note, that in $I_{C_2'}$ the additional position dependent phases take into account the 
relative monodromy w.r.t. $C_2$. This effect appears also in one--loop open string monodromy relations when moving from one boundary to the other \HoheneggerKQY.  Note also the identities $\prod_{i=1}^{n_1}e^{-2\pi i \ap\xi p_iq_1}\prod_{j=1}^{n_2}e^{-2\pi i \ap\xi p_j'q_1}=e^{2\pi i \ap\xi q_1q_2}=1$ and $\prod_{i=1}^{n_1}e^{2\pi i \ap\eta p_iq_2}\prod_{j=1}^{n_2}e^{2\pi i \ap\eta p_j'q_2}=e^{-2\pi i \ap\eta q_1q_2}=1$, i.e. any $\xi,\eta$ dependence drops in the phases of \integrandsi.
Now,  all the theta--functions \integrandsi\  entering the integrand \Contour\ are analytic and single--valued w.r.t. $\xi,\eta$.

Due to \newcoords\ the region $(\sigma^1,\tilde\sigma^2)\in[0,1]^2$ is mapped to a rhombus 
accounting for the $\xi,\eta$--dependence of \integrands, cf. Fig.~4. 
This diamond $\Dc=\Dc_{I}\cup\Dc_{II}\cup\Dc_{III}\cup\Dc_{IV}$ can be divided into the four triangular domains \domains, depicted in Fig.~4.
In these four domains we can study the systems \integrandsi\ yet without specifying the phases $\Pi_{C_2},\Pi_{C_2'}$.  After a careful inspection of the dependence of the local system \integrandsi\ on the regions \domains\ and performing changes of integration variables subject to  (B.10) we find that by shifts in $\xi$ and $\eta$ the two blue triangles 
$\Dc_{II}$ and $\Dc_{IV}$ give:
\eqn\Together{\eqalign{
\int_{\Dc_{II}}\hat I(\xi,\eta)&= \lf(\prod_{i=1}^{n_1}e^{-2\pi i \ap q_1p_i}\ri)\ \int_0^1d\xi    \int_0^{1-\xi} d\eta\ \hat I(\xi,\eta)\ ,\cr
\int_{\Dc_{IV}}\hat I(\xi,\eta)&= \lf(\prod_{i=1}^{n_1}e^{2\pi i \ap q_2p_i}\ri)\  \int_0^1d\xi   
\int_{1-\xi}^1 d\eta\ \hat I(\xi,\eta)\ .}}
Similarly, for the red triangles $\Dc_{I}$ and $\Dc_{III}$ we find:
\eqn\Togetheri{\eqalign{
\int_{\Dc_{I}}\hat I(\xi,\eta)&=  \int_0^1d\xi    \int_0^{\xi} d\eta\ \hat I(\xi,\eta)\ ,\cr
\int_{\Dc_{III}}\hat I(\xi,\eta)&= \lf(\prod_{i=1}^{n_1}e^{-2\pi i \ap (q_1-q_2)p_i}\ri)\   
\int_0^1d\xi   \int_{\xi}^1 d\eta\ \hat I(\xi,\eta)\ .}}
Eventually, after introducing
\eqn\Combph{
\Pi'_{C_2}(\xi,\eta)=\prod_{i=1}^{n_1}e^{-2\pi i \ap  q_1p_i\theta(1-\xi-\eta)}e^{2\pi i \ap q_2p_i\theta(\eta-1+\xi)}+\prod_{i=1}^{n_1}e^{-2\pi i \ap (q_1-q_2)p_i\theta(\eta-\xi)}\ ,}
both \Together\ and \Togetheri\  may be combined to the unit square $[0,1]^2$:
\eqn\Combi{
\int_{\Dc}\hat I(\xi,\eta)=\int_0^1d\xi   \int_{0}^1 d\eta\ 
\ \Pi'_{C_2}(\xi,\eta)\ \hat I(\xi,\eta)\ .}
For  $C_2'$ similar relations can be found leading to \Combi\ with: 
\eqn\Combphi{
\Pi'_{C_2'}(\xi,\eta)=\prod_{j=1}^{n_2}e^{-2\pi i q_1p_j'\theta(1-\xi-\eta)}e^{2\pi i \ap  q_2p_j'\theta(\eta-1+\xi)}+\prod_{j=1}^{n_2}e^{-2\pi i \ap (q_1-q_2)p_j'\theta(\eta-\xi)}\ .}
We are now able to specify the phase factors $\Pi_{C_2},\Pi_{C_2'}$  accounting for the correct branch when moving in the unit interval $(\xi,\eta)\in[0,1]^2$ and crossing some of the branch cuts $\xi,\eta=x_i$. For a generic configuration of $n_o$ open  string vertex positions $x_i,x_j'$ the phase factors $\Pi_{C_2},\Pi_{C_2'}$ account for the correct branch of the integrand \need\ along $C_2$ and $C_2'$, respectively \StiebergerVYA:
\eqn\phases{\eqalign{
\Pi_{C_2}(x,\xi,\eta)&=\prod_{i=1}^{n_1}e^{-2\pi i\ap q_1p_i\theta(x_i-\xi)}\  e^{2\pi i \ap q_2p_i\theta(x_i-\eta)}\ ,\cr
\Pi_{C_2'}(x',\xi,\eta)&=\prod_{j=1}^{n_2}e^{-2\pi i\ap q_1p_j'\theta(x_j'-\xi)}\  e^{2\pi i \ap q_2p_j'\theta(x_j'-\eta)}\ .}}

Eventually, with \Contour\ and \Combi\ the complex integral \need\ over the closed string vertex position $z$ becomes:
\eqn\final{\eqalign{
(1-\varphi)\ \int_{\Cc}d^2z\ I(z,\bar z)=\int_0^1d\xi&\int_0^1d\eta\ 
\lf[\ \Pi_{C_2}(x,\xi,\eta)\  \Pi'_{C_2}(\xi,\eta)\ I_{C_2}(\xi,\eta)\ri.\cr
&\lf.-\Pi_{C_2'}(x',\xi,\eta)\ \Pi'_{C_2'}(\xi,\eta)\ I_{C_2'}(\xi,\eta)\ \ri]\ .}}
Eq. \final\ enables us to express the complex cylinder integral \need\ in terms of two terms  accounting for the contributions from $C_2$ and $C_2'$, respectively.
The closed string has been traded into a pair of two  open strings with (real) vertex positions $\xi$ and $\eta$. On the cylinder this pair of open strings is located at either one of its boundaries. The resulting non--planar cylinder configurations of $n_1\!+\!n_2\!+\!2$ open string positions are depicted in Fig.~6.  
These two contributions, which are given by real
(iterated) integrals at the two boundaries of the cylinder correspond to pure open string one--loop subamplitudes. 
More precisely, while the configuration referring to $C_2$ gives rise to an ordinary non--planar amplitude the open string configuration referring to $C_2'$ can be generated in a one--loop open string monodromy relation  \HoheneggerKQY.
To get the full amplitude \Start\ we have to combine \final\ with the open string part \wehaveE. For $n_c=1$ the latter reads:
\eqn\WehaveE{\eqalign{
E(\{x_i,x_j'\})&\equiv
\prod_{1\leq i<j\leq n_1}\lf|\theta_1(x_{j}-x_{i},\tau)\ri|^{2\ap p_ip_j}
\prod_{1\leq i<j\leq n_2}\lf|\theta_1(x_{j}'-x_{i}',\tau)\ri|^{2\ap p_i'p_j'}\cr
&\times\prod_{1\leq i\leq n_1\atop 1\leq j\leq n_2}\lf|\theta_4(x_{j}'-x_{i},\tau)\ri|^{2\ap p_ip_j'}\ .}}
In total we obtain for the full amplitude
\eqnn\traithenu{
$$\eqalignno{
A_{n_o,1}^{(1)}(\{1,\ldots,n_1\}&|\{1,\ldots,n_2\}||\{q_1,q_2\})=2g_o^{n_o}g_c\ \delta^{(d)}\lf(\sum_{i=1}^{n_1}p_i+\sum_{j=1}^{n_2}p_j'+q_1+q_2\ri)\int d\tau_2\cr 
&\times d\lf(q_1-q_2,\sum\limits_{i=1}^{n_1}p_i\ri)^{-1}\ V_{CKG}^{-1}\ \left(\!\int_{\Ic}\prod_{i=1}^{n_1}dx_i\int_{\Ic'}\prod_{j=1}^{n_2}dx_j'\!\right) E(\{x_i,x_j'\})\cr
&\times \int_0^1d\xi\int_0^1d\eta\  
\Big[\ \Pi_{C_2}(x,\xi,\eta)\  \Pi'_{C_2}(\xi,\eta)\ I_{C_2}(\xi,\eta)\cr
&\hskip1cm-\Pi_{C_2'}(x',\xi,\eta)\ \Pi'_{C_2'}(\xi,\eta)\ I_{C_2'}(\xi,\eta)\ \Big]\ Q(\tau,\{x_i,x_j',\xi,\eta\})\ ,&\traithenu}$$}
with the functions \integrandsi\ and: 
\eqn\kerntri{
d\lf(q_1-q_2,\sum\limits_{i=1}^{n_1}p_i\ri)=1-e^{-2\pi i\ap\sum\limits_{i=1}^{n_1}(q_1-q_2)p_i}\ .}

Again as in the previous section  we may represent \traithenu\ as a sum over open string non--planar subamplitudes involving $N\!=\!n_o+2$ open strings supplemented by the phase factors \Combph\ and \Combphi. More precisely, the open string momenta entering the open string amplitudes   are given by
\eqn\mautt{\eqalign{
p_r&\ ,\ \ \ r=1,\ldots,n-2\ ,\cr
p_{n-1}&=q_1\ ,\ \ \ p_n=q_2\ ,}}
while the $n$ open string insertions $x_s$ along the two boundaries of the cylinder:
\eqn\Mautt{\eqalign{
x_r&\ ,\ \ \ r=1,\ldots,n-2\ ,\cr
x_{n-1}&=\xi\ ,\ \ \ x_n=\eta\ .}}

\subsec{One--loop monodromy relations, closed string insertion and a marked point}

Let us consider an $n$--point one--loop cylinder amplitude $A_{n-1,1}^{(1)}$ with $n_o=n-1$ open strings and one closed string $n_c=1$ with momentum $q$. In the sequel in the expression \thetaLR\ we replace $q:=q_1,\ q_L:=q_{L,1}$ and $q_R:=q_{R,1}$. We distribute the  closed string momentum $q$ into left-- and right--moving momenta according to \split.
Moreover, we consider a massless closed string state ($q^2=0$) and choose $\rho=1$, i.e.
\eqn\Sonneck{
q_L=q\ \ \ ,\ \ \ q_R=0\ ,}
with the total closed string momentum $q=q_L+q_R$. 
 Obviously, for this setup  the non--holomorphic factor \nonholoo\ becomes trivial, i.e.  $I_{nh}=1$. We parameterize the closed string position $z$ as:
\eqn\Kopfkraxen{
z=\sigma^1+i\sigma^2\ \ \ ,\ \ \ \sigma^1\in(0,1)\ ,\ \sigma^2\in(0,\fc{t}{2})\ .}
Due to the conformal Killing symmetry in \Start\ we can cancel the volume $V_{CKG}$ of the conformal Killing group by choosing for one real coordinate the fixing:
\eqn\FIX{
\sigma^1=0\ .}
As a consequence for the case at hand $\Theta_R\equiv 0$ and the expression \Generali\  boils down to:
\eqn\Kaiseralm{
I(\{x_i,x_j',z,\bar z\})=\prod_{1\leq i\leq n_1}\lf[\fc{\theta_1(x_i-i\sigma^2,\tau)}{\theta_1'(0,\tau)}\ri]^{2\ap p_iq}
\prod_{1\leq j\leq n_2}\lf[\fc{\theta_4(x_j'-i\sigma^2,\tau)}{\theta_1'(0,\tau)}\ri]^{2\ap p_j'q}\ .}
The latter assumes a similar structure than the pure open string part \wehaveE. Therefore, it is convenient to introduce the open string data
\eqn\scheffauer{\eqalign{
p_{n_1+1}&:=q\ ,\cr
x_{n_1+1}&:=i\sigma^2\ ,}}
with $n_1+n_2+1=n$ and extend \wehaveE\ to also include the factor \Kaiseralm:
\eqn\UntereWegscheidalm{\eqalign{
\E(\{x_i,x_j'\})&:=(-1)^{2\ap q\sum\limits_{i=1}^{n_1}p_i}\!\!\!\!\!\prod_{1\leq i<j\leq n_1+1}\lf|\fc{\theta_1(x_{j}-x_{i},\tau)}{\theta_1'(0,\tau)}\ri|^{2\ap p_ip_j}
\prod_{1\leq i<j\leq n_2}\lf|\fc{\theta_1(x_{j}'-x_{i}',\tau)}{\theta_1'(0,\tau)}\ri|^{2\ap p_i'p_j'}\cr
&\times\prod_{1\leq i\leq n_1+1\atop 1\leq j\leq n_2}\lf|\fc{\theta_4(x_{j}'-x_{i},\tau)}{\theta_1'(0,\tau)}\ri|^{2\ap p_ip_j'}\ .}}
With this rewriting our  total amplitude $A_{n_o,n_c}^{(1)}$  becomes:
\eqnn\ObereWegscheidalm{
$$\eqalignno{
A_{n-1,1}^{(1)}(&\Ac|\Bc||\{q_L\!\!=\!p_{n_1\!+\!1},q_R\!\!=\!0\})=ig_o^{n-1}g_c\ \delta^{(d)}\lf(\sum_{i=1}^{n_1+1}p_i+\sum_{j=1}^{n_2}p_j'\ri)&\ObereWegscheidalm\cr
&\times \int dt  \int_0^{t/2}d\sigma^2\left(\!\int_{\Ic_\Ac}\prod_{i=1}^{n_1}dx_{\al_i}\int_{\Ic_\Bc}\prod_{j=1}^{n_2}dx_{\bet_j'}\!\right) \lf.\E(\{x_i,x_j'\})\;  Q(\tau,\{x_i,x_j'\})\ri|_{x_{n_1+1}=i\sigma^2},}$$}
with the open string orderings $\Ac=\{1,\ldots,n_1\}$ and $\Bc=\{1,\ldots,n_2\}$ and the integration domains \domain.

Due to \Sonneck\ only the left--moving part of the closed string state contributes in the exponential factor \UntereWegscheidalm.
A typical setup leading to the expression \ObereWegscheidalm\ could be 
an amplitude involving $n-2$ gauge fields and one additional closed string gauge 
 or graviphoton field stemming from a compactification. The vertex operator of such a field is given by 
\eqn\vertex{
V_{A^a}(z,\ov z;\{\xi,q_L,q_R\})=g_c\ \xi_\mu\ \lf[ i\p X^\mu(z)+(q_L \psi)\psi^\mu(z) \ri]\   \tilde\Sigma^a(\ov z)\ e^{iq_L X_L(z)}\ e^{iq_R X_R(\ov z)}\ ,}
 with polarization vector $\xi$.  Subject to \Sonneck\ the right--moving part of \vertex\ contributes only through the current  $\tilde\Sigma^a(\ov z)$. The right--moving sector could describe an internal field within a 
string compactification at zero momentum. Then, only the zero modes of the latter account for a non--vanishing contribution  giving rise to some charge insertion $Q^a\sim\vev{:\!\!\tilde\Sigma^a(\ov z)\!\!:}$, which in turn  comprises into an overall scaling factor of \ObereWegscheidalm. The latter is non--vanishing for anomalous gauge fields or for some orbifold twists. 
Possible candidates for \vertex\ are Kaluza--Klein fields like NS graviphotons or gauge fields \ForgerTU.  Alternatively, a closed string state  propagating along the cylinder providing the contribution that cancels the anomaly could be considered. Furthermore, 
following the setup for collinear limits  \StiebergerLNG\  a dilaton or graviton insertion can be  accounted for by  \ObereWegscheidalm.

With the choice \Sonneck\ and \FIX\ the amplitude \ObereWegscheidalm\ is reminiscent of a pure one--loop $n$--point open string amplitude. More precisely, it corresponds to an object, which appears in the  open string (non--planar) one--loop monodromy relation \HoheneggerKQY.
E.g. for $n_1=2$ and $n_2=n-2$ the latter reads \HoheneggerKQY
\eqnn\MonoRel{
$$\eqalignno{
\h A^{(1)}_n(\{1,n\}|\{2,3,\ldots,n-1\})&+\hat A^{(1)}_n(\{1|n,2,\ldots,n-1\})&\MonoRel\cr
&=\lf(1-e^{-2\ap\pi i p_1p_n}\ri)\ B^{(1)}_n(\{1,n\}|\{2,3,\ldots,n-1\})\ ,}$$}
 involving subamplitudes of $n$ color ordered open strings. 
The objects entering \MonoRel\ are the non--planar subamplitudes $A^{(1)}_n(\{1,n\}|\{2,3,\ldots,n-1\})$  and $\hat A^{(1)}_n(\{1\}|\{n,2,\ldots,n-1\})$  and the boundary term $B^{(1)}_n(\{1,n\}|\{2,\ldots,n-1\})$ further specified in \HoheneggerKQY.
The latter has been first found, explicitly given and demonstrated to  be crucial\foot{The existence and importance of this term had been first shown, emphasized and proven in \HoheneggerKQY\ and then restated once again in \CasaliIHM, while this boundary term  (among other issues) had been completely overlooked in \TourkineBAK. Needless to say, it is actually tried to argue that the boundary term of \HoheneggerKQY\ was absent in a follow-up work by Ochirov and the authors of \TourkineBAK. Unfortunately, that section (Appendix B), where a proof is tried to be brought  suffers from several misprints and flaws. Therefore, it is not clear what actually is intended or tried to be proven by these authors.} to furnish and match the correct $\alpha'$--expansion of \MonoRel\ in Ref. \HoheneggerKQY. This boundary term $B^{(1)}_n$ is given by (with $n_2=n-2$):
\eqnn\BoundaryB{
$$\eqalignno{
B^{(1)}_n(\{1,n\}|\{2,\ldots,n-1\})&=ig_o^n\delta^{(d)}\lf(p_1+p_n+\sum_{j=1}^{n_2}p_j\ri)
\int dt\  V_{CKG}^{-1}\ \left(\int_0^{t/2}d\sigma^2\int_{\Ic_\Ac} dx_1\right) \cr 
&\times\left(\int_{\Ic_\Bc}\prod_{j=2}^{n-1}dx_j\right)Q(\tau,\{x_i\})
\prod_{2\leq i<j\leq n-1}\lf|\fc{\theta_1(x_{j}-x_{i},\tau)}{\theta_1'(0,\tau)}\ri|^{2\ap p_ip_j}\cr
&\times\lf. \lf|\fc{\theta_1(x_{n}-x_{1},\tau)}{\theta_1'(0,\tau)}\ri|^{2\ap p_1p_n}\hskip-0.3cm\prod_{i=1,n\atop 2\leq j\leq n-1}\lf|\fc{\theta_4(x_{j}-x_{i},\tau)}{\theta_1'(0,\tau)}\ri|^{2\ap p_ip_j}\ri|_{x_n=i\sigma^2}\hskip-0.9cm.&\BoundaryB}$$}
In \BoundaryB\ the open string position $x_n$ of the $n$--th gluon  is integrated along the line $\sigma^1\!=\!0$ ($\sigma^1\!=\!1$) where the cylinder is sliced. Note that all other $n\!-\!1$ open string positions $x_1,\ldots,x_{n-1}$ are respectively integrated  along one of the two boundaries of the cylinder subject to the conformal Killing symmetry.
It has already been emphasized  in  \HoheneggerKQY, that from the point of view of pure open string one--loop amplitudes the object \BoundaryB\ 
seems to be unphysical within the set of open string on--shell amplitudes.
However, after carefully inspecting \BoundaryB\ and comparing with \ObereWegscheidalm\ we evidence that (with $g_c=g_o^2$)
\eqn\Wehave{
 g_o\ V_{CKG}\ \times B^{(1)}_n(\{1,n\}|\{2,\ldots,n-1\})\simeq 
 A_{n-1,1}^{(1)}(\Ac|\Bc||\{q_L\!\!=\!p_{n},q_R\!\!=\!0\})\ ,}
with the orderings $\Ac=\{1\}$ and $\Bc=\{2,\ldots,n-1\}$.
Therefore, by extending the setup to the closed string sector we will be able to relate the object \BoundaryB\ to a one--loop amplitude involving $n\!-\!1$ open strings and one closed string.
In other words, one of the open strings mutates into a closed string state when it is integrated along the line $\sigma^1\!=\!0$ ($\sigma^1\!=\!1$).

Note that according to \traithenu\ the amplitude $A_{n-1,1}^{(1)}$ in \Wehave\  corresponds to a linear combination of $n+1$--point open string amplitudes, while $B_n^{(1)}(\{1,n\}|\{2,\ldots,n-1\})$ involves only $n$ open strings. The additional factors $V_{CKG}\simeq\tau_2$ and $g_o$ account for this mismatch in the number of integrated vertex operator insertions. 
As discussed above, for a particular choice of a closed string state with the constraint \Sonneck\ the right--moving part of the closed string completely decouples and gives rise to a zero mode contribution denoted by $Q^a$. As a result the $n+1$--point open string amplitudes appear rather like $n$--point open string amplitudes with one additional  state decoupled and with its  vertex position fixed along the real line at 
a marked point $x_0$ on the cylinder, where the latter is sliced, i.e.:
\eqn\marked{
x_0:=\sigma^1\equiv 0\ .}
With this in mind we may alternatively  write \Wehave\ as
\eqn\wehave{
B^{(1)}_{x_0}(\{1,n\}|\{2,\ldots,n-1\})=A_{n-1,1}^{(1)}(\Ac|\Bc||\{q_L\!\!=\!p_{n},q_R\!\!=\!0\}) \ ,}
with $B^{(1)}_{x_0}$ defined as the extension of the subamplitude $B_n^{(1)}$ to also include one additional open string state completely decoupled from the remaining fields 
and with its  vertex position $x_0$ fixed as \marked:
\eqn\Sonnenstein{
B^{(1)}_{x_0}(\{1,n\}|\{2,\ldots,n-1\})\simeq  g_o\ V_{CKG}\ \times B_n^{(1)}(\{1,n\}|\{2,\ldots,n-1\})\ .}
Eventually, in   \MonoRel\  we may modify all open string subamplitudes by including the marked point $x_0$. The latter represents an additional vertex operator position describing an open string state, which does not interact with the remaining open strings
(e.g. gluon states). Since it is supposed to decouple from the latter the structure of the 
 open string one--loop monodromy relation \MonoRel\ does not alter: 
\eqn\MonoRelx{\eqalign{
\h\ A^{(1)}_{x_0}(\{n,1\}|&\{2,3,\ldots,n-1\})+\hat A^{(1)}_{x_0}(\{1\}|\{2,\ldots,n\})\cr
&=[1-(-1)^{s_{1n}}]\ B^{(1)}_{x_0}(\{n,1\}|\{2,3,\ldots,n-1\})\ .}}
Due to the conformal Killing symmetry except the marked point $x_0$ the remaining $n$ vertex operator positions are integrated.

The additional marked point $x_0$ can be interpreted as the point \marked\ where the cylinder is sliced. Therefore, it could be interesting to clarify whether 
the extra marked point $x_0$ can be related to the auxiliary point, which allows the recursive calculation of genus--one integrals using an extra marked point in a differential equation \BroedelGBA.
We have presented a physical interpretation \Wehave\ of the boundary term $B_n^{(1)}$ appearing in the open string one--loop monodromy relations \MonoRel\ in terms of a string amplitude $A_{n-1,1}^{(1)}$ involving a closed string state. 
The lowest $\alpha'$ order of $B_n^{(1)}$ has been first derived and shown to match \MonoRel\ in \HoheneggerKQY. Recently, in \CasaliKNC\ this field--theory limit has been revisited and it is argued that in this limit the boundary term $B_n^{(1)}$ can be attributed to field theory graphs with contact--terms.  As a consequence the latter should be due to replacing one open string by a  closed string state.
This could give a physical interpretation in string perturbation theory of these
graphs with contact--terms.

%% \subsec{Relations to one--loop open string monodromy relations}

%%  Let us consider $n$ closed strings on the cylinder with Neumann boundary conditions.
%%  Considering the contour \edges\ yields:
%%  \eqn\monoDR{
%% int_0^1d\tilde \sigma_t^2\ [\ \hat I(\sigma^1,-i\tilde\sigma^2)-\hat I'(\sigma^1,-
%% i\tilde\sigma^2)\ ]=0\ .}

%%%%%%%%%%%%%%%%%%%%%%%%%%%%%%%%
\newsec{Concluding remarks}

In this work we have expressed one--loop string scattering amplitudes involving both open and closed strings as sum over pure open string amplitudes. More precisely, one--loop cylinder amplitudes
of $n_o$ open and $n_c$ closed strings can be written as pure open string cylinder amplitudes involving only $n_0+2n_c$ open strings.
These findings generalize the analogous tree--level result to higher loops and extend the 
tree--level observation that in gravitational amplitudes a graviton can be traded for  
two gluons.

Our results are derived from analytic continuation of closed string coordinates on the cylinder world--sheet. Splitting the complex cylinder coordinate into a pair
of two real coordinates gives rise to open string positions located along the two cylinder boundaries
subject to some monodromy phases encoded in the objects \kernetr, \kernelI\ and \kerneli\ (cf. also \kernell\ and \kernel). The latter depend on the loop momentum $\ell$ which flows from one cylinder boundary to the other, cf. Fig.~7. 
The definitions \kernetr, \kernelI\ and \kerneli\ are known from intersection theory for twisted cycles.
It is attributed to the contribution of a single edge, which in turn can be interpreted as 
propagator \MizeraCQS\ entering the inverse KLT kernel.
Hence, it is a central building block of a  one--loop analog of the tree--level KLT kernel \KawaiXQ, see also \doubref\BernSV\BjerrumBohrHN. 
At tree--level the KLT kernel is related to topological features of twisted
homology encoded in intersection numbers of twisted cycles \MizeraCQS.
More precisely, at tree--level the KLT relation  follows from the twisted Riemann period relations, which involves a certain paring of twisted period matrices.
At present, it is not clear how such relations extent on higher Riemann surfaces. 
For $n_c\!=\!1$ our amplitude results \traithen\ and \Traithen\ 
 can be related to the one--loop open string monodromy relations. The latter contain a boundary term \BoundaryB, which is related to non--physical contours on the cylinder. A physical interpretation of the latter in terms of a closed string insertion is attributed in Section 4 by enlarging
the space of basis functions beyond pure open string amplitudes. In fact, thanks to \Wehave\ that boundary term, which is related to non--physical contours, has a physical interpretation as open string amplitude with one additional closed string insertion. Building on \CasaliIHM\ the kernels \kerntri\ (as well as \kernetr, \kernelI\ and \kerneli)  may also play an important r\^ole in intersection theory and  generalized twisted cohomology theory for one--loop string amplitudes.
This way our approach paves the way for a one--loop generalization of the KLT relations.
Further details will be worked out elsewhere.

The cylinder amplitudes \Traithen\ involving one closed string and $n_o$ open strings is written as sum over cylinder amplitudes involving $n_o+2$ open strings.
Similarly, the cylinder amplitudes \Scheffau\ involving $n$ closed strings is written as sum over
cylinder amplitudes involving $2n$ open strings. As a consequence in the field--theory limit these expressions give rise to one--loop EYM amplitudes expressed in terms of pure gluon amplitudes. If the closed strings represent  
graviton states the amplitude result \Scheffau\ expresses the $n$--point one--loop  gravitational  amplitude  as a sum over $2n$--point one--loop gluon amplitudes (in the presence of a D--brane). We can compute its field--theory limit to gain
$n$ graviton one--loop amplitudes expressed in terms of $2n$ gluon amplitudes
in EYM theory. This step provides powerful relations between gravitational and gauge amplitudes at one--loop  generalizing the tree--level results  \StiebergerCEA.
Furthermore, these relations give identities  between gravitational and gauge amplitudes at one--loop level generalizing the all--plus or single--minus helicity gravitational amplitudes \BernSV\ or going beyond rational expressions \NandanODY.

%% The one--loop cylinder scattering of $n_c$ closed strings in the presence of D--branes has %% the interpretation of tree--level closed string scattering of $n_c$ closed strings and two %% off--shell states. In \posORTH and \chargeORTH\ introduces a transverse loop momentum which %%  upon integration provides a string theory setup of the findings \CachazoAOL. 

Finally,  our one--loop results open up to essentially compute any  class of one--loop amplitudes involving closed and open strings in string compactifications with D--branes and orientifolds \BlumenhagenCI\ and express them in terms of pure open string amplitudes. Furthermore, building up on \AntoniadisVW\ various phenomenological interesting connections between one--loop couplings of closed and open strings can be derived.

%%%%%%%%%%%%%%%%%%%%%%%%%%%%%%%%

%%%%%%%%%%%%%%%%%%%%%%%%%%%%%%%%
\bigskip
\noindent
{\it Acknowledgments:} I wish to thank Johannes Broedel for 
%  motivating , interesting, inspiring
stimulating discussions during the startup of this project.
Furthermore, I am indebted to The Kolleg Mathematik Physik Berlin of
Humboldt Universit\"at for warm hospitality and financial support.

\appendix\appA{Open string Green's functions}

\subsec{Open string Green's function on the disk}

The open string world--sheet coordinate $w$ is parameterized by $w=w_1+i w_2$.
The scalar Green function with
\eqn\DIFFGreen{
\square G(w,w')=-2\pi \ap\ \delta(w-w')\ ,}
on the disk  is given in \refs{\NeveuIQ,\HsueRA,\AbouelsaoodGD}:
\eqn\diskH{
G^{(0)}(w_1,w_2)=\ln|w_1-w_2|\pm\ln|w_1-\ov w_2|\ ,}
with the plus sign referring to Neumann boundary conditions $\lf.\p_{\sigma}X\ri|_{\sigma=0}=0$ for the open string ends $\sigma=0,\pi$ and the minus sign referring to Dirichlet boundary conditions 
$\lf.\p_{\tau}X\ri|_{\sigma=0}=0$, respectively. 
After the conformal transformation
\eqn\transrho{
\rho=\fc{w+i}{w-i}\ ,\ {\rm i.e.:}\ \ \  w=-i\ \fc{1+\rho}{1-\rho}}
from the upper half plane $w\in {\bf H}_+$ to the disk $\rho\in{\bf D}_2$
\diskH\ becomes:
\eqn\diskD{
G^{(0)}(\rho_1,\rho_2)=\ln|\rho_1-\rho_2|\pm\ln|1-\rho_1\ov \rho_2|\ ,}
up to the universal term $-2\ln|\h(1-\rho_1)(1-\rho_2)|^2$ in the Neumann case, which drops out of all expectation values. The latter can be put\foot{On the other hand, with the Green's function $\ds{G^{(0)}(z_1,z_2)=\ln\fc{|z_1-z_2|^2}{(1+|z_1|^2)(1+|z_2|)^2}}$ given 
in~\DHokerPDL\ we arrive at $\ds{G^{(0)}(\rho_1,\rho_2)=\ln\fc{|\rho_1-\rho_2|^2}{(1+|\rho_1|^2)(1+|\rho_2|)^2}}$ .} into vertex operator normalizations.

\subsec{Open string Green's function on the annulus}

The world--sheet annulus with the inner radius $a=e^{-2\pi \lambda_1}$ and the outer radius 
$b =e^{2\pi  \lambda_2}$ (with $b>a$) is parameterized by the annulus coordinate $\rho'$ 
\eqn\defrho{
\rho'=e^{2\pi i z'}\ ,\  z'={\sigma_1}'+i{\sigma_2}'\ ,}
with $({\sigma^1}',{\sigma^2}')\in [0,1]\times[-\lambda_2,+\lambda_1]$ and $a\leq|\rho'|\leq b$. 
The flat annulus is related by Weyl scaling to a
cylinder of length $\lambda=\lambda_1+\lambda_2$ and unit circumference with a flat metric,
cf. Fig.~9. 
\iifig\Annulus{Rectangular cylinder with coordinates $z'={\sigma_1}'+i{\sigma_2}'$ mapped to an annulus}{with inner radius $a$ and outer radius $b$ and coordinate $\rho'=e^{2\pi i(\sigma_1'+i\sigma_2')}$.}{\epsfxsize=0.9\hsize\epsfbox{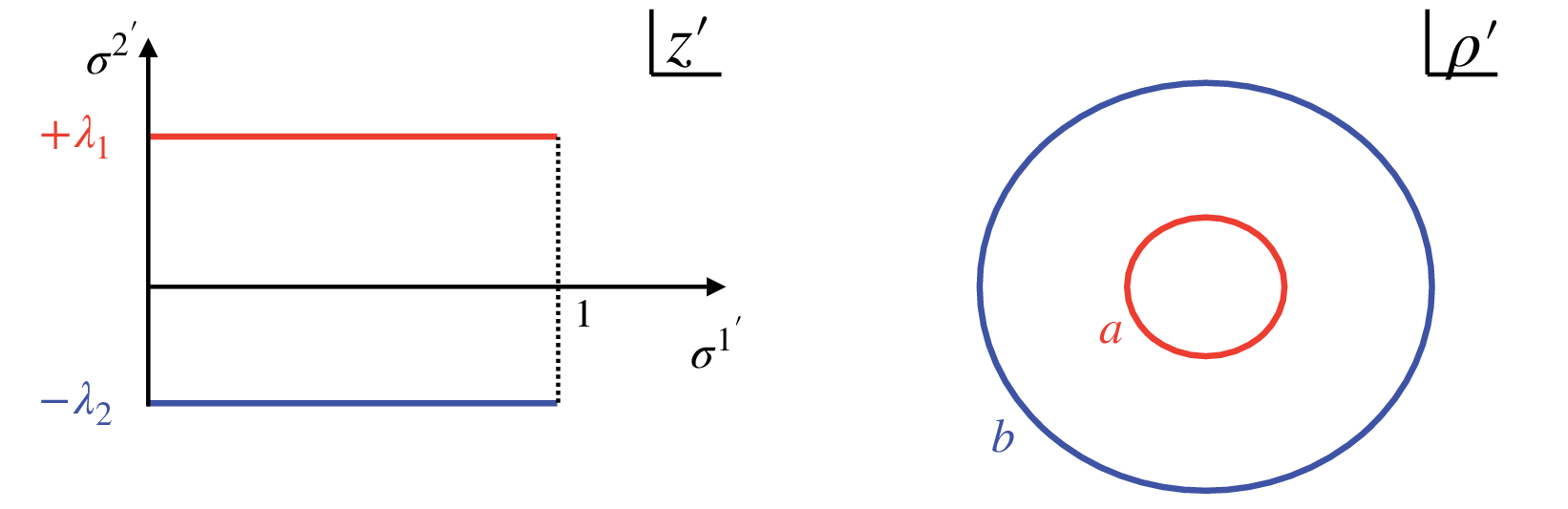}}
\noindent 
The modulus of the annulus is given by $q:=\fc{a}{b}=e^{-2\pi \lambda}$. The Neumann boundary conditions translate into conditions
 at the two boundaries $|\rho|=a,b$. In the following we shall work with the coordinate $\rho:=b^{-1}\rho'$, with $\rho=e^{2\pi i z}$ and $z=\sigma^1+i\sigma^2$ i.e. $0\leq\sigma^2\leq\lambda$ and $q\leq |\rho|\leq 1$.
In terms of these coordinate with \BOSG\ the scalar Green function on  the annulus is given in~\refs{\NeveuIQ,\HsueRA,\AbouelsaoodGD}
\eqn\cyclinderA{
G^{(1)}(z_1,z_2)=G^{(1)}_B(z_1,z_2)\pm G^{(1)}_B(z_1,\ov z_2)+\ln|b|\ ,}
up to the universal term $2\ln 2\pi+\ln|\rho_1\rho_2|+2\ln\prod\limits_{n=1}^\infty(1-q^n)^2$ in the Neumann case, which drops out from all expectation values.

In the limit $q\ra 0$, i.e. $\lambda\ra  \infty$ the expressions \cyclinderA\ reproduce the tree--level Green's functions \diskD\
\eqn\cyclinderAlimit{\eqalign{
G^{(1)}(z_1,z_2)&= \ln|\rho_1-\rho_2|+\ln|1-b^{-2}\rho_1\ov \rho_2|-\fc{2\pi}{\tau_2} (\sigma_1^2)^2-\fc{2\pi}{\tau_2}(\sigma_2^2)^2+\Oc(q)\ ,\cr
G^{(1)}(z_1,z_2)&=\ln|\rho_1-\rho_2|-\ln|1-b^{-2}\rho_1\ov \rho_2|+\fc{4\pi}{\tau_2} \sigma_1^2\sigma^2_2 +\Oc(q)\ ,}}
respectively.

\appendix\appB{Jacobi theta functions and  identities}

A meromorphic function $f(z,\tau)$ of $z$ is defined to be elliptic if it is doubly periodic on a torus, i.e. $f(z+1,\tau)=f(z,\tau)$ and $f(z+\tau,\tau)=f(z,\tau)$.
Elliptic functions having a prescribed set of zeroes and poles in a period parallelogram can be constructed as quotients of Jacobi theta functions.
In this appendix we collect some useful identities for Jacobi theta function.

In the computation of one--loop string amplitudes we encounter Jacobi theta functions which are defined as 
\eqn\DefTheta{
\theta\big[^a_b\big](z,\tau)=\sum_{n=-\infty}^\infty e^{\pi i\left(n-\fc{a}{2}\right)^2\tau}\,e^{2\pi i\left(z-\fc{b}{2}\right)\left(n-\fc{a}{2}\right)}\ ,\ \ \ {\rm for}\ \ a,b=0,1\ .}
Throughout the text we will use the notation
\eqn\Thetaone{\eqalign{
\theta_1(z,\tau)&=\theta\big[^1_1\big](z,\tau)\ ,\ \ \ 
\theta_2(z,\tau)=\theta\big[^1_0\big](z,\tau)\ ,\cr
\theta_3(z,\tau)&=\theta\big[^0_0\big](z,\tau)\ ,\ \ \ 
\theta_4(z,\tau)=\theta\big[^0_1\big](z,\tau)\ .}}
Sometimes, the following representation is useful:
\eqn\Thetaone{\eqalign{
\theta_1(z,\tau)&=2\ e^{\fc{\pi i\tau}{4}}\sum_{n=0}^\infty(-1)^n\ e^{\pi i\tau n(n+1)}\ \sin[\pi (2n+1) z]\ ,\cr
\theta_2(z,\tau)&=2\ e^{\fc{\pi i\tau}{4}}\sum_{n=0}^\infty e^{\pi i\tau n(n+1)}\ \cos[\pi (2n+1) z]\ ,\cr
\theta_3(z,\tau)&=1+2\sum_{n=0}^\infty e^{\pi i\tau n^2}\ \cos(2\pi n z)\ ,\cr
\theta_4(z,\tau)&=1+2\sum_{n=0}^\infty (-1)^n\ e^{\pi i\tau n^2}\ \cos(2\pi n z)\ .}}
Under the modular transformation $(z,\tau)\rightarrow \left(\fc{z}{\tau},-\fc{1}{\tau}\right)$, the theta-function $\theta\big[^a_b\big](z,\tau)$ transforms in the following way
\eqn\TrafoTheta{
\theta\big[^a_b\big]\left(\fc{z}{\tau},-\fc{1}{\tau}\right)=\sqrt{-i\tau}\,e^{\fc{i\pi}{2}\,ab+i\pi\,\fc{z^2}{\tau}}\,\theta\big[^{\ b}_{-a}\big](z,\tau)\ .}
The derivative of the theta function $\theta[^1_1\big](z,\tau)=\theta_1(z,\tau)$ with respect to the first argument can be related to the Dedekind eta function as
\eqn\Dedekind{
\theta'_1(0,it)=2\pi\,\eta^3(it)\ ,}
where
\eqn\ProductDede{
\eta(\tau)=q^{\fc{1}{24}}\ \prod_{n=1}^\infty(1-q^n)\ ,\ \ \ q=e^{2\pi i\tau}\ ,}
which transforms as:
\eqn\TrafoEta{
\eta(-1/\tau)=\sqrt{-i\tau}\,\eta(\tau)\ .}

\def\frac{\fc}
Furthermore, under shifts of the first argument, the Jacobi theta functions transform as \MOS
\eqn\IdentityShiftTheta{
\theta\big[^a_b\big]\left(z+\frac{\epsilon_1}{2}\,\tau+\frac{\epsilon_2}{2},\tau\right)=e^{-\frac{i\pi\tau}{4}\,\epsilon_1^2-\frac{i\pi\epsilon_1}{2}(2z-b)-\frac{i\pi}{2}\,\epsilon_1\epsilon_2}\,\theta\big[^{a-\epsilon_1}_{b-\epsilon_2}\big](z,\tau)\ .}
With $\theta\big[^a_b\big](z\pm\h,\tau)=\theta\big[^a_{b\mp 1}\big]=e^{\mp i\pi a} \theta\big[^a_{b\pm 1}\big]\theta(z,\tau)$ this leads to:
\eqn\halfshifttheta{\matrix{
&\theta_1(z+\h,\tau)=\theta_2(z,\tau), 
&\theta_1(z-\h,\tau)=e^{+ i\pi}\ \theta_2(z,\tau)\ ,\cr
&\theta_2(z+\h,\tau)=e^{- i\pi}\ \theta_1(z,\tau),
&\theta_2(z-\h,\tau)=\theta_1(z,\tau)\ .}}
In particular, we have 
$\theta\big[^a_b\big](z\pm1,\tau)=e^{\mp i\pi a} \theta\big[^a_b\big]$, i.e.:
\eqn\thetashift{\matrix{
&\theta_1(z\pm1,\tau)=e^{\mp i\pi}\ \theta_1(z,\tau),
&\theta_2(z\pm 1,\tau)=e^{\mp i\pi}\ \theta_2(z,\tau)\ ,\cr
&\theta_3(z\pm 1,\tau)=\theta_4(z,\tau),
&\theta_4(z\pm 1,\tau)=\theta_4(z,\tau)\ ,}}
Furthermore, with $\theta\big[^a_b\big](z\pm\fc{\tau}{2},\tau)=e^{-\fc{\pi i\tau}{4}}e^{\mp i\pi (z-\fc{b}{2})} \theta\big[{a\mp 1\atop b}\big]$, we have:
\eqn\shifttheta{\eqalign{
\theta_1(z\pm \fc{\tau}{2},\tau)&=e^{-\fc{i\pi\tau}{4}}\ e^{\mp i\pi z}\ e^{\pm\fc{i\pi}{2}}\theta_4(z,\tau)\ ,\cr
\theta_4(z\pm \fc{\tau}{2},\tau)&=e^{-\fc{i\pi\tau}{4}}\ e^{\mp i\pi z}\ e^{\pm\fc{i\pi}{2}}\theta_1(z,\tau)\ ,}}
while with $\theta\big[^a_b\big](z\pm\tau,\tau)=e^{-\pi i\tau}e^{\mp i\pi (2z-b)} \theta\big[{a\mp 2\atop b}\big]$, we obtain:
\eqn\shiftthetaa{\eqalign{
\theta_1(z\pm \tau,\tau)&=e^{-i\pi\tau}\ e^{\mp 2i\pi z}\ e^{\pm i\pi}\theta_1(z,\tau)\ ,\cr
\theta_4(z\pm \tau,\tau)&=e^{-i\pi\tau}\ e^{\mp 2i\pi z}\ e^{\pm i\pi}\theta_4(z,\tau)\ .}}
Finally, for $\tau=i\tau_2$ we can write $\bar\theta\big[^a_b\big](\bar z,\bar\tau)=\theta\big[^{-a}_{\ b}\big](\bar z,\tau)$, i.e.:
\eqn\complextheta{\eqalign{
\bar\theta_1(\bar z,\bar\tau)&=\theta_1(\bar z,\tau)\ ,\ 
\bar\theta_2(\bar z,\bar\tau)=\theta_2(\bar z,\tau)\ ,\cr
 \bar\theta_3(\bar z,\bar\tau)&=\theta_3(\bar z,\tau)\ ,\ 
 \bar\theta_4(\bar z,\bar\tau)=\theta_4(\bar z,\tau)\ .}}

%%%%%%%%%%%%%%%%%%%%%%%%%%%%%%%%%%%%%%%%%%%%%%%%%%%%%%%%%%
\listrefs
\end